\documentclass[a4paper,12pt]{article}
\pdfoutput=1
\usepackage{geometry}
\usepackage{amsmath,amssymb,amsfonts}
\usepackage{xcolor,graphicx,cite,soul}
\usepackage{array,booktabs,longtable}
\usepackage{caption,microtype}
\usepackage{subcaption}
\usepackage{hyperref}
\usepackage{datetime}
\usepackage{appendix}

\newcommand{\lsim}
{\;\raisebox{-.3em}{$\stackrel{\displaystyle <}{\sim}$}\;}
\newcommand{\gsim}
{\;\raisebox{-.3em}{$\stackrel{\displaystyle >}{\sim}$}\;}

\newcommand\al{\alpha}
\newcommand\be{\beta}
\newcommand\tb{\tan\beta}

\newcommand\CBA{c_{\beta - \alpha}}
\newcommand\SBA{s_{\beta - \alpha}}

\newcommand\ReDiag{\mathop{%
  \raise .5pt\hbox{[}%
  \widetilde{\mathrm{Re}}%
  \raise .5pt\hbox{]}}}
\newcommand\ReOffDiag{\mathop{%
  \raise .5pt\hbox{$\llbracket$}%
  \widetilde{\mathrm{Re}}%
  \raise .5pt\hbox{$\rrbracket$}}}

\newcommand\SW{s_\mathrm{w}}
\newcommand\CW{c_\mathrm{w}}
\newcommand\MW{m_W}
\newcommand\MZ{m_Z}
\newcommand\Mh{m_h}
\newcommand\MH{m_H}
\newcommand\MA{m_A}
\newcommand\MHp{m_{H^\pm}}
\newcommand\mbar{\bar{m}^2}

\newcommand\msq{m_{12}^{2}}

\newcommand\refeq[1]{Eq.~(\ref{#1})}

\newcommand\refta[1]{Tab.~\ref{#1}}
\newcommand\refse[1]{Sect.~\ref{#1}}
\newcommand\refses[1]{Sects.~\ref{#1}}
\newcommand\citere[1]{Ref.~\cite{#1}}

\newcommand{\CP}{{\cal CP}}
\newcommand{\cp}{{\CP}}

\newcommand{\tev}{\,\, \mathrm{TeV}}
\newcommand{\gev}{\,\, \mathrm{GeV}}

\newcommand\HB{\texttt{HiggsBounds}}
\newcommand\HS{\texttt{HiggsSignals}}

\newcommand{\br}{\text{BR}}
\newcommand{\De}{\Delta}

\newcommand{\sig}{\sigma}

\def\order#1{\ensuremath{{\cal O}(#1)}}
\def\reffi#1{\mbox{Fig.~\ref{#1}}}

\def\ga{\gamma}
\def\de{\delta}
\def\la{\lambda}
\newcommand\kala{\ensuremath{\kappa_{\lambda}}}
\newcommand\laSM{\ensuremath{\lambda_{\mathrm{SM}}}}
\newcommand{\lahhh}{\ensuremath{\la_{hhh}}}
\newcommand{\lahhH}{\ensuremath{\la_{hhH}}}
\newcommand{\lahHH}{\ensuremath{\la_{hHH}}}
\newcommand{\lahAA}{\ensuremath{\la_{hAA}}}
\newcommand{\lahHpHm}{\ensuremath{\la_{hH^+H^-}}}

\newcommand{\inter}[2]{\ensuremath{[#1, #2]}}

\definecolor{Orange}{named}{orange}
\definecolor{Purple}{named}{purple}
\definecolor{Lightblue}{cmyk}{0.9,0.1,0.1,0.3}
\definecolor{dgelborange}{cmyk}{0.,0.3,0.5, 0.}
\definecolor{Lila}{rgb}{0.5,0.,1}
\definecolor{Darkgreen}{rgb}{0.,.7,0.2}

\graphicspath{{figs/}}
\allowdisplaybreaks
\begin{document}
\thispagestyle{empty}

\def\thefootnote{\fnsymbol{footnote}}

\begin{flushright}
\mbox{}
IFT--UAM/CSIC-20-30\\
FTUAM-20-3
\end{flushright}

\vspace{0.5cm}

\begin{center}

{\large\sc 
{\bf Exploring sizable triple Higgs couplings in the 2HDM}}\\

\vspace{1cm}

{\sc
F.~Arco$^{1,2}$%
\footnote{email: Francisco.Arco@uam.es}%
, S.~Heinemeyer$^{2,3,4}$%
\footnote{email: Sven.Heinemeyer@cern.ch}%
~and M.J.~Herrero$^{1,2}$%
\footnote{email: Maria.Herrero@uam.es}%
}

\vspace*{.7cm}

{\sl
$^1$Departamento de F\'isica Te\'orica, 
Universidad Aut\'onoma de Madrid, \\ 
Cantoblanco, 28049, Madrid, Spain

\vspace*{0.1cm}

$^2$Instituto de F\'isica Te\'orica (UAM/CSIC), 
Universidad Aut\'onoma de Madrid, \\ 
Cantoblanco, 28049, Madrid, Spain

\vspace*{0.1cm}

$^3$Campus of International Excellence UAM+CSIC, 
Cantoblanco, 28049, Madrid, Spain 

\vspace*{0.1cm}

$^4$Instituto de F\'isica de Cantabria (CSIC-UC), 
39005, Santander, Spain

}

\end{center}

\vspace*{0.1cm}

\begin{abstract}
\noindent
An important task at future colliders is the measurement of the triple Higgs
coupling. Depending on its size relative to the Standard Model (SM) value,
certain collider options result in a higher experimental accuracy. 
Within the framework of Two Higgs Doublet Models (2HDM) type~I and~II we
investigate the allowed ranges for all triple Higgs couplings involving at
least one light, SM-like Higgs boson. We take into account theoretical
constraints (unitarity, stability), experimental constraints from
direct Higgs-boson searches, measurements of the SM-like Higgs-boson
properties, flavor observables and electroweak precision data. We find
that the SM-type triple Higgs coupling w.r.t.\ its SM value,
$\lahhh/\laSM$, can range between $\sim -0.5$ and $\sim 1.5$. Depending on
which value is realized, the HL-LHC can compete with, or is clearly inferior
to the ILC.
We find the coupling $\lahhH$ between $\sim -1.5$ and $\sim 1.5$.
Triple Higgs couplings involving two heavy Higgs bosons, $\lahHH$,
$\lahAA$ and $\lahHpHm$ can reach values up to \order{10}, roughly
independent of the 2HDM type.   
This can lead to potentially strongly enhanced production of two
Higgs-bosons at the HL-LHC or high-energy $e^+e^-$ colliders.
\end{abstract}


\def\thefootnote{\arabic{footnote}}
\setcounter{page}{0}
\setcounter{footnote}{0}

\newpage


\section{Introduction}
\label{sec:intro}

In 2012 the ATLAS and CMS collaborations have discovered a new
particle that -- within theoretical and experimental uncertainties -- is
consistent with the existence of a Standard-Model~(SM) Higgs boson at a mass
of~$\sim 125 \gev$~\cite{Aad:2012tfa,Chatrchyan:2012xdj,Khachatryan:2016vau}.
No conclusive signs of physics beyond the~SM have been found so far at the LHC.
However, the measurements of Higgs-boson couplings, which are known
experimentally to a precision of roughly $\sim 20\%$, leave room for
Beyond Standard-Model (BSM) interpretations. Many BSM models possess
extended Higgs-boson sectors. Consequently, one of the main tasks of the
LHC Run~III and beyond is to determine whether 
the observed scalar boson forms part of the Higgs sector of an extended
model.

A key element in the investigation of the Higgs-boson sector is the
measurement of the trilinear Higgs coupling of the SM-like Higgs boson,
\lahhh\  (for recent reviews on Higgs couplings measurements
  at future colliders see \cite{deBlas:2019rxi, DiMicco:2019ngk}). In
the case of a BSM Higgs-boson sector, equally important is the
measurement of BSM 
trilinear Higgs-boson couplings. Most experimental studies assume the SM value
of \lahhh. However, in BSM models this coupling may differ significantly
from its SM value. The expected achievable precision at different future
colliders in the measurement of \lahhh\ depends on the value realized in
nature. 

A natural extension of the Higgs-boson sector of the SM is the ``Two Higgs
Doublet Model'' (2HDM) (for reviews see,
e.g.,~\cite{Gunion:1989we,Aoki:2009ha, Branco:2011iw}). This model
contains five physical Higgs bosons: the 
light and the heavy $\CP$-even $h$ and $H$, the $\CP$-odd $A$, and the pair of
charged Higgs bosons, $H^\pm$.
The mixing angle $\al$ ($\be$) diagonalizes the $\CP$-even (-odd)
Higgs bosons and $\tb$ is given by the ratio of the two vacuum expectation values, $\tb := v_2/v_1$.
We assume for this work that the
light $\CP$-even Higgs-boson $h$ is SM-like with a mass of 
$\Mh \sim 125 \gev$. All other Higgs bosons are assumed to be heavier.
To avoid flavor changing neutral currents at
tree-level, a $Z_2$~symmetry is imposed~\cite{Glashow:1976nt},
possibly softly broken by the parameter $\msq$.
Depending on how this symmetry is extended to the fermion sector, four
types of the 2HDM can be realized: 
type~I and~II, lepton specific and flipped~\cite{Aoki:2009ha}.
In the 2HDM also the stability conditions for the Higgs
potential change with respect to the SM~\cite{Deshpande:1977rw} (for a
review see~\cite{Bhattacharyya:2015nca}).   

In this paper, focusing on the 2HDM type~I and~II, we
investigate the allowed ranges for all triple Higgs couplings involving at
least one light, SM-like Higgs boson. Specifically: \lahhh, 
\lahhH, \lahHH, \lahAA\ and
\lahHpHm.
The allowed ranges are obtained taking
into account: theoretical 
constraints from unitarity and stability (we use
\cite{Bhattacharyya:2015nca,Akeroyd:2000wc,Barroso:2013awa}),
experimental constraints from direct Higgs-boson searches (we use
\HB~\cite{Bechtle:2008jh,Bechtle:2011sb,Bechtle:2013wla,Bechtle:2015pma},
with data from
\cite{Aaboud:2018sfw,Aad:2019uzh,Aaboud:2018bun,CMS:2019yat,CMS:2018xvc,CMS:ril,ATLAS:2016ixk,CMS:2017epy,CMS:xwa}),
the experimental production and decay rates of the SM-like Higgs
boson (we use \HS~\cite{Bechtle:2013xfa,Bechtle:2014ewa}, where the
experimental data is listed in\cite{higgssignals-www}),
flavor observables (we use
\texttt{SuperIso}~\cite{Mahmoudi:2008tp,Mahmoudi:2009zz},
complemented with~\cite{Li:2014fea,Cheng:2015yfu,Arnan:2017lxi} and data from
\cite{Chen:2001fja,Aubert:2007my,
  Limosani:2009qg,Lees:2012ym,Lees:2012wg,Saito:2014das,Chatrchyan:2013bka,
  Aaltonen:2013as,Aaij:2017vnw,Aaboud:2018mst,Tanabashi:2018oca})  
and electroweak precision observables (EWPO) (we use~$S$, $T$
and~$U$~\cite{Peskin:1990zt,Peskin:1991sw}, complemented with
\cite{Grimus:2007if,Funk:2011ad} and bounds from
\cite{Tanabashi:2018oca}).  
Besides the allowed ranges, in this work we also present a detailed
study of the dependence of the triple Higgs couplings
on the free parameters of the model (to explore the 2HDM parameter space we use \texttt{2HDMC}~\cite{Eriksson:2009ws}).
The main interest in the allowed ranges for the triple Higgs
couplings is that 
they may affect the rates of multiple Higgs boson production at
current and future
colliders. In particular, the production  of Higgs boson pairs like
$hh$, $hH$, $HH$, $hA$, $hH^\pm$, $AA$ and $H^+H^-$ could be
significantly affected by the presence of sizable triple Higgs couplings
within the 2HDM, yet allowed by the present constraints. 

One of the key points of our study when exploring the
parameter space of the 2HDM type~I and~II under the given constraints
is the following: the primary focus of our explorations was to
find allowed parameters that lead to either
large non-SM triple Higgs boson couplings,
or to large deviations from unity in the ratio of the light
triple Higgs-boson coupling w.r.t.\ its SM value, $\lahhh/\laSM$. In
particular, we have explored scenarios with relatively heavy
masses $\MH$, $\MA$ and $\MHp$ near 1~TeV,
but not enforcing the so-called
\textit{alignment limit}, $\cos(\be-\al) \to 0$ (see,
e.g.,~\cite{Bernon:2015qea}).
Furthermore, we have investigated the dependences of the allowed
triple Higgs couplings on
the soft $Z_2$-breaking parameter $\msq$. As we will see, $\msq$
plays a very important role in our search of sizable triple Higgs
couplings. Finding a way to obtain large values for $\msq$,
still being allowed by experimental data and by
the theoretical constraints, turned out to be crucial in the course of
this work. This also constitutes one of the main differences
between our present study 
and  previous studies on constraints in the 2HDM, from LHC physics
\cite{Sirunyan:2018koj,Aad:2019mbh,Kling:2020hmi},  EWPO
\cite{Bertolini:1985ia,Hollik:1986gg,Grimus:2007if}, flavor physics
\cite{Enomoto:2015wbn} and global fits
\cite{Bernon:2015qea,Arbey:2017gmh,Haller:2018nnx, Kraml:2019sis}. The
relevance of $\msq$ in the context of large triple Higgs couplings
in the 2HDM type~II
was  also studied in~\cite{Baglio:2014nea, Barger:2014qva}
(with the then available data).
Furthermore, in this paper, we also explore
special choices for $\msq$ in relation with other 2HDM parameters,
like $\MH$, $\tb$ and $\cos(\be-\al)$. In particular, we explore the
implications of the setting $\msq = \MH^2 \cos^2\al/\tb$
(as considered previously, e.g., in~\cite{Ren:2017jbg}).
        
Our paper is organized as follows. In \refse{sec:2hdm} we discuss details of
the 2HDM and fix our notation. The experimental expectations for the
measurement of \lahhh\ are briefly reviewed in \refse{sec:hhh-exp}. We discuss
in \refse{sec:constr} the theoretical and experimental constraints applied to
our sampling of the 2HDMs. The numerical results are presented in
\refse{sec:numres}. Here we show the maximum deviations of \lahhh\ from the SM
that are still allowed taking into account all constraints. We also present
the values that can be reached for the other triple Higgs couplings
involving at least one $h$. Our conclusions are given in
\refse{sec:conclusions}. 


\section{The Two Higgs Doublet Model}
\label{sec:2hdm}

We assume the $\cp$ conserving 2HDM. The scalar potential of this model
can be written as~\cite{Branco:2011iw}:
\begin{eqnarray}
V &=& m_{11}^2 (\Phi_1^\dagger\Phi_1) + m_{22}^2 (\Phi_2^\dagger\Phi_2) - \msq (\Phi_1^\dagger
\Phi_2 + \Phi_2^\dagger\Phi_1) + \frac{\la_1}{2} (\Phi_1^\dagger \Phi_1)^2 +
\frac{\la_2}{2} (\Phi_2^\dagger \Phi_2)^2 \nonumber \\
&& + \la_3
(\Phi_1^\dagger \Phi_1) (\Phi_2^\dagger \Phi_2) + \la_4
(\Phi_1^\dagger \Phi_2) (\Phi_2^\dagger \Phi_1) + \frac{\la_5}{2}
[(\Phi_1^\dagger \Phi_2)^2 +(\Phi_2^\dagger \Phi_1)^2]  \;,
\label{eq:scalarpot}
\end{eqnarray}
\noindent
where $\Phi_1$ and $\Phi_2$ denote the two $SU(2)_L$ doublets.
To avoid the occurrence of tree-level flavor
changing neutral currents (FCNC), a $Z_2$ symmetry is imposed on the 
scalar potential of the model under which the 
scalar fields transform as:
\begin{align}
  \Phi_1 \to \Phi_1\;, \quad \Phi_2 \to - \Phi_2\;.
  \label{eq:2HDMZ2}
\end{align}
This $Z_2$, however, is softly broken by the $\msq$ term in
the Lagrangian. The extension of the $Z_2$ symmetry to the Yukawa
sector forbids
tree-level FCNCs. 
This results in four variants of 2HDM, 
depending on the $Z_2$ parities of the 
fermions. \refta{tab:types} lists the couplings 
for each type of fermion 
allowed by the $Z_2$ parity in four different types of 2HDM.

\begin{table}[htb!]
\begin{center}
\begin{tabular}{lccc} 
\hline
  & $u$-type & $d$-type & leptons \\
\hline
type~I & $\Phi_2$ & $\Phi_2$ & $\Phi_2$ \\
type~II & $\Phi_2$ & $\Phi_1$ & $\Phi_1$ \\
type~III (lepton-specific) & $\Phi_2$ & $\Phi_2$ & $\Phi_1$ \\
type~IV (flipped) & $\Phi_2$ & $\Phi_1$ & $\Phi_2$ \\
\hline
\end{tabular}
\caption{Allowed fermion couplings in 
the four types of 2HDM.}
\label{tab:types}
\end{center}
\end{table}

Taking the electroweak symmetry breaking (EWSB) minima to be
neutral and $\cp$-conserving, the scalar fields after EWSB
can be parameterized as:%
\footnote{We follow here the notation for the field components and field
rotations as in \cite{Arnan:2017lxi}}%
\begin{eqnarray}
\Phi_1 = \left( \begin{array}{c} \phi_1^+ \\ \frac{1}{\sqrt{2}} (v_1 +
    \rho_1 + i \eta_1) \end{array} \right) \;, \quad
\Phi_2 = \left( \begin{array}{c} \phi_2^+ \\ \frac{1}{\sqrt{2}} (v_2 +
    \rho_2 + i \eta_2) \end{array} \right) \;,
\label{eq:2hdmvevs}
\end{eqnarray}
where $v_1, v_2$ are the real vevs acquired by the fields
$\Phi_1, \Phi_2$, respectively, with $\tb := v_2/v_1$ and they satisfy the 
relation $v = \sqrt{(v_1^2 +v_2^2)}$ where $v\simeq246\gev$ is the SM vev.
The eight degrees of freedom above, $\phi_{1,2}^\pm$, $\rho_{1,2}$ and
$\eta_{1,2}$, give rise to three Goldstone bosons, $G^\pm$ and $G^0$,
and five massive physical scalar fields: two $\cp$-even scalar fields,
$h$ and $H$, one $\cp$-odd one, $A$, and one charged pair, $H^\pm$. These
are defined by:
\begin{eqnarray}
\phi_1^\pm&=&\cos \be \, G^\pm -\sin \be \, H^\pm,\nonumber \\
\phi_2^\pm&=&\sin \be \, G^\pm +\cos \be \, H^\pm, \nonumber \\
\eta_1&=&\cos \be \, G^0 -\sin \be \, A, \nonumber \\
\eta_2&=&\sin \be \, G^0 +\cos \be \, A, \nonumber \\ 
\rho_1&=&\cos \al \, H -\sin \al \, h, \nonumber \\
\rho_2&=&\sin \al \, H +\cos \al \, h, \nonumber
\end{eqnarray}
where the mixing angle diagonalizing the $\cp$-even sector is denoted as $\al$. 

From \refeq{eq:scalarpot}, one can see that there are altogether
8 independent parameters in the model,
\begin{equation}
m_{11}^2 \; , \quad m_{22}^2 \;, \quad \msq \;,
\quad \la_{i,~~i=1,5} \;.
\label{eq:original_inputs}
\end{equation}
However, one can use the 
two minimization conditions of the potential at the vacuum to substitute
the bilinears $m_{11}^2$ and $m_{22}^2$ 
for $v$ and $\tb$:
\begin{eqnarray}
	m_{11}^{2}&=m_{12}^{2}\tan\be-\frac{v^{2}}{2}\left[\la_{1}\cos^{2}\be+\left(\la_{3}+\la_{4}+\la_{5}\right)\sin^{2}\be\right],\\m_{22}^{2}&=m_{12}^{2}\cot\be-\frac{v^{2}}{2}\left[\la_{2}\sin^{2}\be+\left(\la_{3}+\la_{4}+\la_{5}\right)\cos^{2}\be\right].
\end{eqnarray}

Furthermore, the couplings $\la_i$ in \refeq{eq:scalarpot} can be replaced by the
physical scalar masses and mixing angles:
\begin{eqnarray}
	v^{2}\la_{1}&=&\frac{1}{\cos^{2}\be}\left(m_{h}^{2}\sin^{2}\al+m_{H}^{2}\cos^{2}\al-\bar{m}^2\sin^2\be\right), 
	\label{l1} \\
	v^{2}\la_{2}&=&\frac{1}{\sin^{2}\be}\left(m_{h}^{2}\cos^{2}\al+m_{H}^{2}\sin^{2}\al-\bar{m}^2\cos^2\be\right),
	\label{l2} \\
	v^{2}\la_{3}&=&\frac{\sin2\al}{\sin2\be}\left(m_{H}^{2}-m_{h}^{2}\right)+2m_{H^{\pm}}^{2}-\bar{m}^2,
	\label{l3} \\
	v^{2}\la_{4}&=&m_{A}^{2}-2m_{H^{\pm}}^{2}+\bar{m}^2,
	\label{l4} \\
	v^{2}\la_{5}&=&\bar{m}^2-m_{A}^{2},
	\label{l5}
\end{eqnarray}
where $\Mh \le \MH$ denote the masses of the $\cp$-even Higgs-bosons, 
$\MA$, $\MHp$ denote 
the masses of the physical $\cp$-odd and charged Higgs bosons
respectively and, for later convenience, we have defined a new mass
squared parameter $\bar{m}^2$, derived from $\msq$, given by: 
\begin{eqnarray} 
 \bar{m}^2&=&\frac{\msq}{\sin\be\cos\be} \,.
\label{eq:mbar}
\end{eqnarray} 
We will study the 2HDM in the physical basis, where the free parameters
of the model, which we use as input, are chosen as:
\begin{equation}
c_{\be-\al} \; , \quad \tb \;, \quad v \; ,
\quad \Mh\;, \quad \MH \;, \quad \MA \;, \quad \MHp \;, \quad \msq \;.
\label{eq:inputs}
\end{equation}
From now on we use sometimes the short-hand notation $s_x = \sin(x)$, $c_x = \cos(x)$.
In our analysis we will identify the lightest $\cp$-even Higgs boson,
$h$, with the one observed at $\sim 125 \gev$.

The couplings of the Higgs bosons to SM particles are modified
w.r.t.\ the SM Higgs-coupling predictions due to the mixing in the Higgs
sector. It is convenient to express the couplings of the neutral scalar
mass eigenstates $h_i$ normalized to the corresponding SM couplings.
We therefore introduce the coupling coefficients $c_{h_i V V}$ such that the couplings to the massive vector bosons
are given by:
\begin{equation}
\left(g_{h_i W W}\right)_{\mu\nu} =
\mathrm{i} g_{\mu\nu} \left(c_{h_i V V}\right) g m_W
\quad \text{and } \quad
\left(g_{h_i Z Z}\right)_{\mu\nu} =
\mathrm{i} g_{\mu\nu} \left(c_{h_i V V}\right) \frac{g m_Z}{\CW} \, ,
\end{equation}
where $g$ is the $SU(2)_L$ gauge coupling, $\CW$ the cosine of weak
mixing angle, $\CW = \MW/\MZ, \SW = \sqrt{1 - \CW^2}$,
and $\MW$ and $\MZ$ the masses of the $W$ boson
and the $Z$ boson, respectively. 
For the $\cp$-even boson couplings we have that $c_{h V V}=\SBA$ and $c_{H V V}=\CBA$ whereas the $\cp$-odd is $c_{A V V}=0$.

In the Yukawa sector, the discrete $Z_2$ symmetry leads to the following
Lagrangian: 
\begin{multline}
	\mathcal{L}_\mathrm{Yuk}=-\sum_{f=u,d,l}\frac{m_f}{v}\left[\xi_h^f\bar{f}fh + \xi_H^f\bar{f}fH +i \xi_A^f\bar{f}\gamma_5fA \right] \\
	-\left[\frac{\sqrt{2}}{v}\bar{u}\left(m_uV_{\mathrm{CKM}}\xi_A^uP_L+V_{\mathrm{CKM}}m_d\xi_A^dP_R\right)dH^+ 
	+\frac{\sqrt{2}m_l}{v}\xi_A^l\bar{\nu}P_RlH^+ + \mathrm{h.c.} \right],
\end{multline}
where the coefficients $\xi_{h_i}^f$ are defined in \refta{tab:yukcoupling} for type~I and~II. The parameters $\xi_{h,H}^f$ can be interpreted as the ratio of the Higgs coupling with the fermions w.r.t. the SM coupling.
\begin{table}
\begin{center}
\begin{tabular}{c|c|c}
 & type~I  & type~II\tabularnewline
\hline 
$\xi_{h}^{u}$  & $s_{\be-\al}+c_{\be-\al}\cot\be$  & $s_{\be-\al}+c_{\be-\al}\cot\be$\tabularnewline
$\xi_{h}^{d,l}$  & $s_{\be-\al}+c_{\be-\al}\cot\be$  & $s_{\be-\al}-c_{\be-\al}\tan\be$\tabularnewline\hline
$\xi_{H}^{u}$  & $c_{\be-\al}-s_{\be-\al}\tan\be$  & $c_{\be-\al}-s_{\be-\al}\tan\be$\tabularnewline
$\xi_{H}^{d,l}$  & $c_{\be-\al}-s_{\be-\al}\tan\be$  & $c_{\be-\al}+s_{\be-\al}\tan\be$\tabularnewline\hline
$\xi_{A}^{u}$  & $-\cot\be$  & $-\cot\be$\tabularnewline
$\xi_{A}^{d,l}$  & $\cot\be$  & $-\tan\be$\tabularnewline
\end{tabular}
\caption{Yukawa couplings relative to the SM for $h$ (upper part),
    $H$ (middle part) and $A$ (lower part) in the 2HDM type~I~(II) in
    the middle (right) column.}
\label{tab:yukcoupling}
\end{center}
\end{table}

The potential of the 2HDM produces new interactions in the scalar
sector. In this paper we will study in detail the couplings of the
lightest $\cp$-even Higgs boson with the other BSM bosons, concretely \lahhh, 
\lahhH, \lahHH, \lahAA\  and \lahHpHm.
We define
these $\la_{h h_i h_j}$ couplings such that the Feynman rules are given by:
\begin{equation}
	\begin{gathered}
		\includegraphics{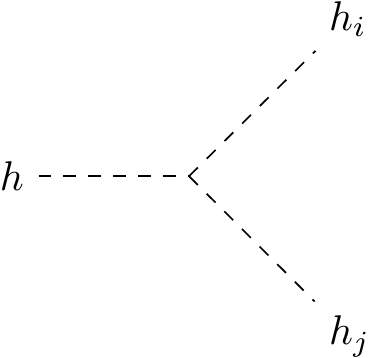}
	\end{gathered}
	=- i\, v\, n!\; \la_{h h_i h_j}
\label{eq:lambda}
\end{equation}
where $n$ is the number of identical particles in the vertex. The
explicit expressions for the couplings $\la_{hh_ih_j}$ are shown in
Appendix \ref{appendix:FR}. We adopt this notation so the light Higgs
trilinear has the same definition as in the SM, i.e. $-6iv\laSM$ with
$\laSM=\Mh^2/2v^2\simeq0.13$. 

It should be noted that all the couplings of the $\cp$-even Higgs bosons
strongly 
depend on $\CBA$. In particular, if $\CBA=0$ one can recover all the
interactions of the SM Higgs boson for the $h$ state, what is known as
the \textit{alignment limit}. This limit is very interesting because, as
we will discuss in \refse{sec:collider}, the Higgs measurements in
colliders seem to overall agree with the SM values. However, in the
alignment limit in general one can still have BSM physics related to the
Higgs sector, like $hH^+H^-$ or $ZHA$ interactions for example. On the
other hand, the parameter $\msq$ may have a relevant impact on the
triple Higgs boson couplings. In the alignment limit,  it does not
affect the couplings $\lahhh$ and  $\lahhH$, but there are
potentially relevant effects on the other couplings, $\lahHH$,
$\lahAA$ and $\lahHpHm$. Outside the alignment limit  (i.e.\ for
$|\SBA| \lsim 1$) the effect of $\msq$ can also enter in a
relevant way into $\lahhh$ and $\lahhH$.


\section{Experimental expectations for \boldmath{\lahhh}}
\label{sec:hhh-exp}

A determination of \lahhh\ (at different degrees of precision) will be
able at future 
collider experiments. Various production cross sections show different
dependences on \lahhh, making those channels complementary to each other. Most
evaluations of the anticipated experimental precision in \lahhh\ focus on the
SM value. However, as we will analyze below, substantially different values of
\lahhh\ are possible in the 2HDM (and other BSM models). The potential for the
measurement of \lahhh\ at a future collider experiment thus strongly depends
on the value of $\kala := \lahhh/\laSM$ that is realized in nature. 

\begin{figure}[t]
\begin{center}
\includegraphics[width=0.7\textwidth]{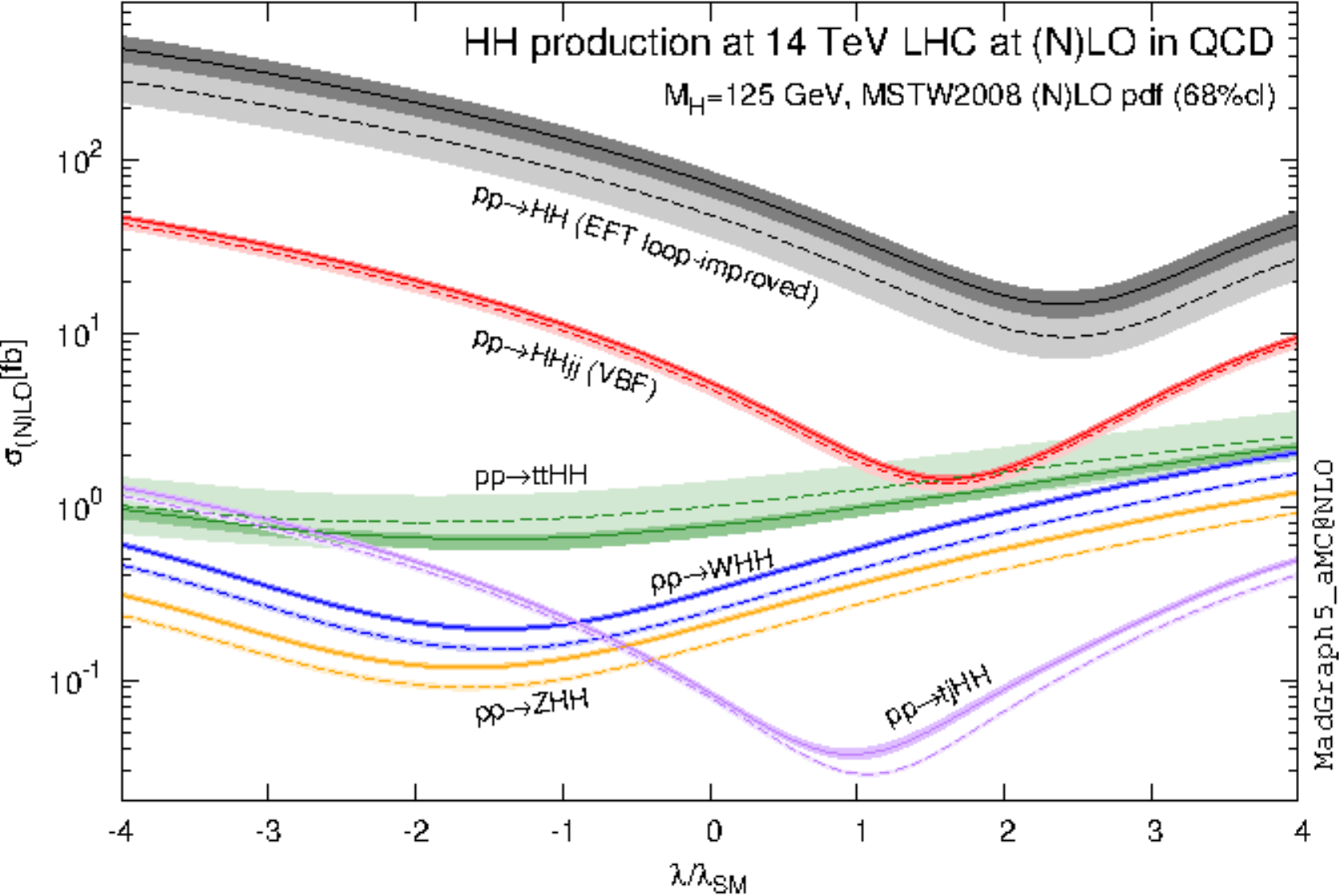}
\caption{\label{fig:HH-lamhhh-LHC}
Production cross sections for a pair of SM Higgs bosons as a function of
\lahhh/\laSM\ at the LHC~\cite{Frederix:2014hta}.
}
\end{center}
\end{figure}

In \reffi{fig:HH-lamhhh-LHC} we show the the various double Higgs production
cross sections in the SM in $pp$ collisions with $\sqrt{s} = 14 \tev$ at
(next-to) leading order ((N)LO)
QCD, see \citere{Frederix:2014hta} for details. The largest cross section is 
given by $gg \to hh$,%
\footnote{
We denote here the Higgs boson at $\sim 125 \gev$ with $h$. In
\reffi{fig:HH-lamhhh-LHC}, taken from \citere{Frederix:2014hta}, it is
denoted as~$H$.}%
~which will be most relevant for the measurement of
$\lahhh$ at the HL-LHC. One can see that the production cross section has a
minimum around $\kala \sim 2$. Consequently, if such a value was
realized, it is expected (see below) that the future experimental
precision would be worse than for, e.g.,  
$\kala = 1$, where a determination at the level of $\sim 50\%$ is
anticipated~\cite{Cepeda:2019klc}.
Largest production cross sections, on the other hand, are
for negative $\kala$. Consequently, a BSM model with very small or even
negative values of $\kala$ is expected to result in a better
determination of \lahhh. A similar behavior is observed for the second
largest production channel, the WBF channel $pp \to hh$jj (where j
denotes a jet), with a minimum around $\kala \sim 1.5$. Different
dependences are observed for the other, less relevant channels.

\begin{figure}[t]
\begin{center}
\includegraphics[width=0.45\textwidth]{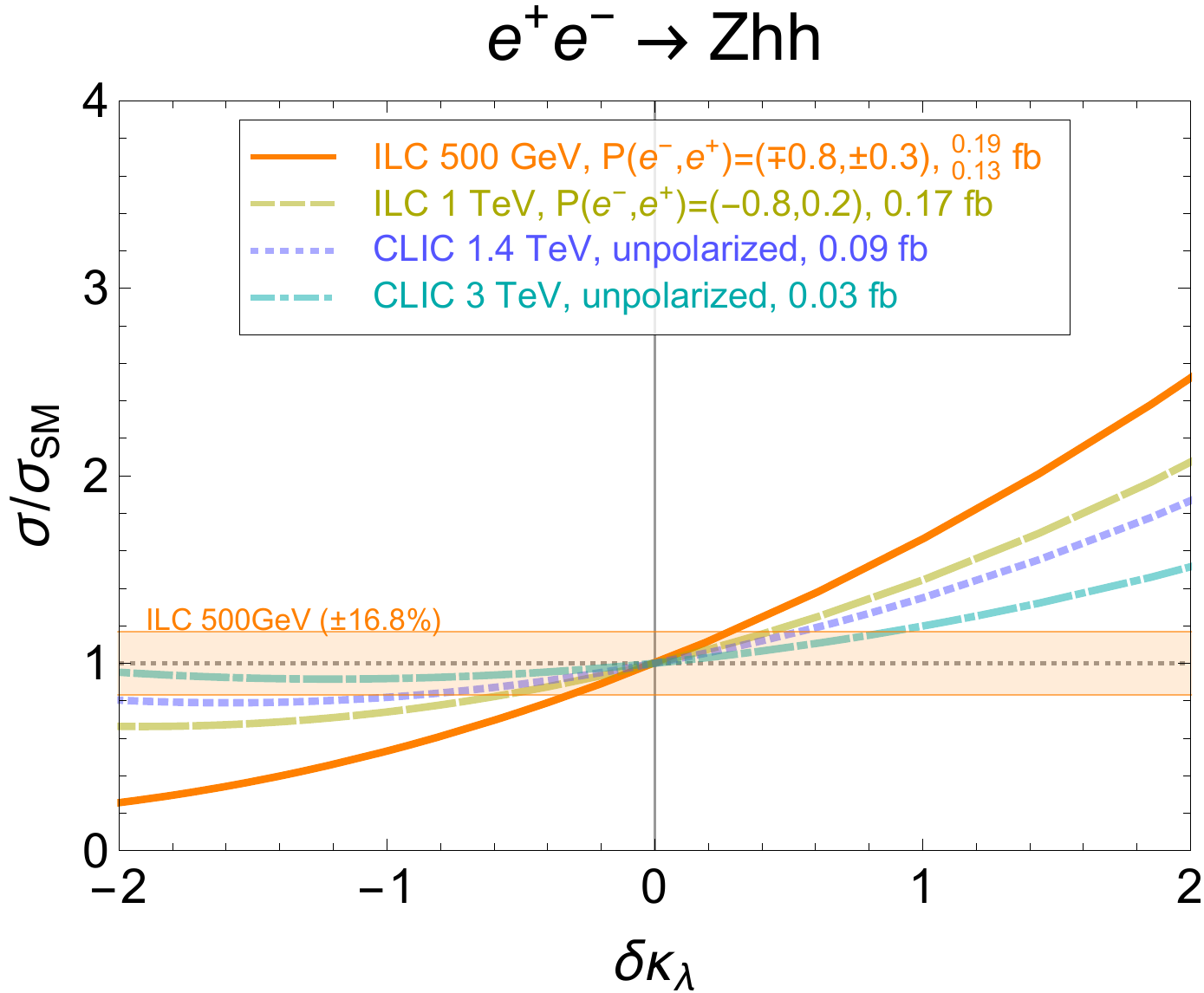}
\includegraphics[width=0.45\textwidth]{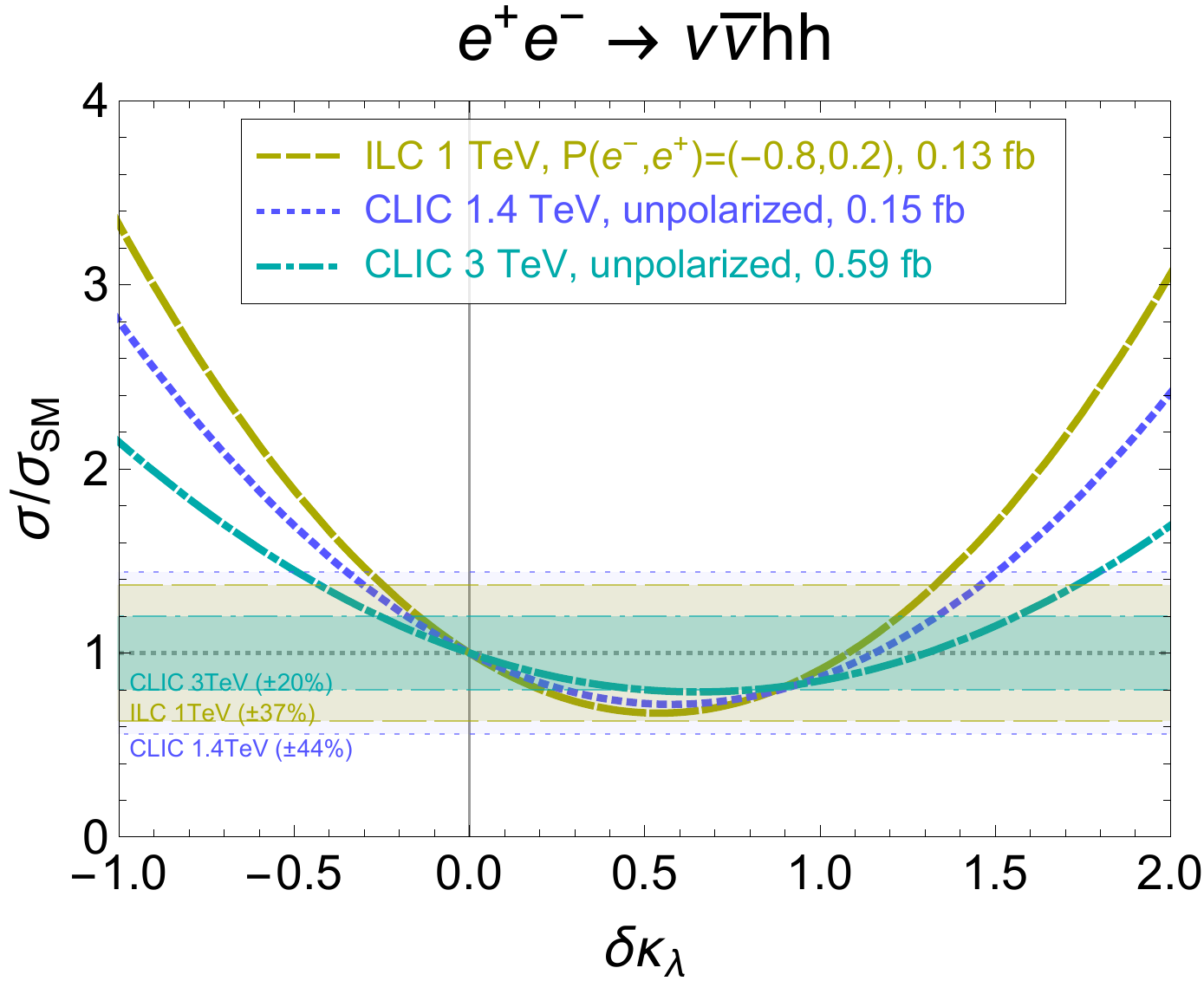}
\caption{\label{fig:HH-lamhhh-ee}
Higgs-strahlung (left) and WBF production (right) of a pair of SM Higgs bosons
as a function of \lahhh\ at the ILC and CLIC~\cite{DiVita:2017vrr}.
It should be noted that the experimental precision on the total cross section
indicated by the horizontal bands is valid only for the SM case. 
}
\end{center}
\end{figure}

Similarly, in \reffi{fig:HH-lamhhh-ee} we show the dependence on
$\de\kala := \kala - 1$ for the
Higgs-strahlung process, $e^+e^- \to Zhh$ (left) and the weak-boson
fusion (WBF) channel, $e^+e^- \to \nu\bar\nu hh$ (right) for various
center-of-mass energies, $\sqrt{s}$, at the ILC and
CLIC~\cite{DiVita:2017vrr}. Also 
indicated as horizontal colored bands are the anticipated experimental
accuracies at the ILC500 (left) and ILC\,1TeV, CLIC\,1.4TeV and CLIC\,3TeV
(right). As for the HL-LHC, also at $e^+e^-$ colliders the different
production channels exhibit a different dependence on \lahhh. For the
Higgs-strahlung process smaller (larger) cross sections are obtained for
smaller (larger) \kala. Higher values of $\sqrt{s}$ yield a weaker
dependence on \lahhh, as well as a smaller absolute cross section (as
typical for $s$-channel processes).
Consequently, a determination of \lahhh\ based
(only) on the Higgs-strahlung channel is expected to be best at lower
$\sqrt{s}$ (e.g.\ at the ILC500) and for larger values of
\kala.  The WBF channel exhibits a minimum at $\de\kala \sim 0.5$. As
for the Higgs-strahlung channel the dependence becomes weaker for larger
values of $\sqrt{s}$, whereas the absolute values of the cross
section increase with $\sqrt{s}$ (as typical for $t$-channel
processes). Consequently, a case-by-case study is necessary to take into
account the different, opposing effects.

\begin{figure}[t]
\begin{center}
\includegraphics[width=0.49\textwidth]{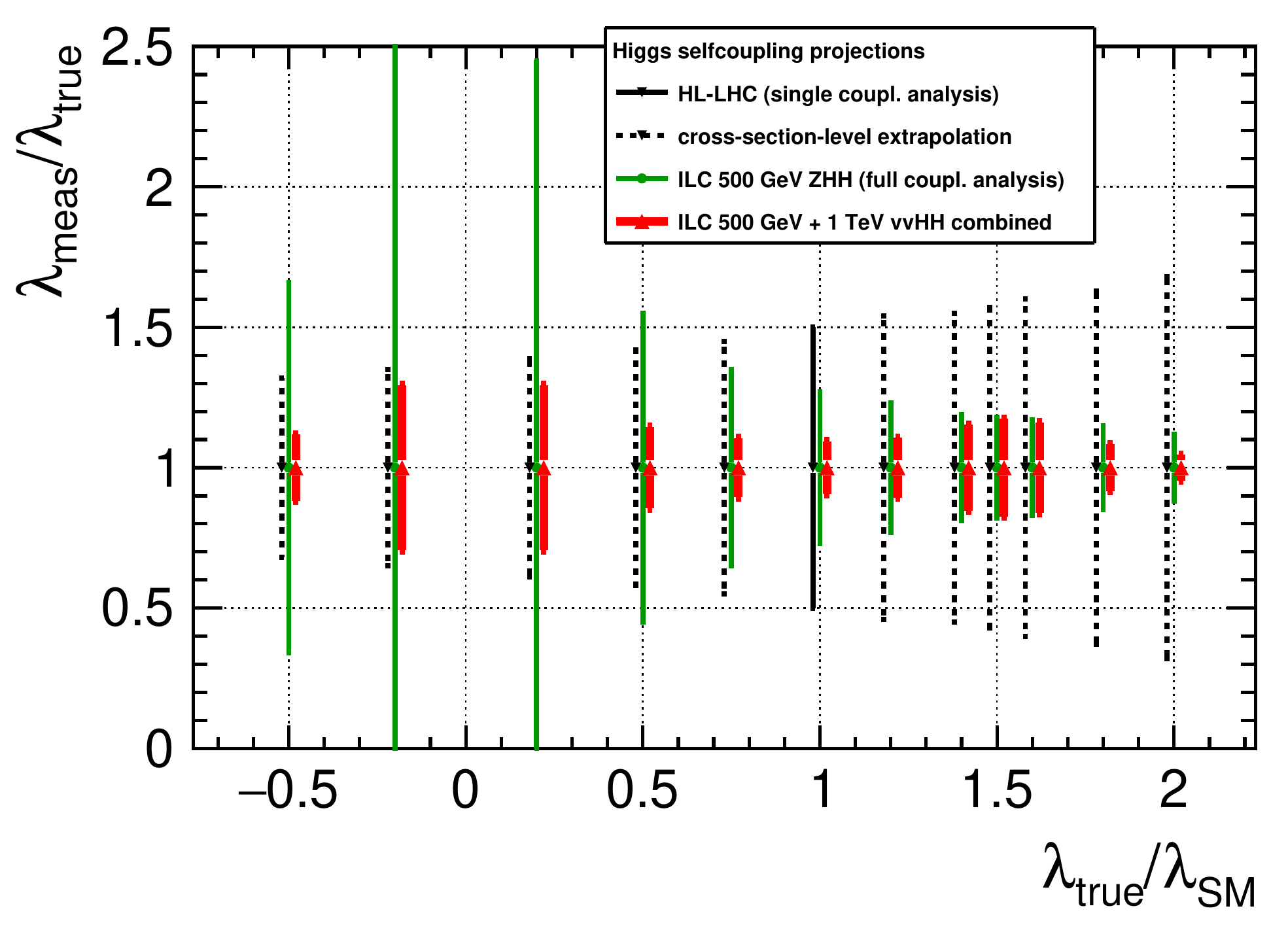}
\includegraphics[width=0.49\textwidth]{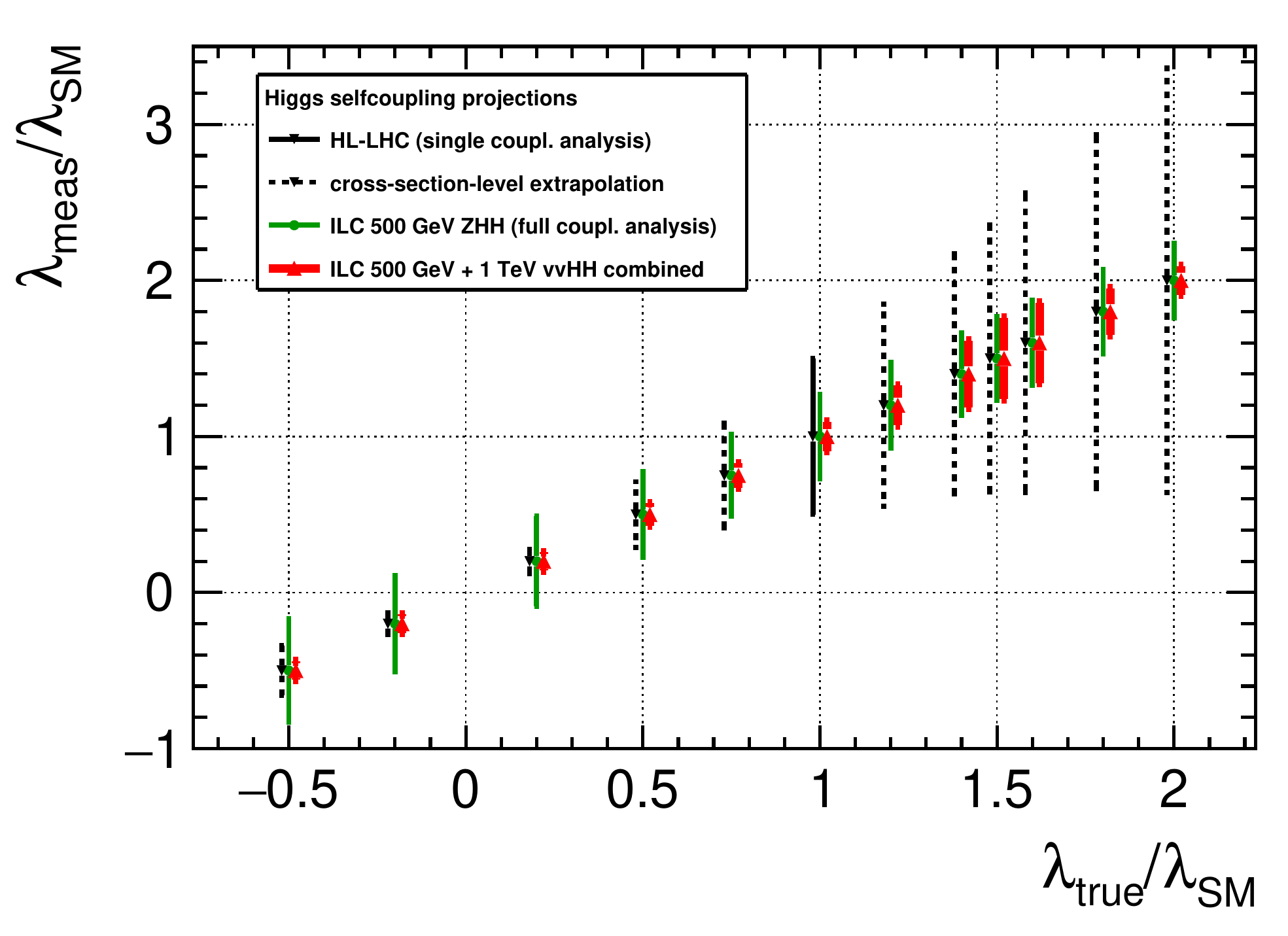}
\caption{\label{fig:lahhh-HLLHC-ee}
Anticipated precision in the experimental determination of \lahhh\ as a
function of $\lahhh/\lahhh^{\rm SM}$~\cite{lahhh-exp}, relative (left)
and absolute (right).
}
\end{center}
\end{figure}

The results of such a case-by-case study are shown in
\reffi{fig:lahhh-HLLHC-ee}~\cite{lahhh-exp}. Depicted are the relative
(left) and absolute (right) 
accuracies of a determination of \lahhh\ (``$\la_{\rm meas}/\la_{\rm true}$'')
as a function of \kala\ (``$\la_{\rm true}/\la_{\rm SM}$'') in the range of
-0.5\,\ldots\,2. Compared are the anticipated HL-LHC precision (based on
a scaling of the results for $\kala = 1$), the ILC500 precision
(i.e.\ using only the Higgs-strahlung channel) and the ILC500+\,1\,TeV
accuracy (i.e.\ also using the WBF channel results). 
Here it should be kept in mind that the HL-LHC analysis assumes that the
other Higgs-boson couplings take their SM value, whereas for the ILC
analysis it has been shown that the inclusion of the variation of the
other Higgs-boson couplings does not lead to a degradation of the
anticipated precision. It is worth mentioning that in all these
analyses, other possible channels that might contribute to double Higgs
production in extensions of the SM, like for instance the 2HDM, are not
considered. 

The achievable
precisions follow the cross section dependences discussed above. At the
HL-LHC the most (im)precise determination is expected for smaller
(larger) values of \kala. A $\sim 35 (70) \%$ relative precision is
anticipated for $\kala = -0.5 (2.0)$, as can be seen in the left plot of
\reffi{fig:lahhh-HLLHC-ee}. Using only ILC500 results better (worse)
experimental determinations are expected for larger (smaller) values of
\kala, ranging from $\sim 65\%$ at $\kala = -0.5$ to $\sim 15\%$ at
$\kala = 2.0$. The large {\em relative} uncertainties close to $\kala = 0$
are caused exactly by the smallness of the triple Higgs coupling. As can
be seen in the right plot of \reffi{fig:lahhh-HLLHC-ee}, the absolute
determination of $\kala$ continuously improves with smaller $\kala$. 
The combination with the WBF measurements at
$\sqrt{s} = 1 \tev$ yields a substantially better determination for all
values of $\kala$, but no monotonous behavior is found, owing to the
different opposing effects, as discussed above. Future precisions between
$\sim 5\%$ and $\sim 30\%$ are expected, depending on the value of
\kala\ realized in nature. Again the largest {\em relative}
uncertainties of up to $30\%$ are found close to $\kala = 0$, whereas
the absolute determination exhibits a nearly constant very precise
determination of $\kala$ in the interval $\inter{-0.5}{1.0}$.

These results clearly show that the physics
potential of a future collider experiment strongly depends on the
actual value of \lahhh\ realized in a BSM model. This motivates the
analysis presented in the following sections showing which values of
\lahhh\ (and other triple Higgs couplings) can be realized in 2HDMs,
taking into account all existing experimental and theoretical
constraints.


\section{Experimental and theoretical constraints}
\label{sec:constr}

In this section we will describe the various theoretical and
experimental constraints considered in our scans.


\subsection{Constraints from electroweak precision data}
\label{sec:stu}

Constraints from the electroweak precision observables (EWPO)
can, in a simple 
approximation, be expressed in terms of the oblique parameters $S$, $T$ and
$U$~\cite{Peskin:1990zt,Peskin:1991sw}. This approximation holds if
the BSM effects enter mainly via corrections to
gauge boson self-energies, as it is the case for extended Higgs sectors.
Under these assumptions, the corrections are independent of the Yukawa
sector of the 2HDM, and therefore the same for all types.

In 2HDMs there is a strong correlation between $T$ and
$U$, and it is known that $T$ is by far more constraining than $U$~\cite{Funk:2011ad}. Hence, $U$ can safely be dropped in the present analysis. 
Specifically, our criterion to accept a point in the 2HDM parameter
space, as being in agreement with the EWPO data, is as follows. For a
given choice of input parameters in \refeq{eq:inputs} to be allowed by
the experimental observation, we require that the prediction of the $S$
and the $T$ parameter are in agreement with their experimental
values $S=0.02\pm0.10$ and $T=0.07\pm0.12$\cite{Tanabashi:2018oca}. In
this section we will study and compare the requirement of agreement at
the  $1\,\sig$ and $2\,\sig$ level. In our posterior numerical
analysis in \refse{sec:numres} we will require agreement at $2\,\sig$. 

In the 2HDM, as mentioned above, the most constraining oblique
parameter is $T$, thus, we will focus in the following of this
section on the constraints from the $T$~parameter. In the
forthcoming analysis in \refse{sec:numres} we have checked that once
the allowed regions by $T$ are set, these are also allowed by $S$ and $U$,
i.e.\ effectively it is sufficient to require agreement of $T$ with
its experimental value.
One peculiarity of the $T$ parameter in the 2HDM is that it depends on
the relative mass squared differences of the scalar Higgs bosons. This
can be seen in the explicit expression for the $T$ parameter in the
$\cp$ conserving 2HDM that is given by~\cite{Grimus:2007if}: 
\begin{equation}
\begin{split}
	T&=\frac{g^2}{64\pi^2m_W^2} 
	 \left\lbrace F\left( m_A^2, m_{H^\pm}^2 \right) 
	+ s_{\be-\al}^2 \left[ F \left( m_H^2, m_{H^\pm}^2 \right) - F \left( m_H^2, m_A^2 \right)  \right] \right. \\ 
	&+ c_{\be-\al}^2 \left[ F \left(m_{H^\pm}^2,m_h^2 \right) - F \left( m_A^2, m_h^2 \right)  \right] 
	 + 3s_{\be-\al}^2 \left[ F \left( m_H^2, m_Z^2 \right) - F \left( m_H^2, m_W^2 \right)  \right]  \\
	&+ 3 c_{\be-\al}^2 \left[ F \left( m_h^2,m_Z^2 \right) - F \left( m_h^2,m_W^2 \right) \right] 
	 - \left. 3\left[ F \left(m_{h_\textrm{SM}}^2, m_Z^2 \right) - F \left( m_{h_\textrm{SM}}^2, m_W^2 \right) \right]  \right\rbrace,
\end{split}
\label{eq:Tparam}
\end{equation}
where $F\left( x,y \right)= \frac{x+y}{2} - \frac{xy}{x+y}\log{\frac{x}{y}}$, 
and it satisfies that $F\left(x,x\right)=0$. Therefore, 
the contributions to $T$ become small when either the mass of $H$ or $A$
is sufficiently close to the mass of the charged Higgs boson
$H^\pm$~\cite{Bertolini:1985ia,Hollik:1986gg}. 
This motivates us to define three different simplified scenarios to
explore the parameter space that is allowed by the EWPO
in the 2HDM:
scenario~A, where $m_A=m_{H^\pm}$;
scenario~B, where $m_H=m_{H^\pm}$ and
scenario~C where the masses of all the BSM Higgs bosons are equal,
$m_H=m_A=m_{H^\pm}$. 
One can see from \refeq{eq:Tparam} that in scenario A the main
contributions to the $T$~parameter vanish for any value of
$\CBA$, whereas in scenario~B a contribution proportional to
$\CBA^2 F\left(\MA^2,\MHp^2\right)$ still survives, that will remain
small close to the alignment limit.

\begin{figure}[t!]
\begin{center}
\includegraphics[width=0.5\textwidth]{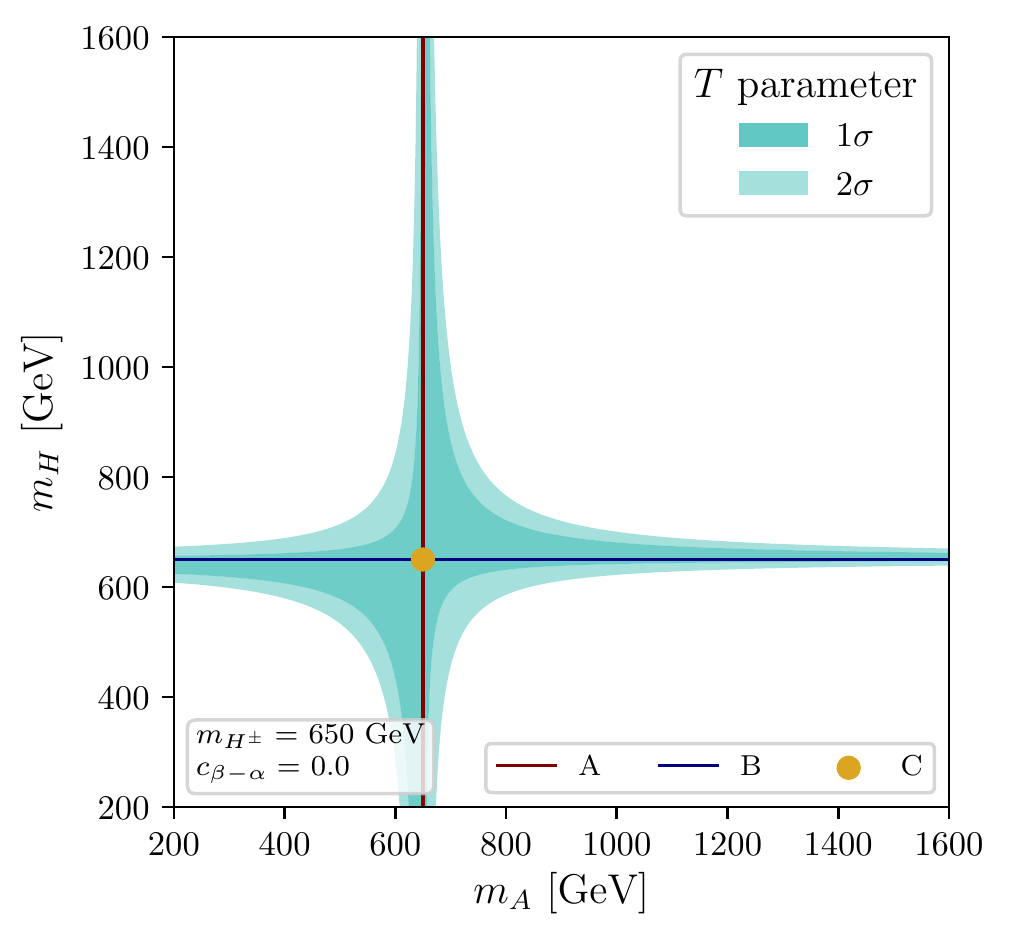}\includegraphics[width=0.5\textwidth]{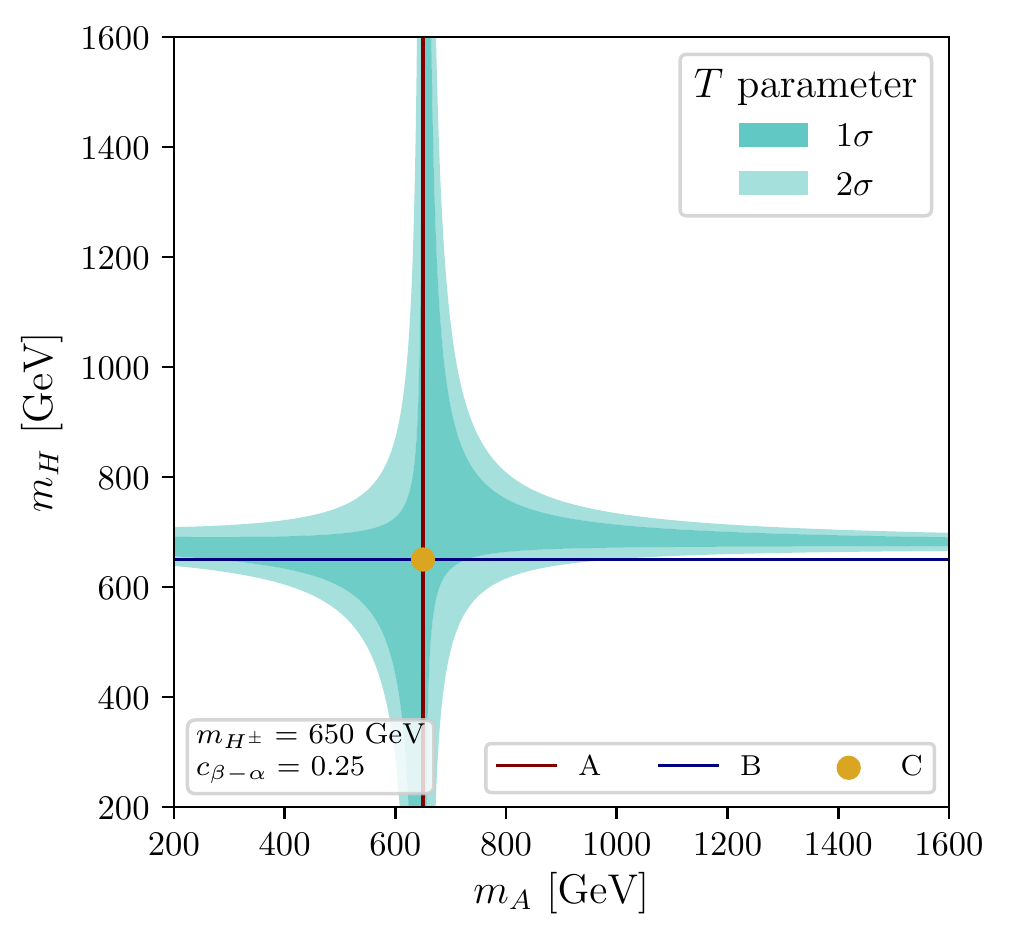}
\includegraphics[width=0.5\textwidth]{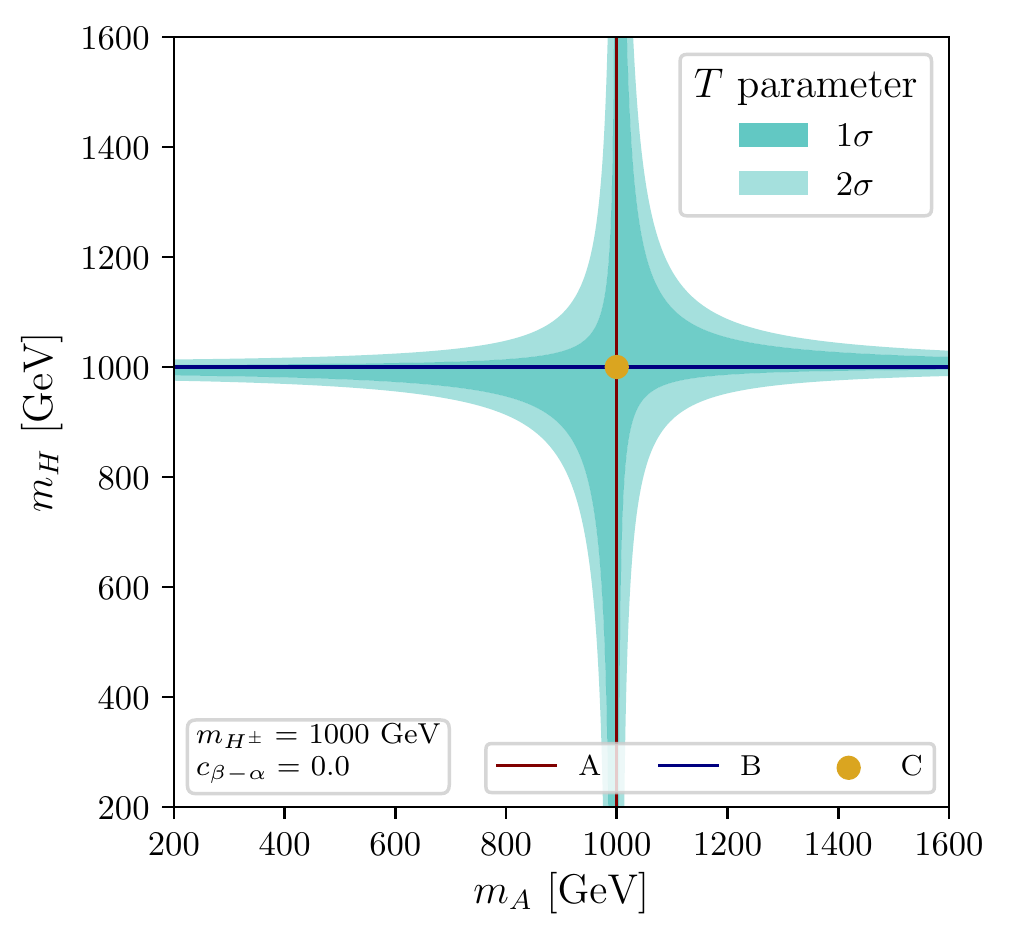}\includegraphics[width=0.5\textwidth]{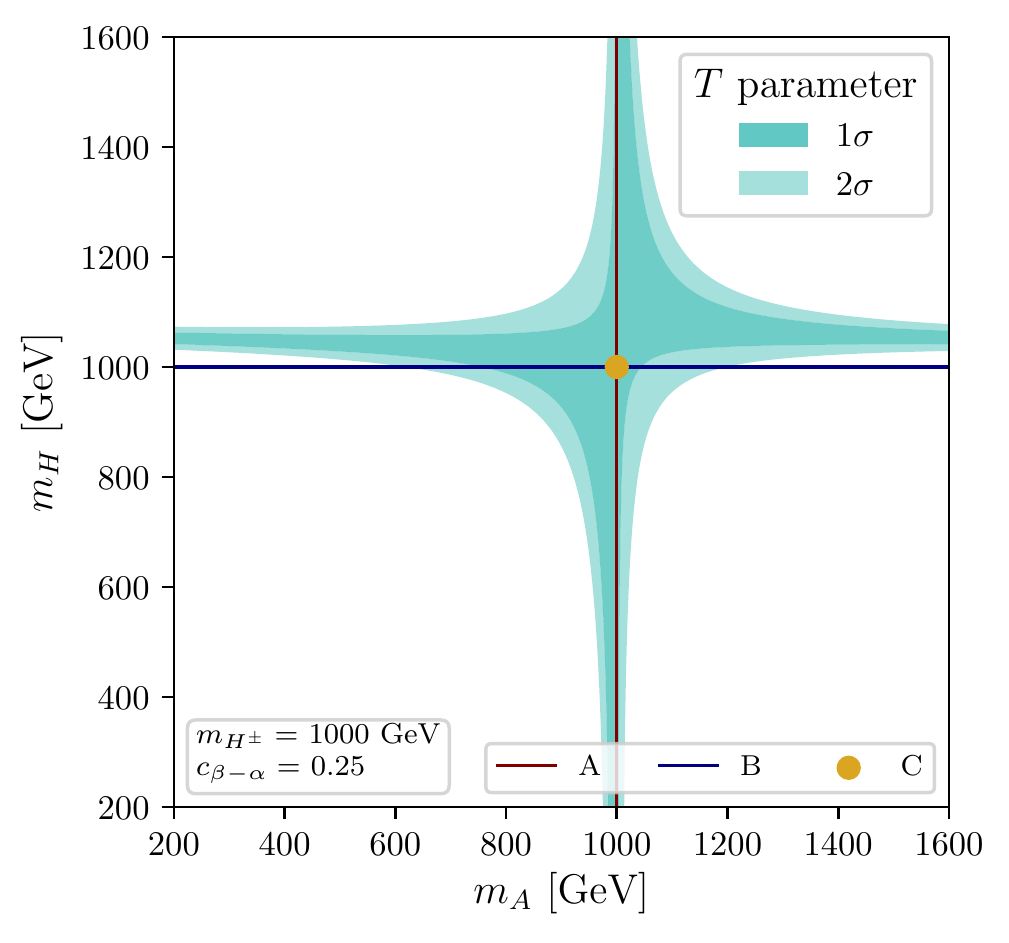}
\caption{\label{fig:EWPO}
$T$ parameter for $\MHp = 650 \gev$ (top) and $\MHp = 1000 \gev$
  (bottom) in the alignment limit, $\CBA = 0$, (left) and for
  $\CBA = 0.25$ (right). In scenario~A $\MHp=\MA$ (red lines),
  in scenario~B $\MHp = \MH$ (blue lines),
  and in scenario~C $\MHp=\MH=\MA$ (yellow points).
}
\end{center}
\end{figure}

Our study of the impact of these scenarios in the prediction for the $T$
parameter for different values of $\CBA$ and $\MHp$ is summarized in
\reffi{fig:EWPO}. The 2HDM parameter space is explored with the
\texttt{2HDMC} code~\cite{Eriksson:2009ws}.
For a given set of input parameters 
and a given Yukawa type of the 2HDM, the code
computes as output the mass spectrum, decay widths and branching ratios
of all the Higgs bosons. It furthermore calculates 
the $S$, $T$ and $U$ parameters and contribution to the anomalous
magnetic moment of the muon, $\left(g-2\right)_\mu$.
In \reffi{fig:EWPO} it can be seen that for scenario~A any mass
splitting between $\MH$ and $\MA=\MHp$ is allowed inside the $2\sigma$
region even far from the alignment limit. However, this is not the
case for scenario~B, where it can be seen that the prediction for $T$ is
only inside the $2\sigma$ region close to the alignment limit.
If one goes to higher values of $\CBA$ (plots on the right) there
are some values of $\MA$ that are disallowed, for example when
$\MHp=650\gev$ and $\CBA = 0.25$ the allowed
region is $\MA-\MHp < 350 \gev$ (upper right plot).
This effect becomes stronger for larger values of $\MHp$.
For instance, for $\MHp=1000 \gev$ and $\CBA = 0.25$ (lower left plot)
the allowed region shrinks to $-380 \gev < \MA-\MHp < 200 \gev$.
In general, scenario~A and C (as a subset of scenario~A) is broadly
allowed by $T$ for any value of $\CBA$ and mass splitting among the
Higgs bosons, whereas scenario~B can lead to a large deviation if $\CBA$
and $\MHp$ increases (which is taken into account in
\refse{sec:numres} as discussed above).


\subsection{Theoretical constraints}
\label{sec:theo}

Like all models with extended scalar sectors, the 2HDM
also faces important constraints coming from tree-level
perturbartive unitarity and stability of the vacuum. We briefly describe these constraints below
(for a discussion of higher-order effects and other considerations regarding the alignment limit, see,
  e.g.,~\cite{Goodsell:2018fex,Chen:2019pkq}). 

\begin{itemize}

\item \textbf{Tree-level perturbative unitarity}

Perturbative unitarity is achieved by demanding that the eigenvalues of
the lowest partial wave scattering matrices of the $2 \to 2$
processes in the scalar sector of the 2HDM, at the tree level, remain
below $16 \pi$. This leads to the following constraints
\cite{Akeroyd:2000wc,Bhattacharyya:2015nca}:  
\begin{eqnarray}
	\left|\la_{3}\pm\la_{4}\right|\le16\pi, \label{u1}\\
	\left|\la_{3}\pm\la_{5}\right|\le16\pi,\\
	\left|\la_{3}+2\la_{4}\pm3\la_{5}\right|\le16\pi,\\
	\left|\frac{1}{2}\left(\la_{1}+\la_{2}\pm\sqrt{\left(\la_{1}-\la_{2}\right)^{2}+4\la_{4}^{2}}\right)\right|\le16\pi,\\
	\left|\frac{1}{2}\left(\la_{1}+\la_{2}\pm\sqrt{\left(\la_{1}-\la_{2}\right)^{2}+4\la_{5}^{2}}\right)\right|\le16\pi,\\
	\left|\frac{1}{2}\left(3\la_{1}+3\la_{2}\pm\sqrt{9\left(\la_{1}-\la_{2}\right)^{2}+4\left(2\la_{3}+\la_{4}\right)^{2}}\right)\right|\le16\pi.
\end{eqnarray}
It should be noted that the above requirement of tree level perturbative
unitarity, 
limiting the maximum size of the given combinations of $\la_i$'s,
also ensures indirectly that the potential remains perturbative up to
very high scales. Hence, in the present paper we do not incorporate
additional constraints from other alternative criteria to require
perturbativity that are based on limiting the size of the separate
$\la_i$'s  which could be {\it a priori} more restrictive than the
one applied here. 

\item  \textbf{Stability} 

First, we require the boundedness from below criterion. This criterion demands 
that the potential does not go to minus infinity when the field values
approach infinity. This is fulfilled if the following conditions are
satisfied \cite{Deshpande:1977rw,Barroso:2013awa,Bhattacharyya:2015nca}: 
\begin{eqnarray}
\la_1\geq0, \label{eq:stab1} \\
\la_2\geq0, \label{eq:stab2} \\
\la_3+\sqrt{\la_1 \la_2}\geq0, \label{eq:stab3} \\
\la_3+\la_4-\left|\la_5\right|+\sqrt{\la_1 \la_2}\geq0. \label{eq:stab4} 
\end{eqnarray}
Besides those inequalities, we will also demand that the minimum of the
theory is a global minimum of the potential that can be achieved if
\cite{Barroso:2013awa} 
\begin{equation}
	m_{12}^{2}\left(m_{11}^{2}-m_{22}^{2}\sqrt{\frac{\la_{1}}{\la_{2}}}\right)\left(\tan\be-\sqrt[4]{\frac{\la_{1}}{\la_{2}}}\right)\geq0. 
\label{eq:TrueMin}
\end{equation}

\end{itemize}

\begin{figure}[t!]
\begin{center}
	\includegraphics[width=0.33\textwidth]{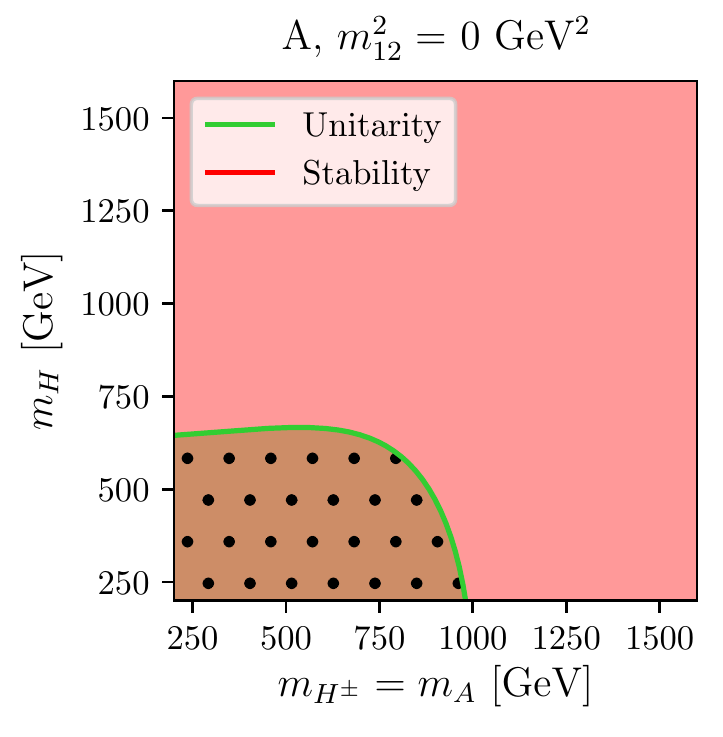}\includegraphics[width=0.33\textwidth]{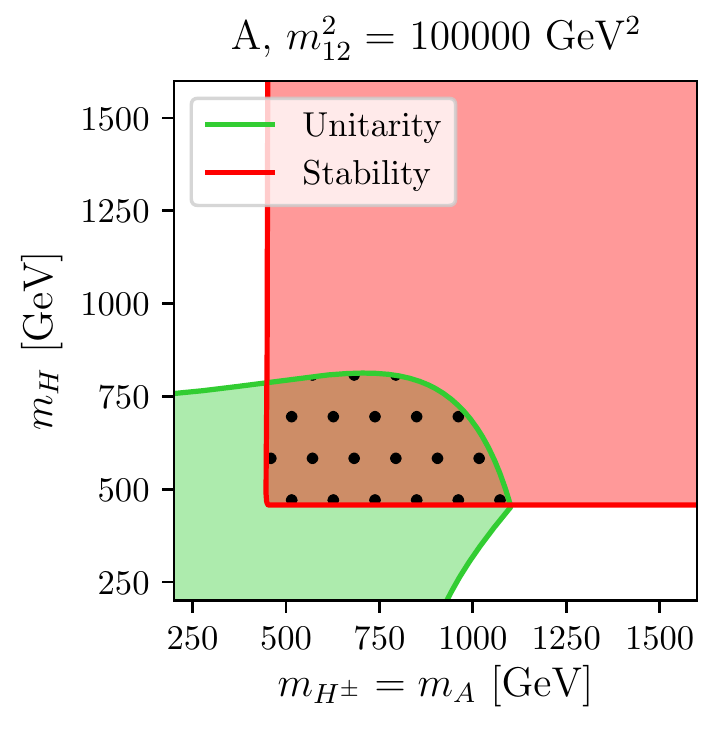}\includegraphics[width=0.33\textwidth]{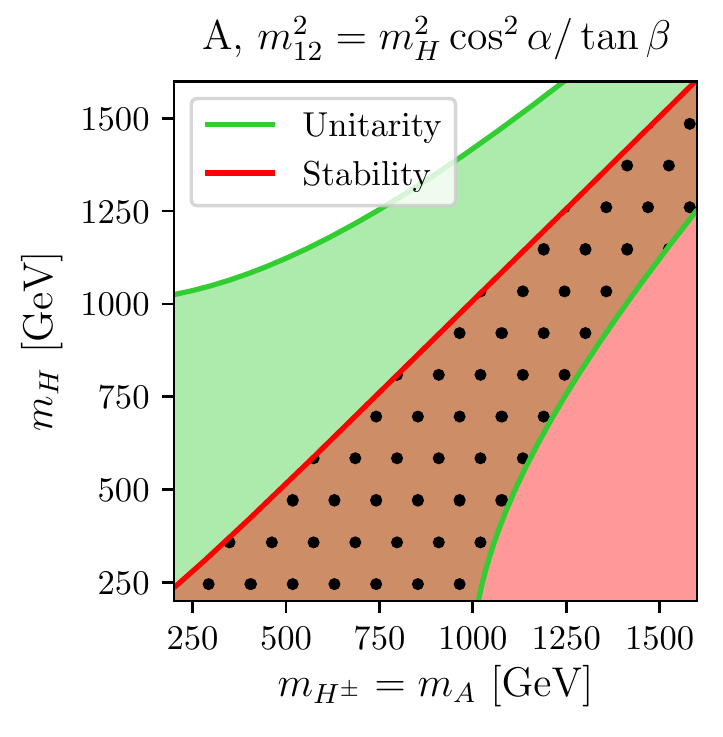}
	\includegraphics[width=0.33\textwidth]{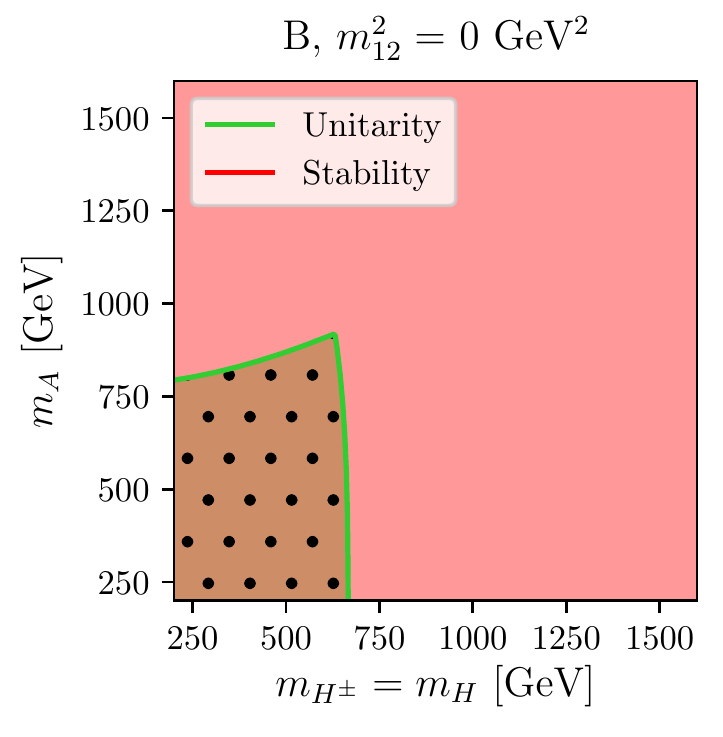}\includegraphics[width=0.33\textwidth]{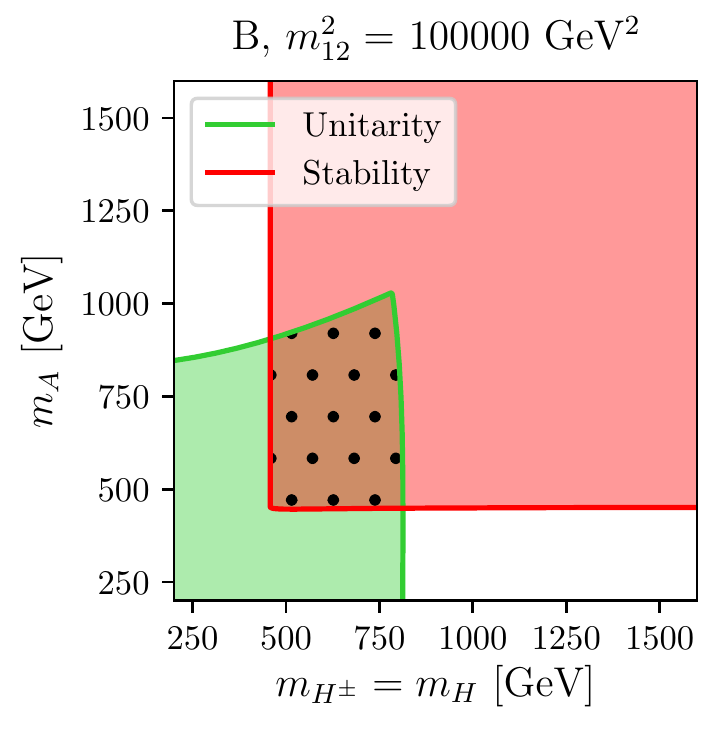}\includegraphics[width=0.33\textwidth]{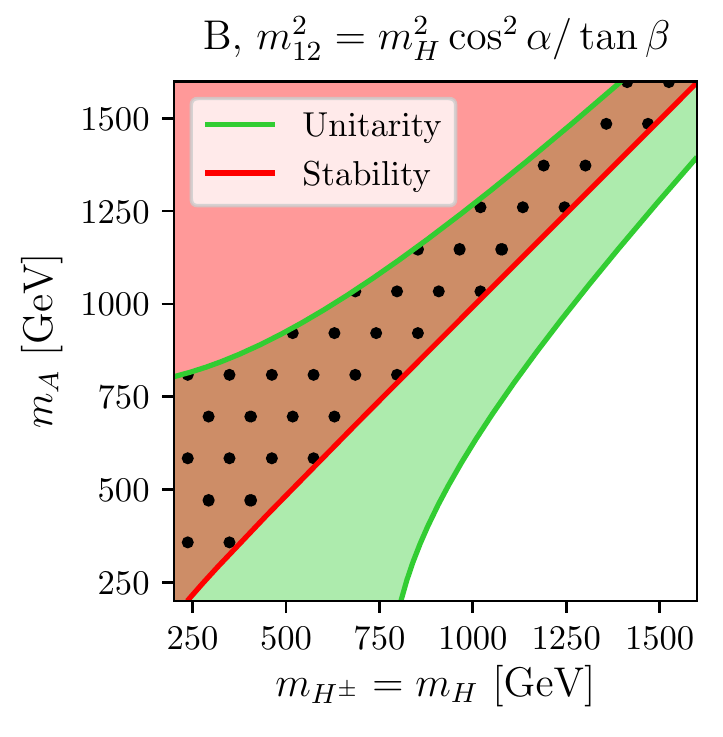}
\caption{\label{fig:THEO}
Allowed areas in two selected Higgs masses of the 2HDM parameter
  space, delimited by the theoretical constraints from unitarity (green
  areas), stability (red areas), and both together (dotted areas),
  obtained from equations (\ref{u1}) to (\ref{eq:TrueMin}).  The
  alignment limit is assumed and  $\tb$ is fixed to $\tb=1.5$ for
  scenario~A (top plots) and scenario~B (bottom plots). 
$\msq$ is set to 0 (left plots), 100000$\gev^2$
(middle plots) and $\msq = \MH^2 \cos^2\al/\tb$ (right plots).
}
\end{center}
\end{figure}

According to equations (\ref{l1}) to (\ref{l5})  the size of the triple
couplings $\la_i$ are closely related to the size of the
masses of the Higgs bosons and $\msq$. In general, the size of the
triple Higgs couplings involving one $h$ and two heavy Higgs bosons grow
with the corresponding heavy Higgs mass and, therefore, they can be
large for large heavy masses, near the TeV scale. Consequently,
unitarity sets limits on the maximum allowed size of these large heavy
masses, as can be seen in  \reffi{fig:THEO}. One finds that only in
scenarios where the heavy masses are large but nearly degenerate that
these unitary bounds  can be relaxed. The parameter $\msq$ also plays an
important role in that concern. The plots on the right in
\reffi{fig:THEO} show that by setting the value of this parameter to
$\msq = \MH^2 \cos^2\al/\tb$, a diagonal corridor opens up
allowing for larger values of these heavy masses  above 1500 GeV and
with a considerable splitting. 
On the other hand, $\msq$ enters with a negative sign in some of the stability conditions (Eqs. (\ref{eq:stab1}) - (\ref{eq:stab4})) and \refeq{eq:TrueMin} imposes $\msq\geq0$. Therefore if $\msq$ is large the Higgs boson masses should be also large to compensate those negative contributions.
In fact,
setting $\msq$  to large values reduces considerably the allowed region
by stability and shrinks it to the upper right corner in these two
dimensional mass plots. This reduces as well the intersection area with
the unitarity allowed region (dotted areas), as can be seen in the two
plots in the middle with $\msq = 100000 \gev^2$.
Here $\la_1$ plays an important role, as it contains a negative
  contribution $\propto \msq$ that grows with $\tb$, see
  \refeq{l1}. This can drive $\la_1$ to negative values and yield
  disagreement with the stability condition in \refeq{eq:stab1}.
One way to minimize this effect on $\la_1$ is
to fix $\msq$ such that  the two last terms  in \refeq{l1} cancel each
other.  
This condition leads to the above commented  equation allowing for the
diagonal corridor in the right plots of \reffi{fig:THEO} where the
intersection region (dotted area) is clearly expanded.  Therefore, to
enlarge the allowed region by unitarity and stability in our forthcoming
analysis we will consider this as a special interesting case where to
explore the maximum allowed size of the triple Higgs couplings. This
condition on $\msq$ has been considered previously \cite{Ren:2017jbg}
and can also be  translated into a condition on $\bar m^2$, using
\refeq{eq:mbar}, 
\begin{equation}
  \msq = \frac{\MH^2\cos^2\al}{\tb}, \quad
  \bar{m}^2 = \frac{\MH^2\cos^2\al}{\sin^2\be}~.
\label{eq:m12special}
\end{equation}
Regarding the comparison of the allowed regions for the two considered
scenarios~A and~B, we show in \reffi{fig:THEO} some specific examples,
for $\tb=1.5$,  where one can clearly see the impact of
$\msq\neq 0$ and compare it with imposing \refeq{eq:m12special}. In the
case when $\msq=0$ (left) all masses are allowed by stability but they
are restricted by unitarity, and the final allowed dotted region is, in
scenario~A, for masses $\MHp=\MA \lesssim1000\gev$ and $\MH \lesssim 650
\gev$ and, in scenario~B, for masses roughly below 750 GeV. When $\msq$
increases (center) the allowed region by unitarity is similar to the
previous situation, but due to the large value for $\msq$, now to get
stability, the masses should be larger than approximately
$500\gev$ in both scenarios~A and~B. The situation is completely
different in the right plots where \refeq{eq:m12special} is adopted. In
these cases masses can get very large values as well as $\msq$ and also
splitting between the two free masses is allowed. This splitting
stretches in both scenarios~A and~B as the masses grow and the final
allowed region by stability and unitarity is confined to a diagonal
corridor which is narrower in scenario~B than in scenario~A. It should
be noted that
in cases where \refeq{eq:m12special} is satisfied, in order to cope with
the theoretical constraints scenario~A demands that $\MHp=\MA\geq\MH$
and scenario~B that $\MA\geq\MHp=\MH$. 
The allowed region by both theoretical constraints in the left and
center columns would dramatically shrink for a larger value of $\tb$
because of the size of $\la_1$, but the right plots would remain
similar. In some sense, \refeq{eq:m12special} gives an upper limit for
$\msq$ for large masses and large $\tb$.


\subsection{Constraints from direct searches at colliders}
\label{sec:collider}

\begin{figure}[t]
\begin{center}
	\includegraphics[width=0.33\textwidth]{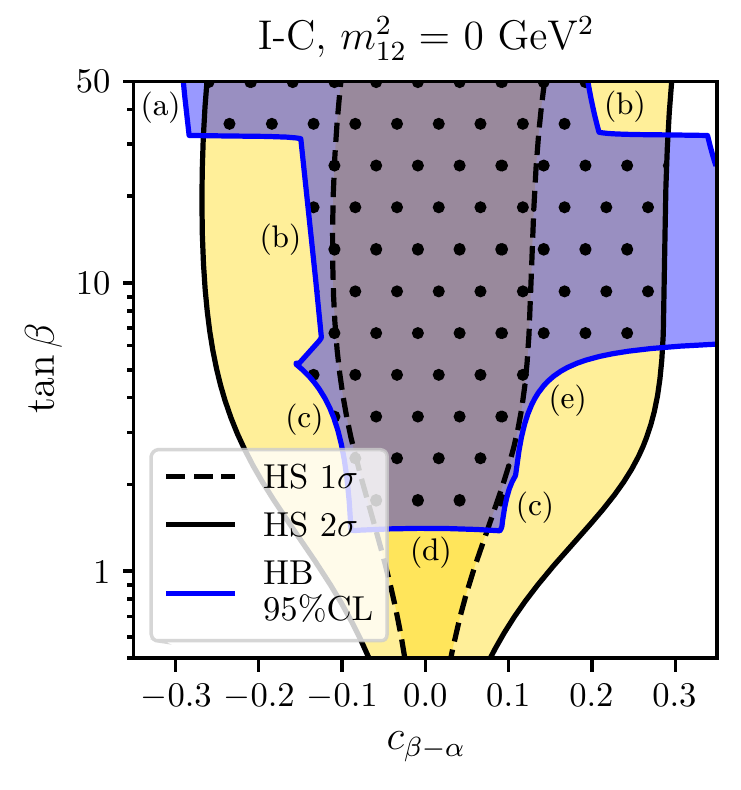}\includegraphics[width=0.33\textwidth]{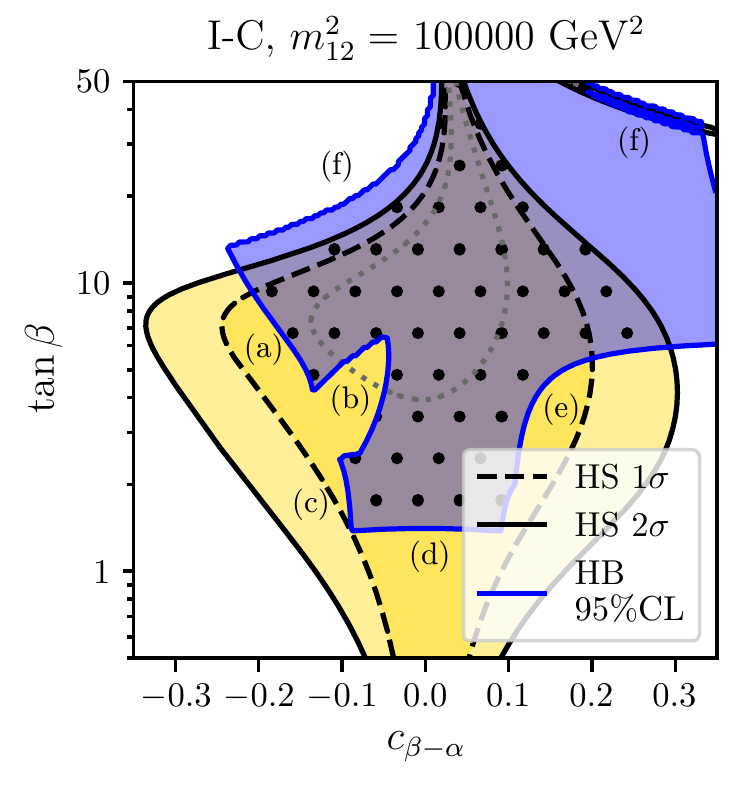}\includegraphics[width=0.33\textwidth]{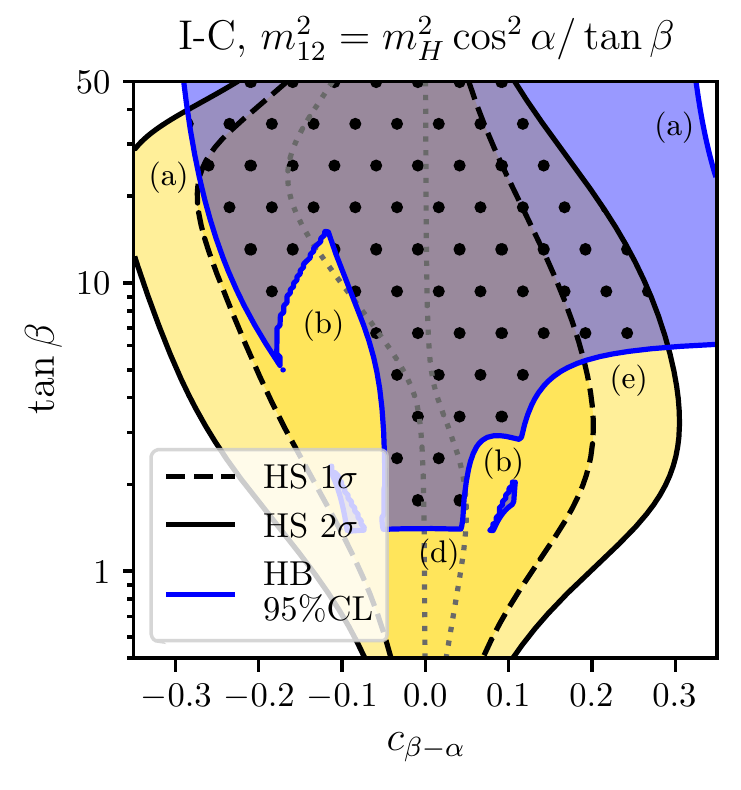}
	\includegraphics[width=0.33\textwidth]{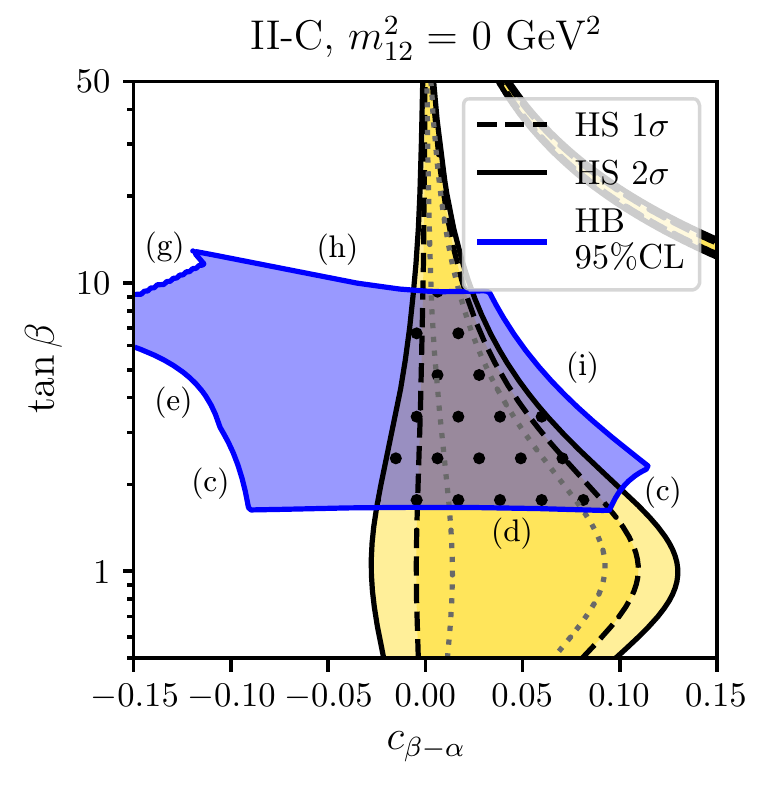}\includegraphics[width=0.33\textwidth]{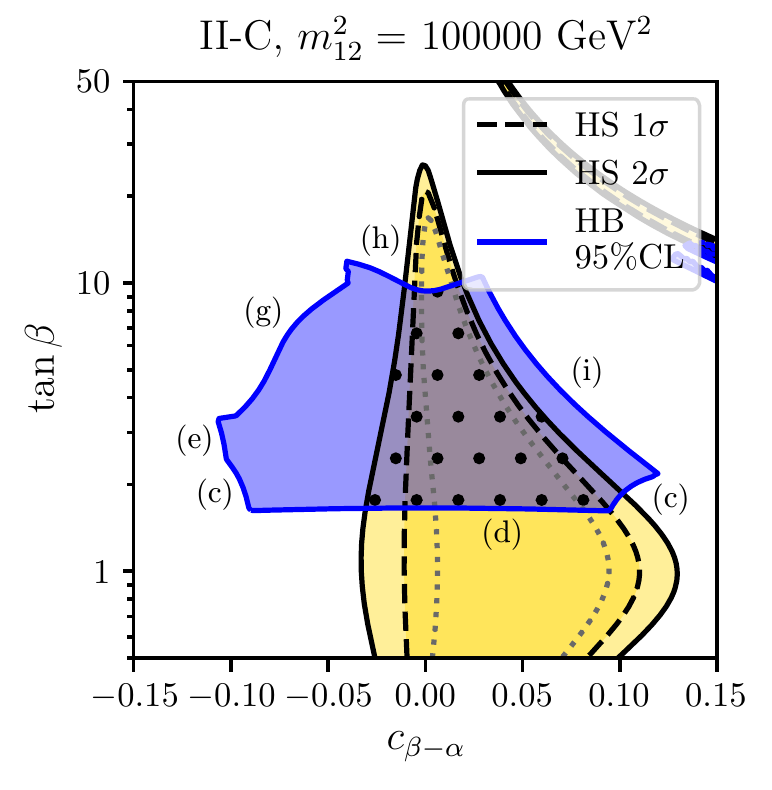}\includegraphics[width=0.33\textwidth]{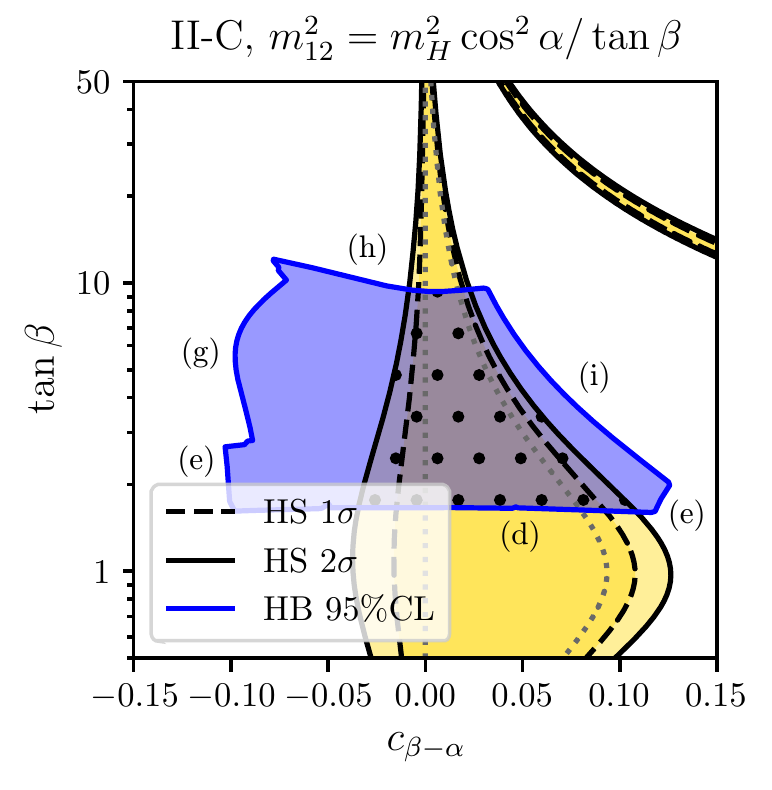}
\end{center}
	\caption{\label{fig:COLLIDER}
Allowed regions of the 2HDM in the $(\CBA, \tb)$ plane from BSM Higgs
bosons searches and direct measurements of the $125\gev$ Higgs from
\HB\ (blue regions) and \HS\ (yellow regions) in scenario~C with
$\MH=\MA=\MHp=650\gev$ for Yukawa type~I (top) and~II (bottom) and
different values of $\msq$. The dotted grey lines correspond to
  contours with the same $\chi ^2$ for the $125 \gev$ Higgs-boson
 rate measurements at the LHC as found in the SM.}
\end{figure}

The $95\%$ confidence level
exclusion limits of all important searches for BSM Higgs bosons
are included in the public code
\HB\,\texttt{v.5.3.2}~\cite{Bechtle:2008jh,Bechtle:2011sb,Bechtle:2013wla,Bechtle:2015pma},
including Run~2 data from the LHC. Given a set of theoretical
predictions in a particular model, \HB\ determines which is the most
sensitive channel and determines, based on this most sensitive
channel, whether the point is allowed or not at the $95\%$~CL.
As input the code requires some specific predictions from the model,
like branching ratios or Higgs couplings, that we computed with the
help of the \texttt{2HDMC} code (see \refse{sec:stu}).
In \reffi{fig:COLLIDER} plotted in blue are shown the allowed
regions of the 2HDM in the $(\CBA, \tb)$ plane 
for the case where all the masses of the heavy Higgs bosons are set to
$650\gev$, i.e.\ in the simplest scenario~C. 
In the upper (lower) row we show the results for the 2HDM
type~I~(II) with $\msq = 0, 100000 \gev^2$ and set via
\refeq{eq:m12special} in the left, middle and right column, respectively.
The particular exclusion channel that sets a bound limiting this
blue region is specified with a Latin letter and corresponds to
one of the following channels: 

(a) $pp\to H\to hh\to (b\bar{b})(\tau^+\tau^-)$~~\cite{Aaboud:2018sfw},

(b) $pp\to H\to hh\to (b\bar{b})(b\bar{b}/\tau^+\tau^-/W^+W^-/\gamma\gamma)$~~\cite{Aad:2019uzh},

(c) $pp\to H\to VV$~~\cite{Aaboud:2018bun},

(d) $pp\to H^\pm tb\to (tb)tb$~~\cite{CMS:2019yat},

(e) $gg\to A\to Zh\to (l^+l^-)(b\bar{b})$~~\cite{CMS:2018xvc},

(f) $pp\to hX\to\gamma\gamma X$~~\cite{CMS:ril},

(g) $pp\to H\to hh\to (b\bar{b})(b\bar{b})$~~\cite{ATLAS:2016ixk},

(h) $pp\to H\to\tau^+\tau^-$~~\cite{CMS:2017epy},

(i) $pp\to h\to ZZ\to (l^+l^-)(l^+l^-)$~~\cite{CMS:xwa}.

In broad terms, the 2HDM type~I seems to be less constrained than type~II by the searches of heavy Higgs bosons. Both types have a lower bound on $\tb\sim1.4$ from channel (d), whereas type~II also has an upper bound given by channel (h). 
In type~I negative values of $\CBA$ are constrained by channels (a), (b)
and (c) while for positive values the more relevant channel is (e). On
the other hand, in type~II for a negative $\CBA$ channels (e) and (g)
become the most restrictive ones, and in the positive $\CBA$ region
channel (i) is the most sensitive one. 

It is also worth to notice that for $\msq=100000\gev^2$ (center plots)
there are more stringent bounds than in the other cases coming from
channel (f) in type~I and from channel (g) in type~II. This is an
example of how $\msq$ can be relevant in some situations when the
contributions from the scalar sector are important. Clearly, the
experimental bounds on BSM Higgs searches strongly depend on the masses
of such particles, so the allowed contours and the exclusion channels
shown in \reffi{fig:COLLIDER} will change for a different value of the
masses. In general, for smaller values of the input masses the parameter
space would be more constrained. 


\subsection{Constraints from the SM-like Higgs-boson properties}
\label{sec:SMlike}

Any model beyond the SM has to accommodate the SM-like Higgs boson,
with mass and signal strengths as they were measured at the
LHC~\cite{Aad:2012tfa,Chatrchyan:2012xdj,Khachatryan:2016vau}.
In our scans the compatibility of the $\cp$-even scalar $h$ with a mass
of $125.09\gev$ with the measurements of signal strengths at Tevatron and LHC
is checked with the code
\texttt{HiggsSignals v.2.2.3}~\cite{Bechtle:2013xfa,Bechtle:2014ewa}. 
\texttt{HiggsSignals} provides a
statistical $\chi^2$ analysis of the SM-like Higgs-boson predictions of
a certain model compared to the measurement of Higgs-boson signal rates
and masses from Tevatron and LHC. Again, the predictions of the 2HDM
have been obtained with the {\tt{2HDMC}} code. The complete list
of implemented experimental data can be found in~\citere{higgssignals-www}.
Here and in our posterior analysis we will require that for a  parameter
point of the 2HDM to be allowed, the corresponding $\chi^2$ is within
$2\,\sig$ ($\De\chi^2 = 6.18$)
from the SM fit: $\chi_\mathrm{SM}^2=43.6$.

In \reffi{fig:COLLIDER} we present the results of \HS\ for scenario~C
with $\MH=\MA=\MHp=650\gev$ as a function of $\tb$ and $\CBA$, which are
the most relevant parameters to determine the couplings of the $h$ boson
to the SM particles. In yellow are shown the allowed regions from
\HS. (In blue are shown the allowed regions from \HB, as discussed
in the previous subsection). 
In this figure we show the contours from \HS\ corresponding to a
$1\sigma$ (dashed lines) and $2\sigma$ (solid lines) distance from the
SM fit and the contours that have the same fit as the SM (dotted grey
lines). In consequence, the regions inside these dotted grey lines have
a better agreement with the experimental results that the SM. 
It can be seen that the  parameter space is strongly constrained
for $\CBA$ to be close to the alignment limit, such that $h$
behaves sufficiently SM-like. In particular, the $2\sig$ allowed region
for the Yukawa type~II (bottom) is substantially smaller compared to
type~I (top). 
In particular for type~II, we find that negative values of $\CBA$ are
very disfavored. The maximum deviation from the alignment limit takes
place for $\tb\sim1$, where values between $\CBA=0.13$ and $\CBA=-0.03$
can be found inside the $2\sigma$ region from the SM. However, as $\tb$
increases the model is forced to be very close to the alignment limit to
agree with the experimental data. This is caused by an enhancement of
the coupling of $h$ to $b$-quark (see \refta{tab:yukcoupling}). It
should be noted
that in the type~II fits a new allowed branch appears in the upper right
part of the plot which corresponds to $\xi_h^d=-1$, known as the
\textit{wrong sign} Yukawa region. 
For type~I the constraints are weaker, specially for $\tb>3$, where we
can accommodate inside the $2\sigma$ region values for $\CBA$ up to $\pm 0.3$. 
\reffi{fig:COLLIDER} also captures the role of $\msq$ in the fits. In
type~I $\msq$ barely changes the fits for $\tb\lesssim3$
region. However, the increment of $\msq$ narrows the $1\sigma$,
$2\sigma$ contours around the alignment limit, notably for
$\msq=100000\gev^2$ (upper center) where the fit forces $\CBA\sim0$ when
$\tb$ is large. In the case of type~II the fits seems to be
roughly independent of $\msq$, except again for $\msq=100000\gev^2$ where the
model is completely outside the $2\sigma$ region for $\tb>25$. 

In addition, it can be seen that an extensive region exists for both
types that gives a better fit to the experimental data than the SM
i.e.\ $\chi^2<\chi_\mathrm{SM}^2$, even though for type~I $\msq$ is
required to be different from zero. Such regions are expected due
to the additional freedom in the 2HDM to accommodate the LHC measurements.
For the sake of completeness, we would like to comment that the impact
of $\MH$, $\MA$ and $\MHp$ could be important for the fit when they are
low, because only then they could give sizable contributions to the
light Higgs measurements, specially for the $H$ boson. 

Other recent studies from LHC data analysis
\cite{Sirunyan:2018koj,Aad:2019mbh,Kling:2020hmi}, also set similar
constraints on the ($\CBA,\tb$) plane, since these are the most
relevant 2HDM parameters (entering the Higgs-boson couplings) at the LHC. One of the main
differences to our study is that, as emphasized in the
introduction, we have a strong focus
on the role played by the $\msq$ parameter, which
turns out to be relevant in our search of sizable triple Higgs
couplings.


\subsection{Constraints from flavor physics}
\label{sec:flavor}

Constraints from flavor physics have proven to be very significant
in the 2HDM mainly because of the presence of the charged Higgs boson.
Various flavor observables like rare $B$~decays, 
$B$~meson mixing parameters, $\br(B \to X_s \gamma)$, 
LEP constraints on $Z$ decay partial widths
etc., which are sensitive to charged Higgs boson exchange, provide
effective constraints on the available 
parameter space~\cite{Enomoto:2015wbn,Arbey:2017gmh}. 
Here we will take into account the decays $B \to X_s \gamma$ and
$B_s \to \mu^+ \mu^-$, which we find to be the most constraining
ones and whose experimental values are (we use the average from~\cite{Tanabashi:2018oca}):
\begin{equation*}
\begin{split}
	\br(B \to X_s \gamma)=\left(3.1\pm1.1\right)\times10^{-4}, \\
	\br(B_s \to \mu^+ \mu^-)=\left(2.7^{\ +0.6}_{\ -0.5}\right)\times 10^{-9}.
\end{split}
\end{equation*}
We will set our bounds in the $2\sig$ region from the central value
according to the experimental value. 

In order to compute the theoretical predictions in the 2HDM we have used
the public code
\texttt{SuperIso}~\cite{Mahmoudi:2008tp,Mahmoudi:2009zz} with the
model input given by {\tt{2HDMC}}. Moreover, we have included in
\texttt{SuperIso} the contributions to the Wilson coefficient $C_P$ from
the Higgs-penguin diagrams, that are missing in the public version and
that can be relevant for the $\br(B_s \to \mu^+ \mu^-)$
prediction~\cite{Li:2014fea,Arnan:2017lxi,Cheng:2015yfu}.

\begin{figure}[t!]
\begin{center}
	\includegraphics[width=0.33\textwidth]{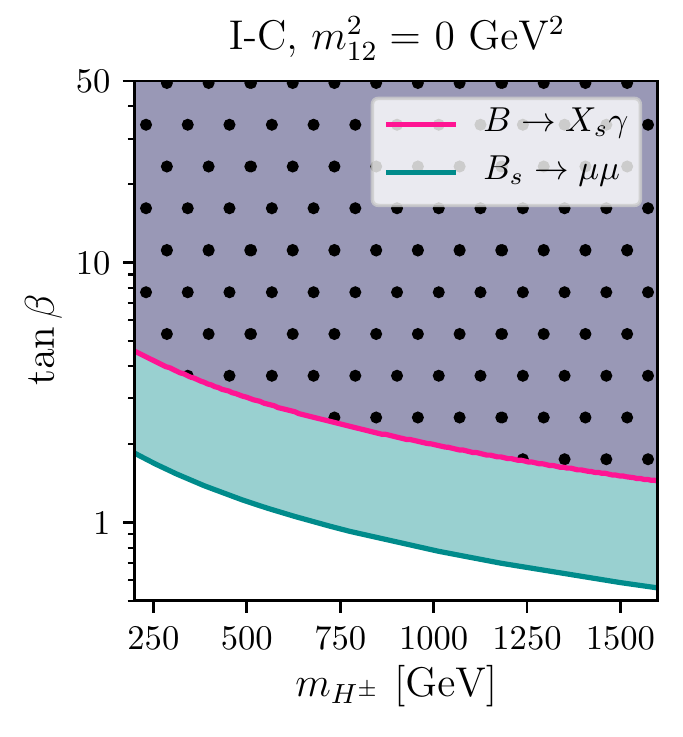}\includegraphics[width=0.33\textwidth]{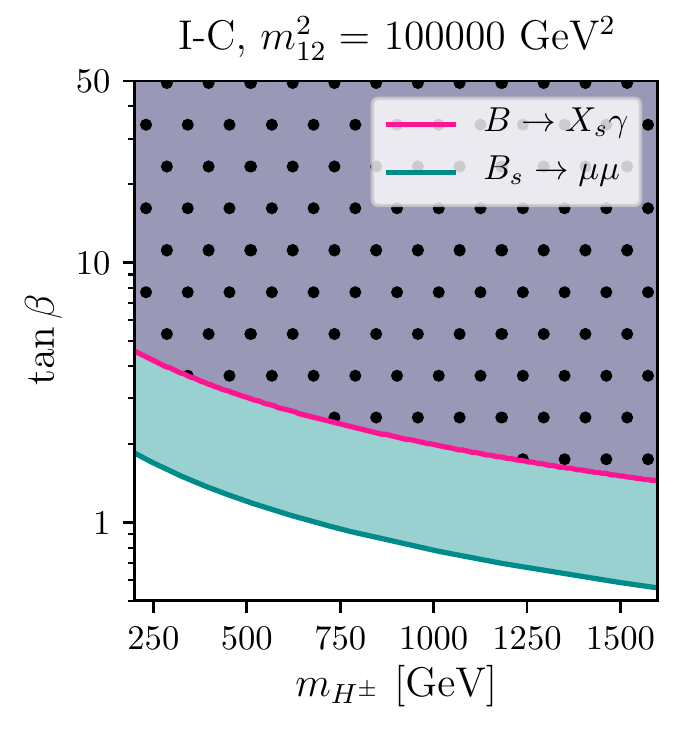}\includegraphics[width=0.33\textwidth]{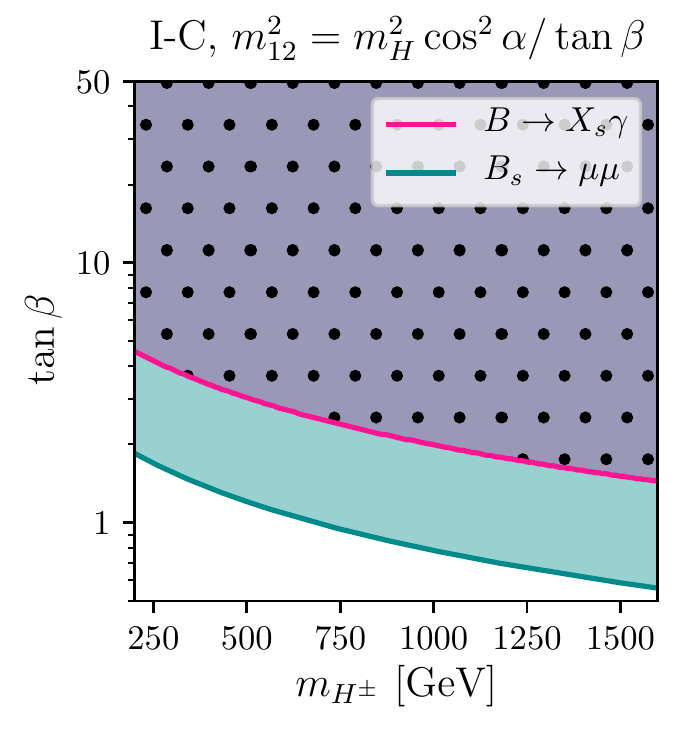}
	\includegraphics[width=0.33\textwidth]{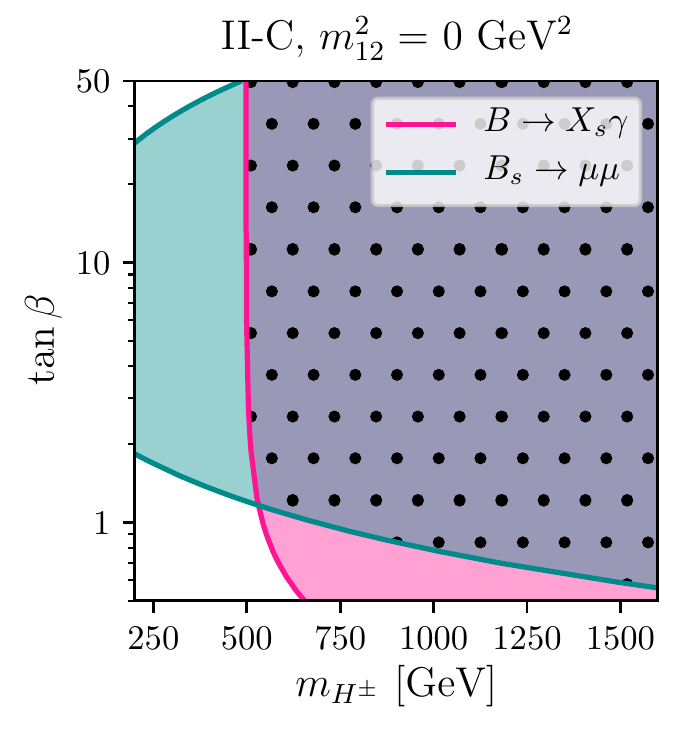}\includegraphics[width=0.33\textwidth]{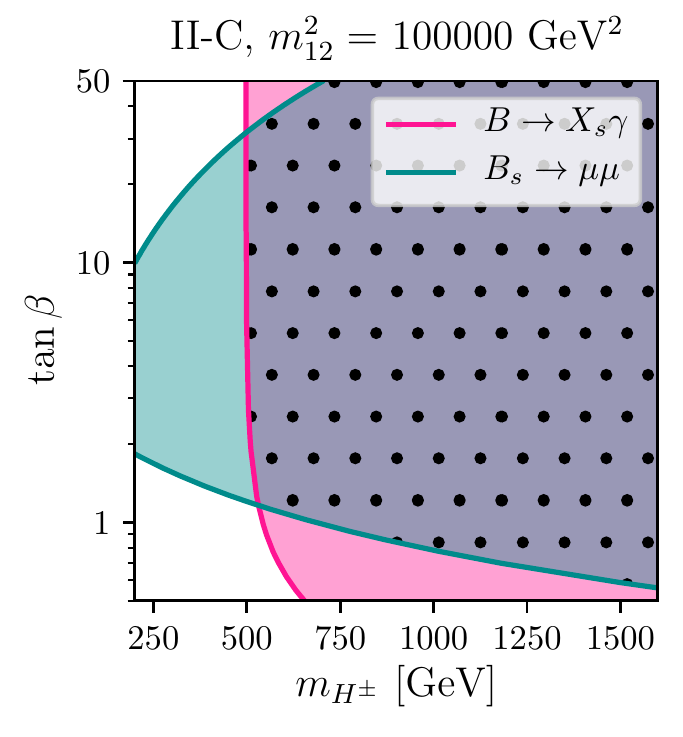}\includegraphics[width=0.33\textwidth]{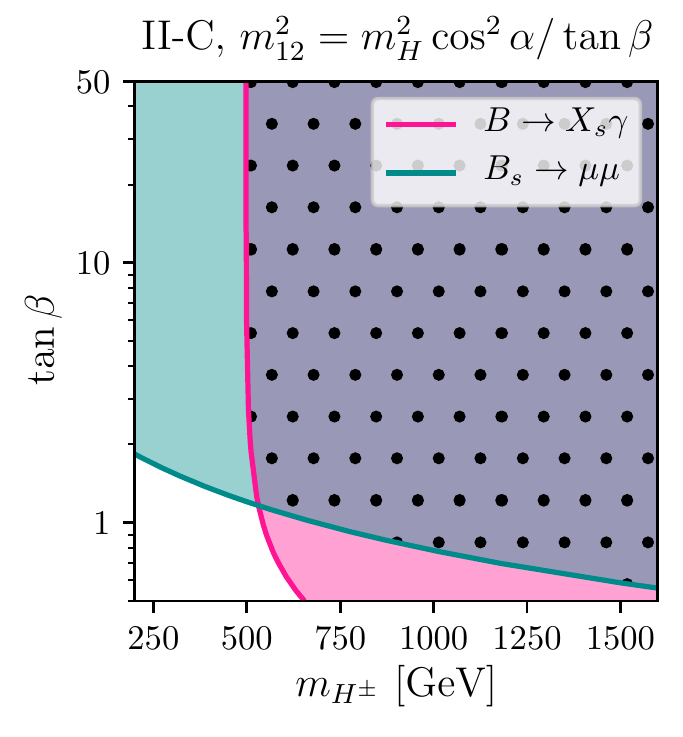}
	\caption{\label{fig:FLAV}
Allowed regions in the $(\MHp,\tb)$ plane of the 2HDM for
scenario~C with $m_H=m_A=m_{H^\pm}$ and for Yukawa types I (upper
row) and~II (lower row) for $\msq = 0, 100000 \gev$ and set via
\refeq{eq:m12special} in the left, middle and right column, respectively.
The alignment limit, $\CBA=0$, has been fixed. Pink areas are the
allowed regions by $B \to X_s \gamma$ and teal areas those allowed
by $B_s \to \mu^+ \mu^-$. Dotted areas are the intersections of these
two allowed regions.
}
\end{center}
\end{figure}

In \reffi{fig:FLAV} we present the allowed regions from the flavor
constraints in the $(\MHp,\tb)$ plane in the alignment limit for
scenario~C (all masses of BSM bosons degenerated) for Yukawa types~I
(upper row) and~II (lower row) for
$\msq = 0, 100000 \gev$ and set via \refeq{eq:m12special} in the left,
middle and right column, respectively. We show the regions allowed
by $B \to X_s \gamma$ (pink areas) and by $B_s \to \mu^+ \mu^-$ (teal
areas). Dotted areas are the intersections of these two allowed
regions. The 2HDM contribution to the process $B\to X_s\gamma$ depends on
the couplings of the $b$ and $s$ quarks with the other $u$-type quarks
through a charged Higgs boson. As the Yukawa couplings of the
charged Higgs bosons in the 2HDM scale like the ones of the $\cp$-odd
Higgs boson, this coupling is given by a combination of $\xi_A^{u,d}$
(see \refta{tab:yukcoupling}) and the quark masses. In the case of
model type~I those couplings are enhanced for large values of
$\cot\be$, in consequence the region of low $\tb$ is forbidden in the
top plots of \reffi{fig:FLAV} and softly fades as the mass of the
charged Higgs increases. On the contrary, in type~II it is found a well
known $\tb$ independent constraint of $\MHp>500\gev$. The BSM
contributions to $B\to X_s\gamma$ are induced from the Yukawa coupling and
therefore neither $\CBA$ or $\msq$ affects the bounds, as it can be
seen in the figure. 
Focusing on $B_s\to\mu^+\mu^-$ one finds a similar
constraint for low $\tb$ on both model types due to analogous arguments
discussed before for $B\to X_s\gamma$. Nevertheless, in model type~II
there is a disallowed region for large $\tb$ and low masses. This is due
to the contributions from the Higgs-penguin diagrams (mediated by $H$ and $h$) to the process
$B_s\to\mu^+\mu^-$ which are sensitive to $\msq$, via $\la_{HH^+H^-}$ and $\la_{hH^+H^-}$ from the loops involving charged Higgs bosons,  and that are enhanced at large 
$\tb$ (see also~\cite{Cheng:2015yfu}). The largest effect from $m_{12}^2$ on $B_s\to\mu^+\mu^-$ is from 
$\la_{HH^+H^-}$ since the $H$-penguin diagram goes as $\tan^3\beta$, and this leads to relevant constraints in the large $\tan\beta$ and low $m_{H^+}$ region. If,
however,  $\msq$ is fixed to \refeq{eq:m12special} and if the alignment
limit is taken, then the coupling $\la_{HH^+H^-}$
vanishes and in consequence the Higgs penguins contributions are not
large enough to give a bound in that region.


\section{Numerical results}
\label{sec:numres}

In this section we analyze numerically which intervals (or extreme
values) of the various triple Higgs boson couplings are still allowed, taking
into account all experimental and theoretical constraints as discussed
in \refse{sec:constr}. In the case of $\lahhh$ this will give a
guideline to which collider option may be needed to perform a precise
experimental determination. For the triple Higgs couplings involving heavy
Higgs bosons this will indicate in which processes large effects,
e.g.\ possibly enhanced production cross sections, can be expected due to
large triple Higgs couplings. 

We perform our evaluation in both type~I and type~II models (and leave
the other types for future investigations). We start our exploration with the
``simplest'' scenario~C, but later also explore scenario~A and~B. In the
headlines of our plots we indicate which type and which scenario are chosen.
The other parameters are chosen such as to maximize either the
deviations of $\lahhh$ from it SM value (where the plots below show
$\kala := \lahhh/\laSM$), or to maximize (positive or
negative) the size of the triple Higgs couplings involving the heavy Higgs
bosons (where the plots below show the triple Higgs couplings as defined in
\refeq{eq:lambda}).


\subsection{Scenario C}
\label{sec:scenC}

We start with scenario~C, i.e.\ $\MHp = \MH = \MA$%
\footnote{Here and in the following we will denote this common mass as
$\MHp$.}, and $\Mh = 125 \gev$. 
In \reffi{fig:C1-cba-tb} we show the $(\CBA, \tb)$ plane in the 2HDM
type~I, where $\msq$ is fixed by \refeq{eq:m12special} to maximize
the regions allowed by unitarity and stability of the potential, see
\refse{sec:theo}. The common Higgs boson mass scale is set to 
$\MHp = 1000 \gev$. Dotted areas always refer to the intersections of
the allowed regions by the various analysis involved. 
The first three panels of \reffi{fig:C1-cba-tb}(A) indicate the restrictions
from three sets of constraints. The upper left panel shows the areas allowed
by \HB\ and \HS, as discussed in \refses{sec:collider} and
\ref{sec:SMlike}. One can see that a wide area roughly centered around
$\CBA = 0$ (i.e.\ the alignment limit) is allowed by the direct BSM
Higgs-boson searches as well as by the requirement that the Higgs-boson
at $\sim 125 \gev$ is in agreement with the LHC rate measurements.
The upper right plot shows the constraints from flavor physics, as
discussed in \refse{sec:flavor}. Following the explanations given there,
in this realization of the type~I scenario the two constraints result in
lower limits on $\tb$, where $B \to X_s \ga$ gives the stronger
constraint.
The last set of constraints is given in the middle left plot, showing
the effects of requiring unitarity and stability of the potential as
discussed in \refse{sec:theo}.
The middle right plot indicates the intersection set of the three other
panels. Being in scenario~C the electroweak precision constraints, see
\refse{sec:stu} are automatically fulfilled. In the $(\CBA$,$\tb)$
plane this intersection defining the total allowed area starts at $\tb
\sim 2$ up to the highest 
investigated values, where we stopped at $\tb = 50$. $\CBA = 0$, is
allowed for all $\tb$ values, with a roughly triangular shape, extending
up to $\CBA \sim 0.2$.

\begin{figure}[p]
\begin{center}
	{\small 2HDM type I, scenario C, $\msq = (\MH^2\cos^2\al)/(\tb)$}
	
	\includegraphics[height=0.25\textheight]{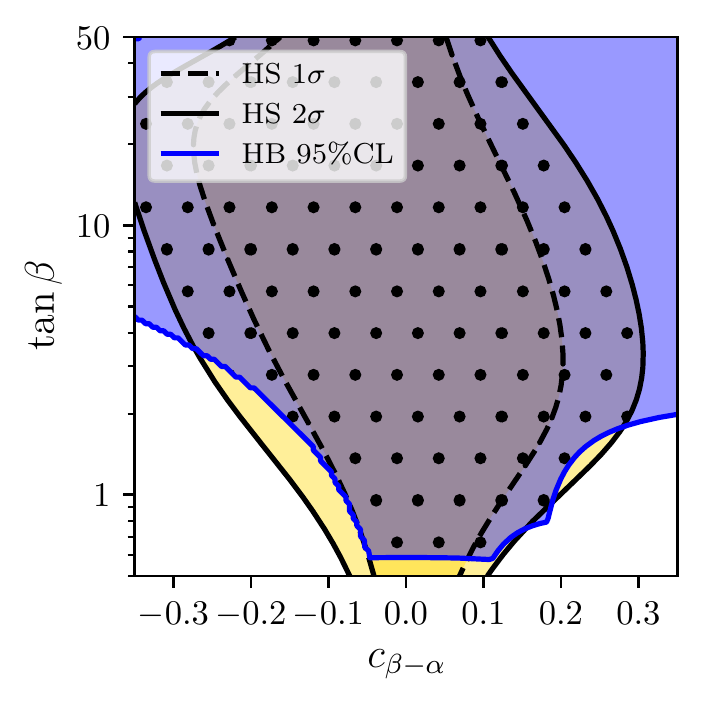}\includegraphics[height=0.25\textheight]{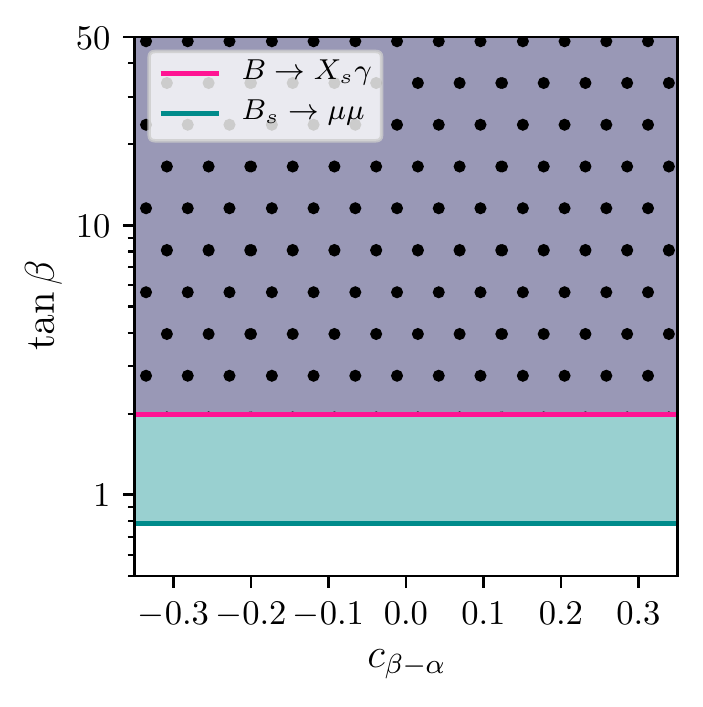}\vspace{-0.5em}
	\includegraphics[height=0.25\textheight]{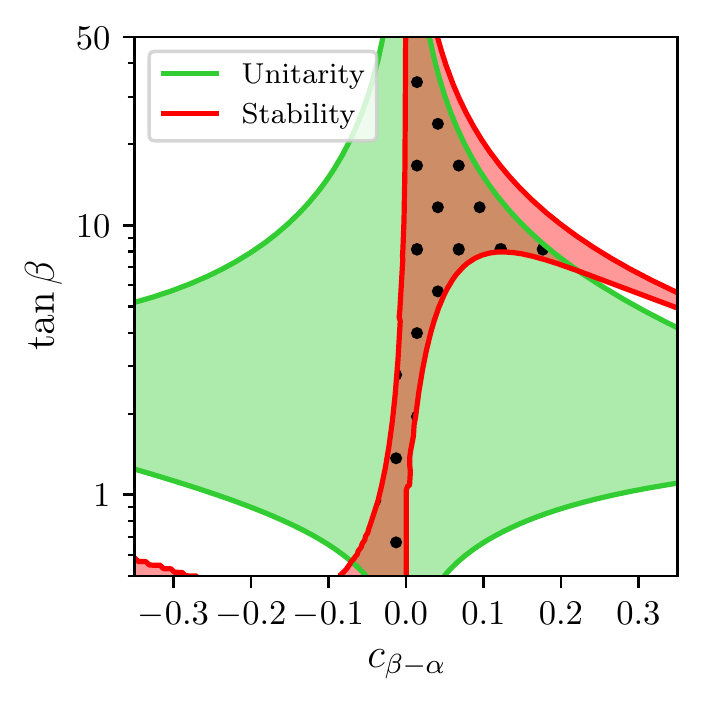}\includegraphics[height=0.25\textheight]{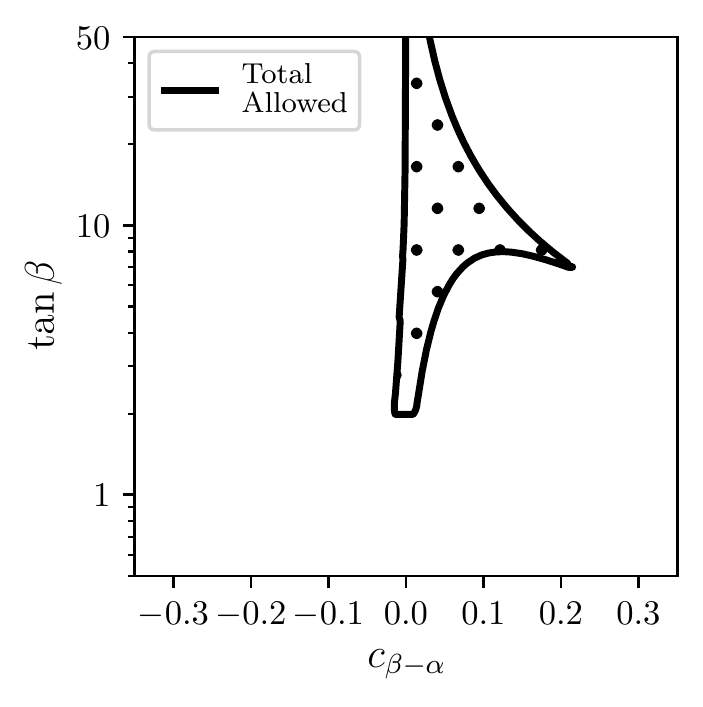}\vspace{-0.5em}
	\includegraphics[height=0.4\textheight]{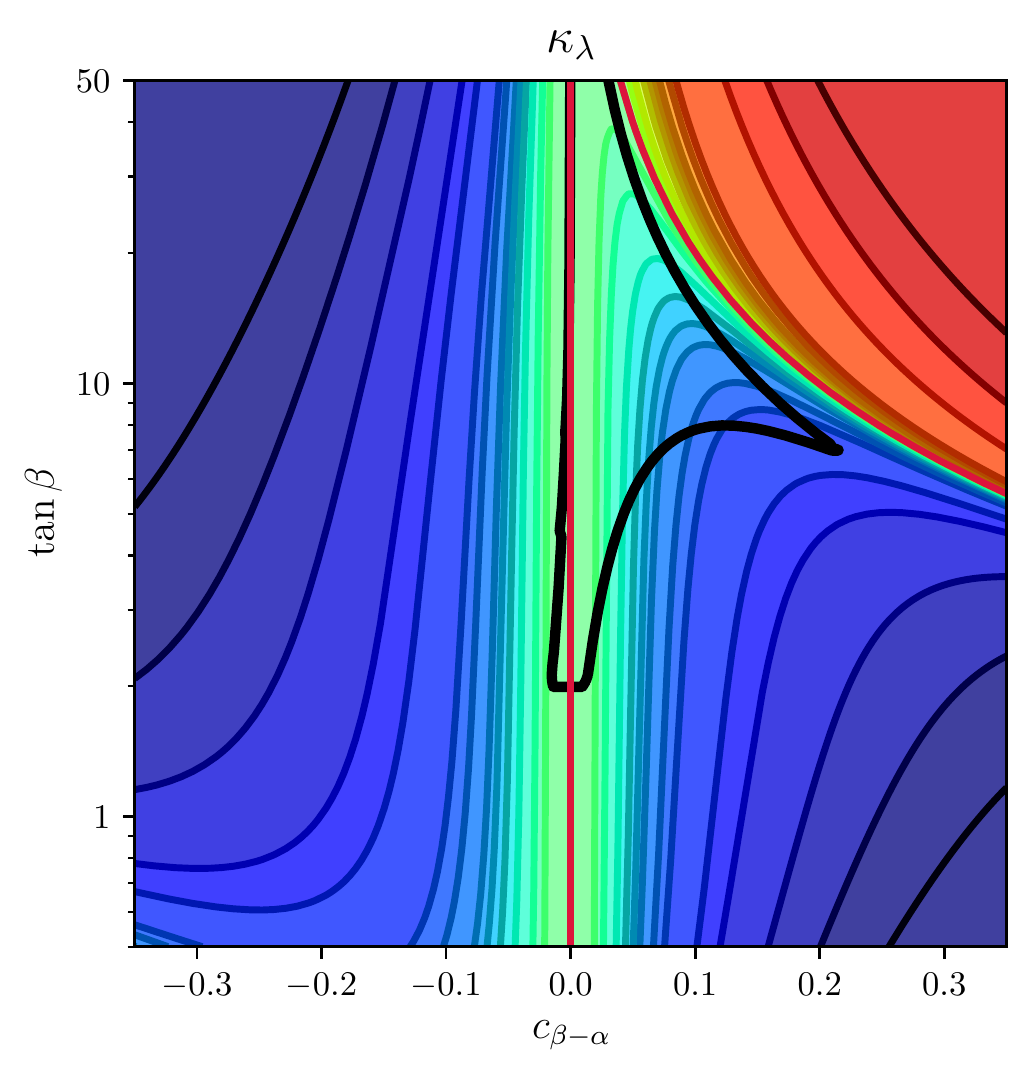}\includegraphics[height=0.4\textheight]{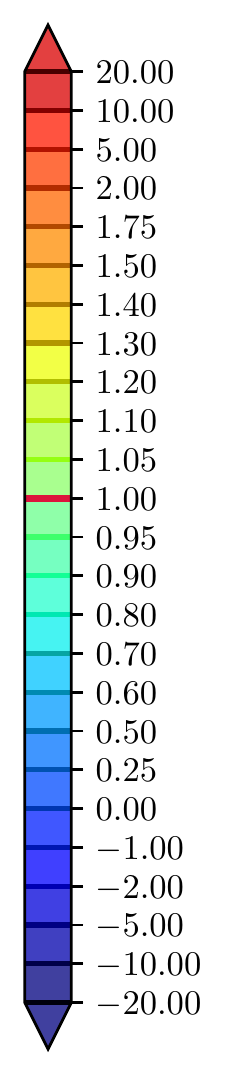}
\caption{
  {\bf (A)} \footnotesize{
Predictions for $\kala = \lahhh/\laSM$ in the 
2HDM type~I, scenario~C,  for $\MH = \MA = \MHp =1000 \gev$ and $\msq =
(\MH^2\cos^2\al)/(\tb)$ in the $(\CBA,\tb)$ plane.    
\emph{Upper left plot:} Allowed areas by  direct searches at colliders (blue), constraints from the SM-like Higgs boson properties (yellow) and both (dotted).
\emph{Upper right plot:} Allowed areas by flavor physics from
$B \to X_s \gamma$ (pink),
$B_s \to \mu^+ \mu-$ (teal) and both (dotted). 
\emph{Middle left plot:} Allowed areas by the theoretical constraints from unitarity (green), stability  (red) and  both (dotted).
\emph{Middle right plot:} Total allowed area (dotted). 
\emph{Lower big plot:} Contour lines of $\kala = \lahhh/\laSM$. 
Red contours correspond to $\kala=1$.
The thick solid contours is the boundary of the total allowed area.} 
}
\label{fig:C1-cba-tb}        
\end{center}
\end{figure}

\begin{figure}[t]\ContinuedFloat
\begin{center}
	{\small 2HDM type I, scenario C, $\msq = (\MH^2\cos^2\al)/(\tb)$}
	
	\begin{subfigure}[b]{0.7\textwidth}
		\includegraphics[height=0.25\textheight]{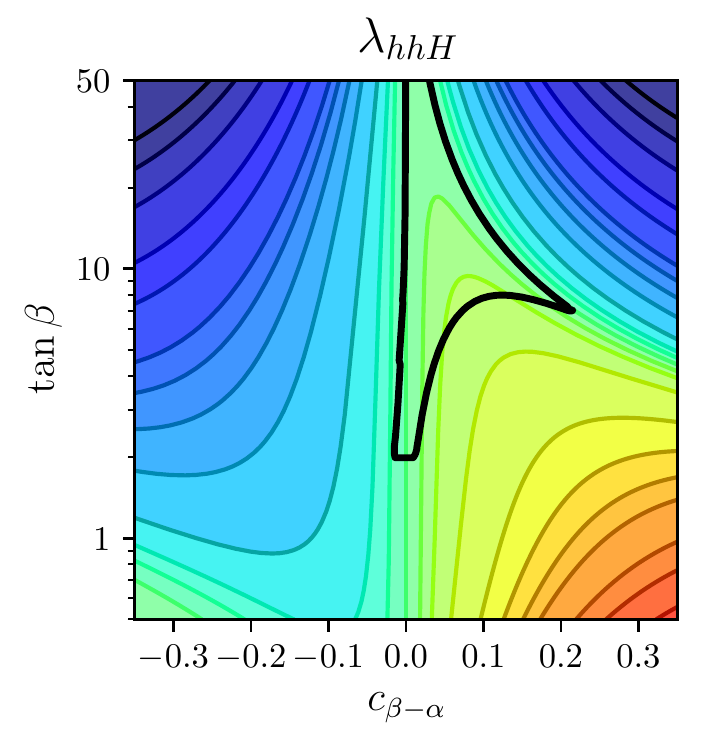}\includegraphics[height=0.25\textheight]{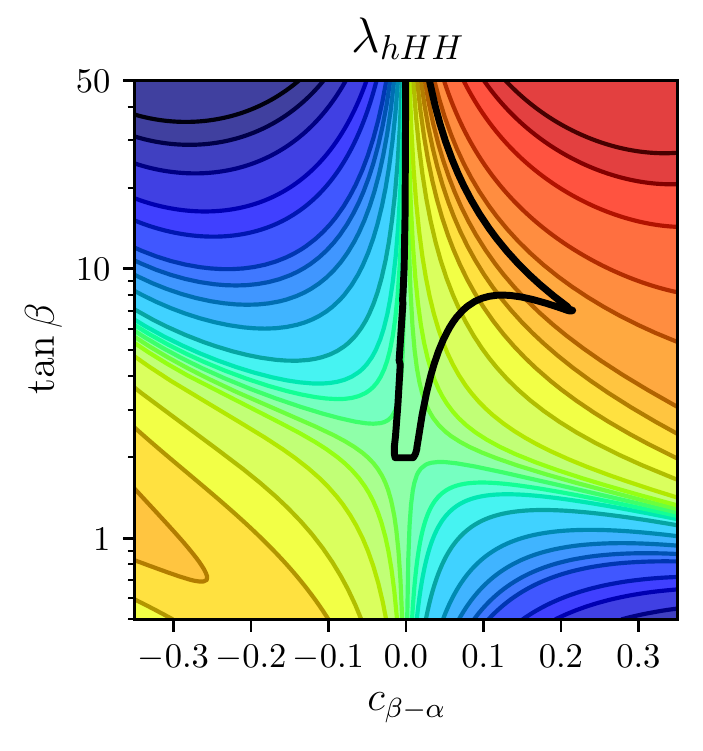}
		\includegraphics[height=0.25\textheight]{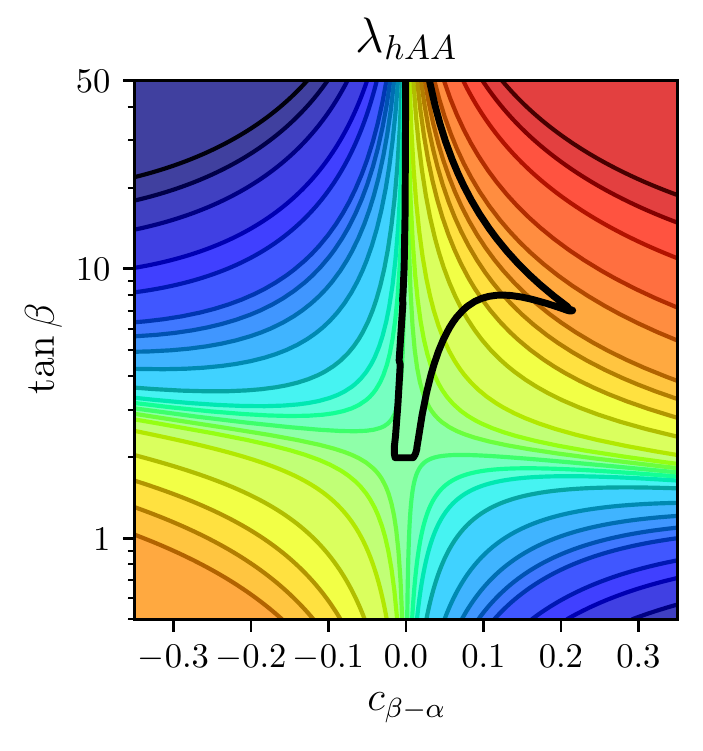}\includegraphics[height=0.25\textheight]{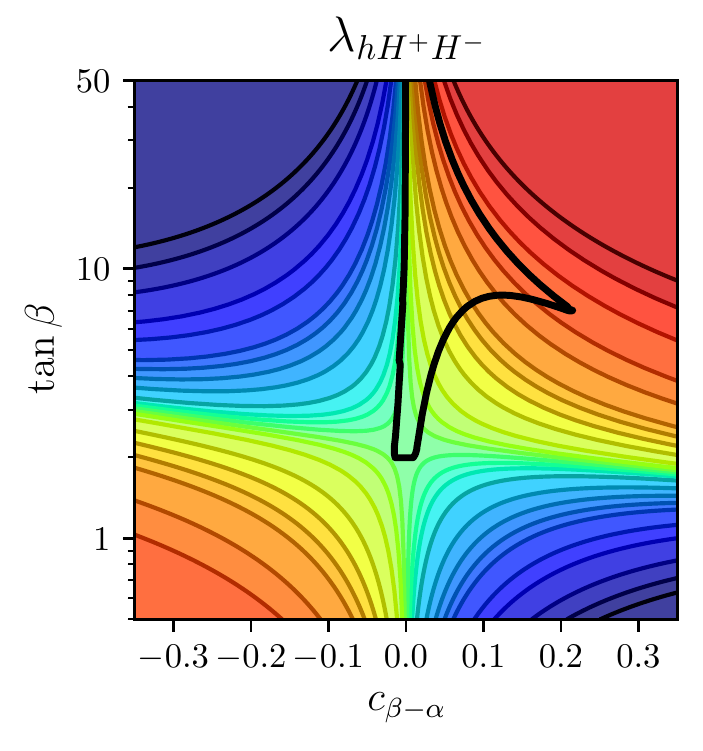}
	\end{subfigure}	
	\begin{subfigure}[b]{0.1\textwidth}
		\includegraphics[height=0.48\textheight]{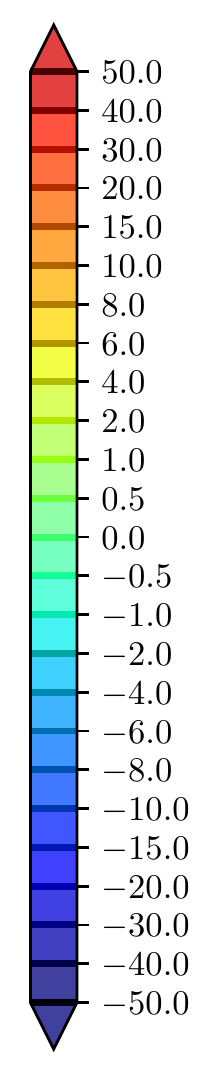}
	\end{subfigure}
\caption{
{\bf (B)} \footnotesize{
Contour lines for triple Higgs couplings in the 2HDM type~I, scenario~C,
for $\MH = \MA = \MHp =1000 \gev$ and $\msq = (\MH^2\cos^2\al)/(\tb)$
in the $(\CBA,\tb)$ plane.
\emph{Upper left:} $\lahhH$,
\emph{upper right:} $\lahHH$,
\emph{lower left:} $\lahAA$,
\emph{lower right:} $\lahHpHm$. 
The thick solid contour is as in \protect\reffi{fig:C1-cba-tb}(A). 
}}
\label{fig:C1-cba-tb-heavy}
\end{center}
\end{figure}

The results for $\kala=\lahhh/\laSM$ are presented in the
lower plot of \reffi{fig:C1-cba-tb}(A), with the total allowed area
discussed above being now marked  by the bounding black solid
line. The red solid line indicates $\kala \equiv 1$.  
This is either the alignment limit for $\CBA = 0$, or the ``wrong sign
limit'' in the upper right corner. For the latter, see the
  discussion in \refse{sec:SMlike}.
The color code shows the values 
reached by $\kala$. In the area allowed by all experimental and
theoretical constraints, values of $\kala \lsim 1$ are realized, going
down to $\kala \sim -0.4$ in the ``tip'' to the right of the allowed area. The
corresponding implications will be discussed in \refse{sec:impl}.

We now turn to the triple Higgs couplings involving at least one heavy
Higgs boson. In \reffi{fig:C1-cba-tb-heavy}(B) we show the results for
$\lahhH$, $\lahHH$, $\lahAA$ and $\lahHpHm$ in the upper left, upper
right, lower left and lower right plot, respectively. As before, the area
allowed by all experimental and theoretical constraints is indicated by
a black solid line, and the color code shows the values reached by the
triple Higgs couplings. In all four cases we find positive couplings with the minimum values reached
for $\CBA = 0$. The larger values are
found in the right edge of the allowed area, with largest values
(as in the case of $\lahhh$) in the ``tip''  to the right of the allowed area. 
$\lahhH$ is found to be larger 
around $\tb \sim 8$ and $\CBA \sim 0.1$. 
The maximum values found for the rest of the triple Higgs couplings
in this case are $\lahHH \sim 12$, $\lahAA \sim 12$
and $\lahHpHm \sim 24$. It should be noted that here and in the
following $\lahHpHm$ 
always reaches the maximum values of all the considered triple Higgs
boson couplings. The corresponding phenomenological 
implications will be discussed in \refse{sec:impl}.

\begin{figure}[p]
\begin{center}
	{\small 2HDM type~I, scenario C}
	
	\includegraphics[height=0.25\textheight]{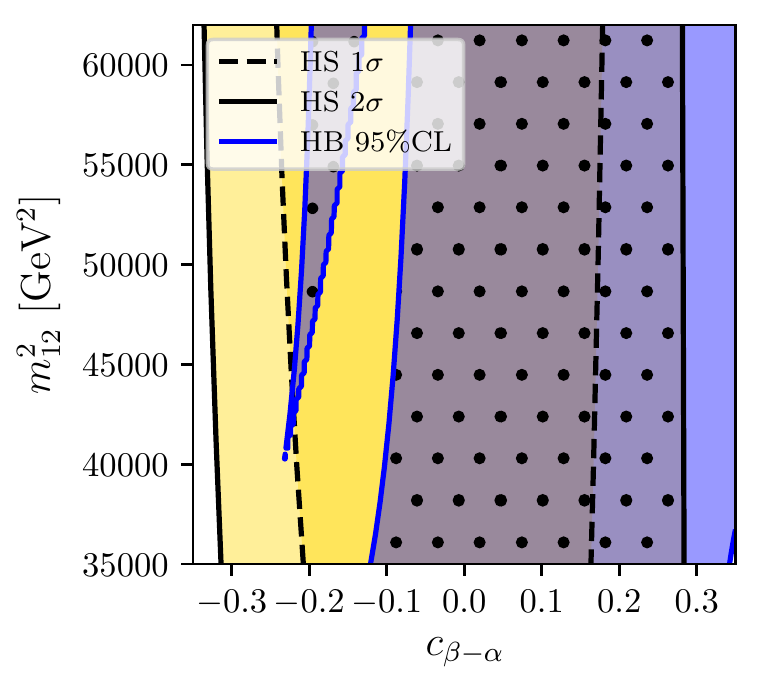}\includegraphics[height=0.25\textheight]{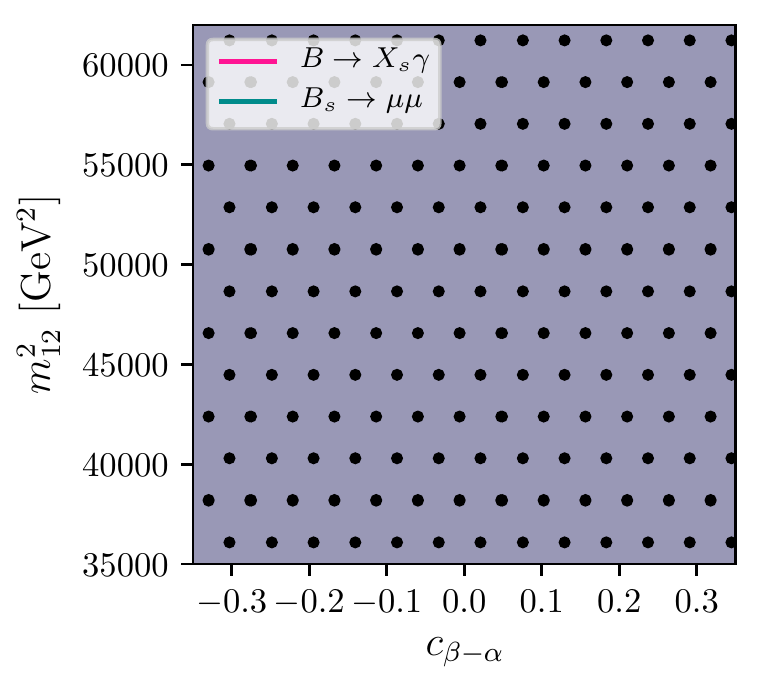}
	\includegraphics[height=0.25\textheight]{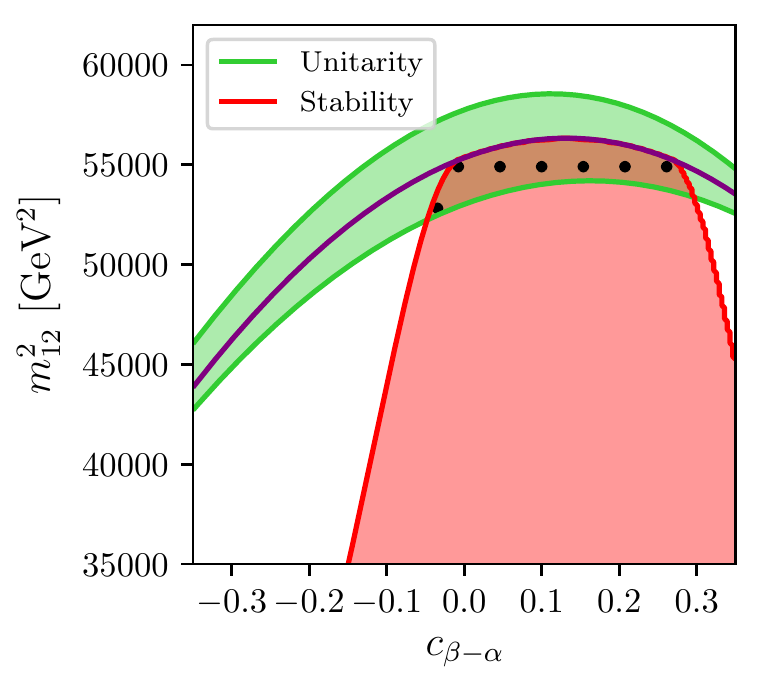}\includegraphics[height=0.25\textheight]{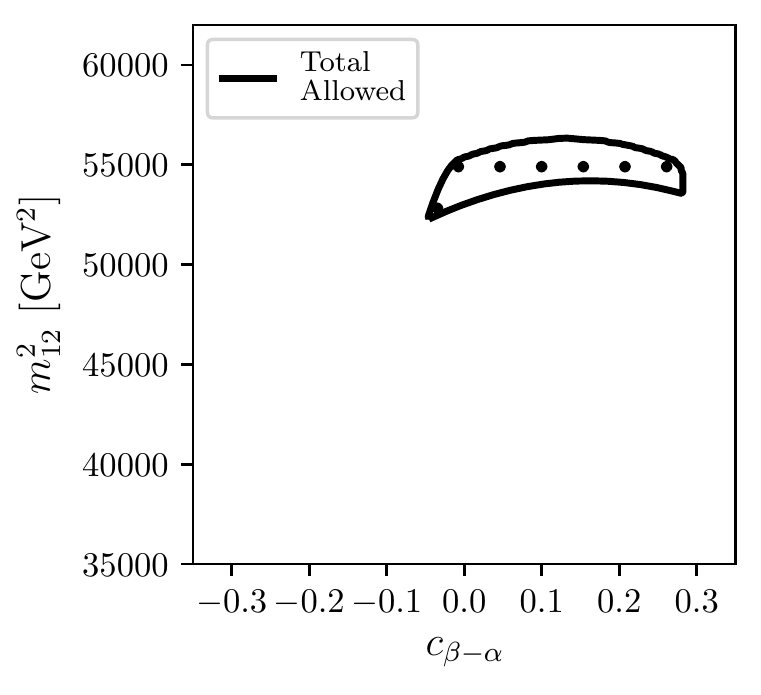}
	\includegraphics[height=0.4\textheight]{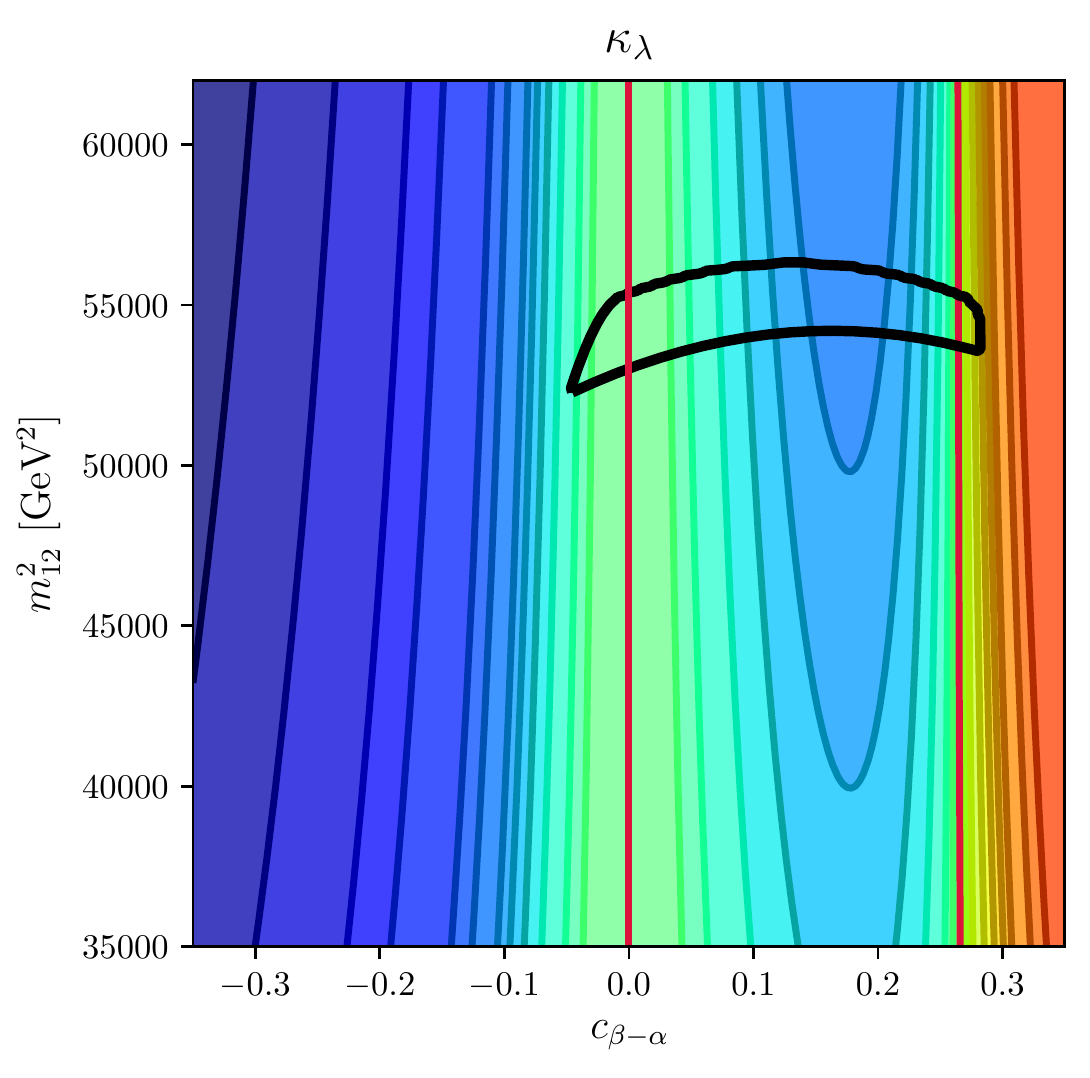}\includegraphics[height=0.4\textheight]{hhhcolorbar}
	\caption{
{\bf (A)} \footnotesize{
Predictions for $\kala = \lahhh/\laSM$ in the  2HDM type~I, scenario~C,
in the $(\CBA, \msq)$ plane 
for  $\MH = \MA = \MHp = 650 \gev$ and $\tb = 7.5$.}
The description of the allowed regions is as in \protect\reffi{fig:C1-cba-tb}(A).
Purple contour in the middle left plot satisfies the condition $\msq=(\MH^2\cos^2\al)/(\tb)$.
}
\label{fig:C1-cba-m122}
\end{center}
\end{figure}
\begin{figure}[t]\ContinuedFloat
\begin{center}
	{\small 2HDM type I, scenario C}
	
	\begin{subfigure}[b]{0.7\textwidth}
		\includegraphics[height=0.25\textheight]{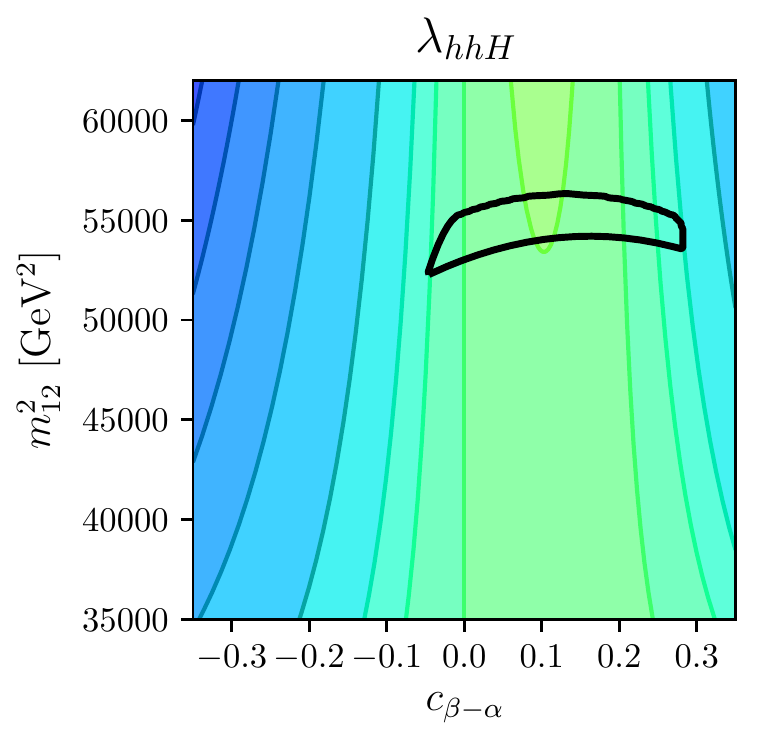}\includegraphics[height=0.25\textheight]{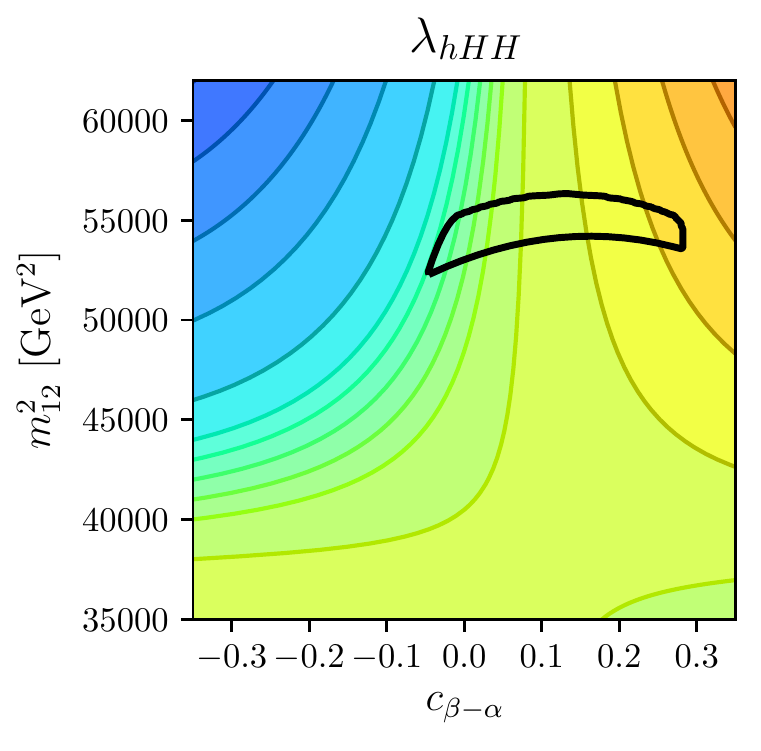}
		\includegraphics[height=0.25\textheight]{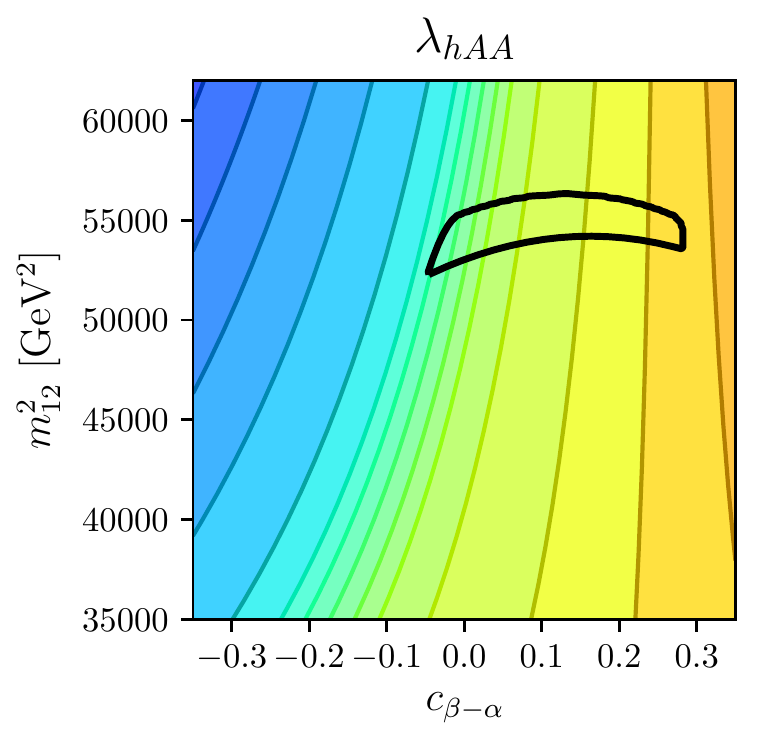}\includegraphics[height=0.25\textheight]{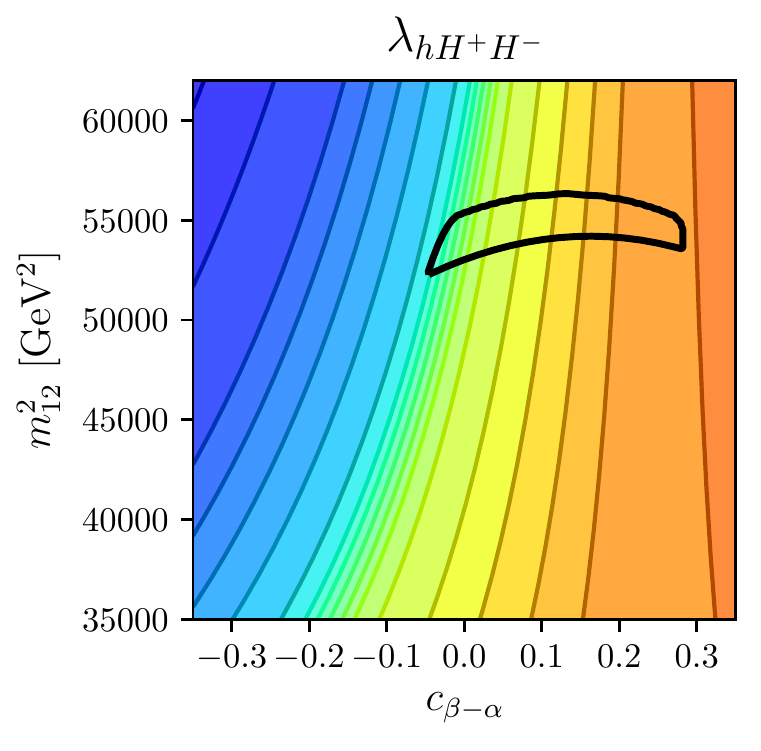}
	\end{subfigure}	
	\begin{subfigure}[b]{0.1\textwidth}
		\includegraphics[height=0.48\textheight]{heavycolorbar}
	\end{subfigure}
\caption{
{\bf (B)} \footnotesize{
Contour lines for triple Higgs couplings in the  2HDM type~I, scenario~C,
in the $(\CBA, \msq)$ plane 
for  $\MH = \MA = \MHp = 650 \gev$ and $\tb = 7.5$.
\emph{Upper left:} $\lahhH$,
\emph{upper right:} $\lahHH$,
\emph{lower left:} $\lahAA$,
\emph{lower right:} $\lahHpHm$. 
The thick solid contour is as in \protect\reffi{fig:C1-cba-m122}(A). 
}}
\label{fig:C1-cba-m122-heavy}
\end{center}
\end{figure}

\medskip
We continue the exploration of scenario~C, type~I in the $(\CBA, \msq)$
plane for $\MHp = \MH = \MA = 650 \gev$ and $\tb = 7.5$, as shown in
\reffi{fig:C1-cba-m122}. The sequence and the color coding of the plots is the
same as in 
\reffi{fig:C1-cba-tb}. The overall allowed area is restricted, particularly
by the requirement of unitarity and stability, to be within a curved
band around $\msq = 55000 \gev^2$, 
ranging from $\CBA \sim 0$ to $\CBA \sim 0.28$.
Here the purple solid line in the middle left plot indicates that 
\refeq{eq:m12special} is satisfied. The lower plot in
\reffi{fig:C1-cba-m122}(A) presents the results for $\kala$, which show a weak
dependence on $\msq$. Values of $\kala \sim 1$ are found around $\CBA = 0$
(as required by the alignment limit), but also around $\CBA \sim 0.26$. 
The lowest value of $\kala \sim 0.5$ is realized for $\CBA = 0.2$, 
whereas the highest value of $\kala \sim 1.2$ are found for 
$\CBA \sim 0.28$. Contrary to the $(\CBA, \tb)$ plane shown in
\reffi{fig:C1-cba-tb}, we now also encounter values of $\kala$ larger
than~1. However, these are realized for the largest departure of the
alignment limit, and thus will be under scrutiny by the next round of
Higgs-boson rate measurements at the LHC.

\begin{figure}[p]
\begin{center}
	{\small 2HDM type I, scenario C, $\msq = (\MH^2\cos^2\al)/(\tb)$}
	
	\includegraphics[height=0.25\textheight]{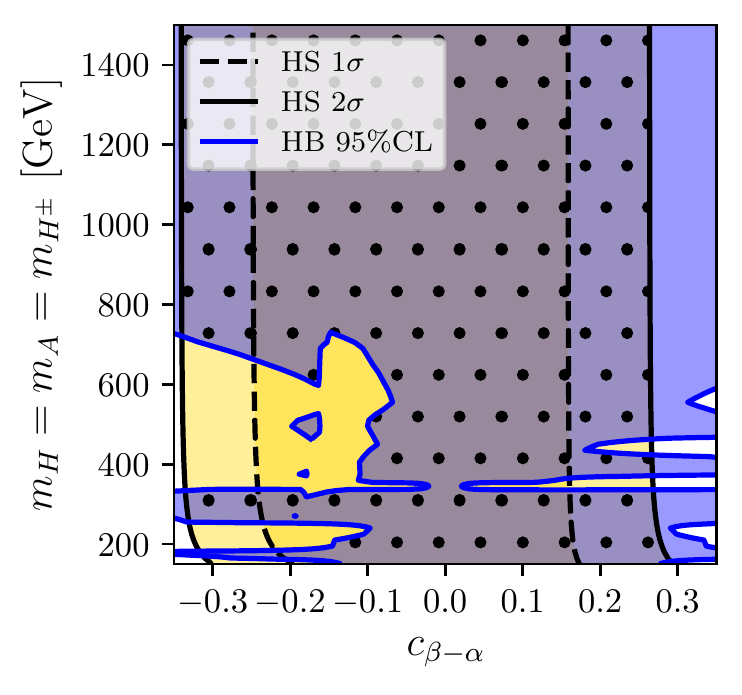}\includegraphics[height=0.25\textheight]{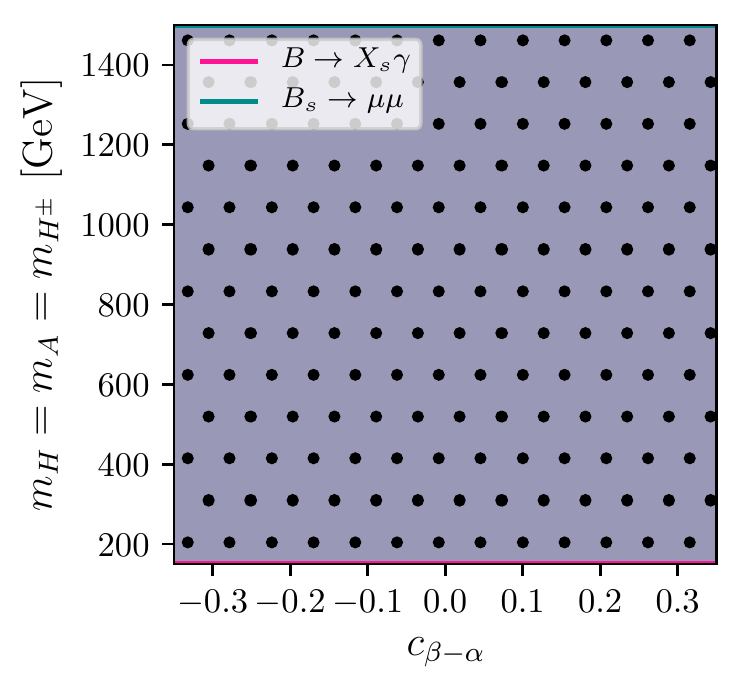}
	\includegraphics[height=0.25\textheight]{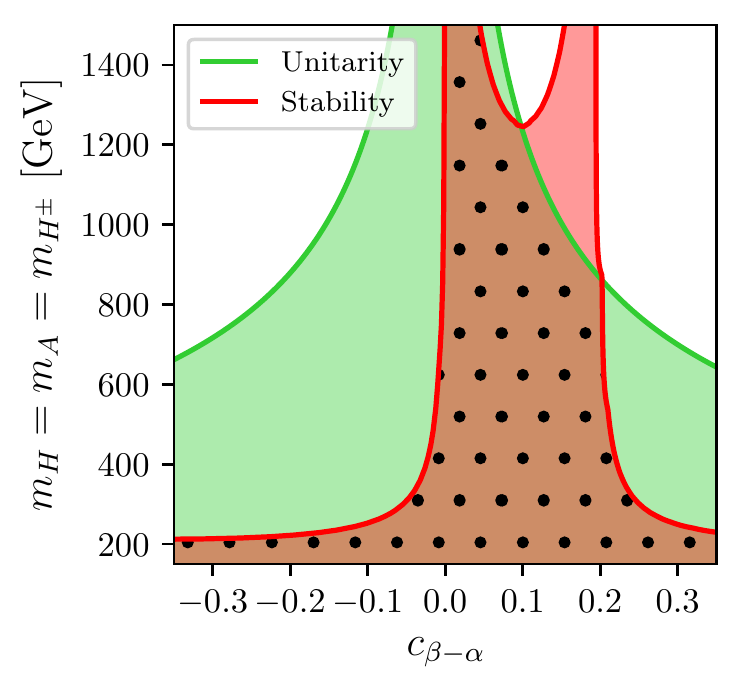}\includegraphics[height=0.25\textheight]{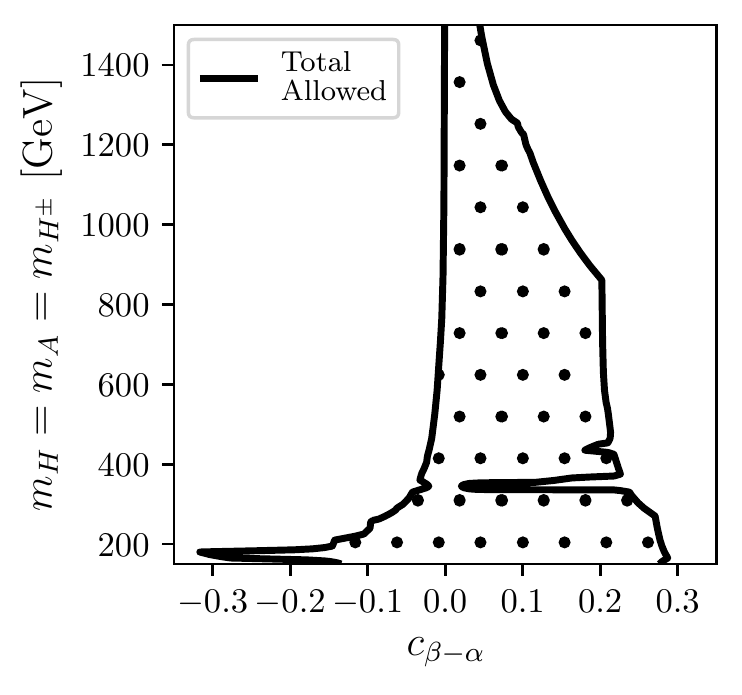}
	\includegraphics[height=0.4\textheight]{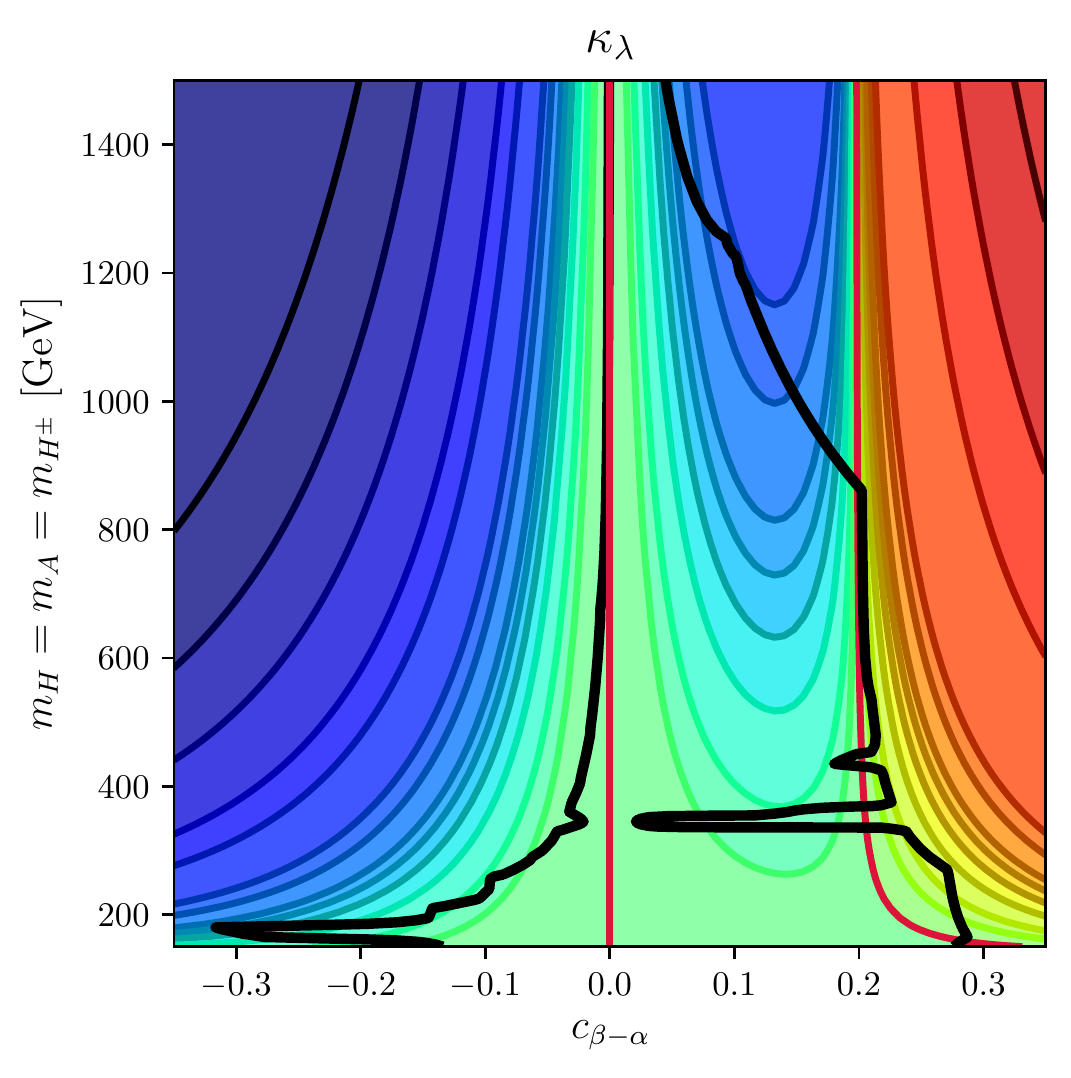}\includegraphics[height=0.4\textheight]{hhhcolorbar}
	\caption{
{\bf (A)} \footnotesize{
Predictions for $\kala = \lahhh/\laSM$ in the  2HDM type~I, scenario~C,
in the $(\CBA, m)$ plane with $m = \MH = \MA = \MHp$,
$\msq = (\MH^2\cos^2\al)/(\tb)$ and $\tb = 10$.}
The description of the allowed regions is
as in \protect\reffi{fig:C1-cba-tb}(A).
}
\label{fig:C1-cba-MHp}
\end{center}
\end{figure}
\begin{figure}[t]\ContinuedFloat
\begin{center}
	{\small 2HDM type I, scenario C, $\msq = (\MH^2\cos^2\al)/(\tb)$}
	
	\begin{subfigure}[b]{0.7\textwidth}
		\includegraphics[height=0.25\textheight]{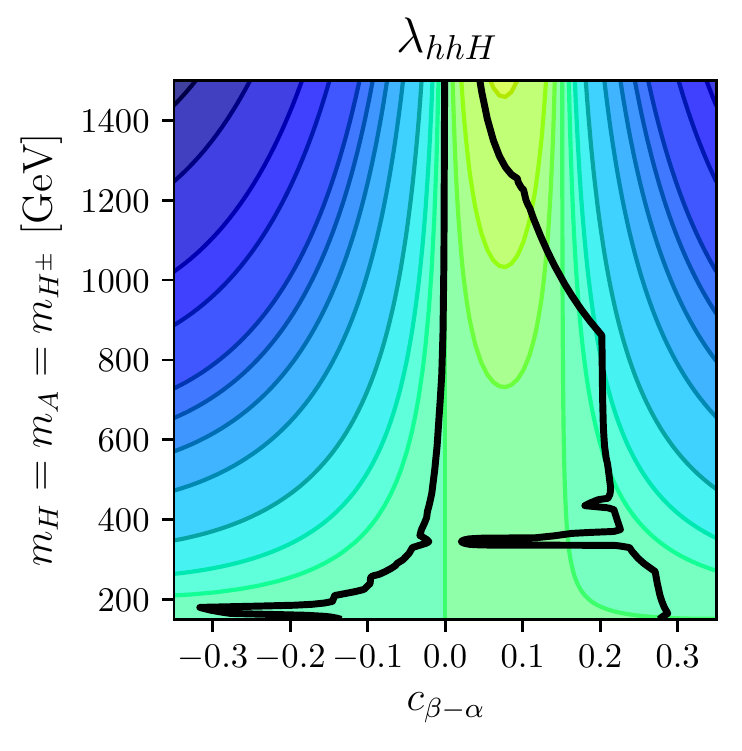}\includegraphics[height=0.25\textheight]{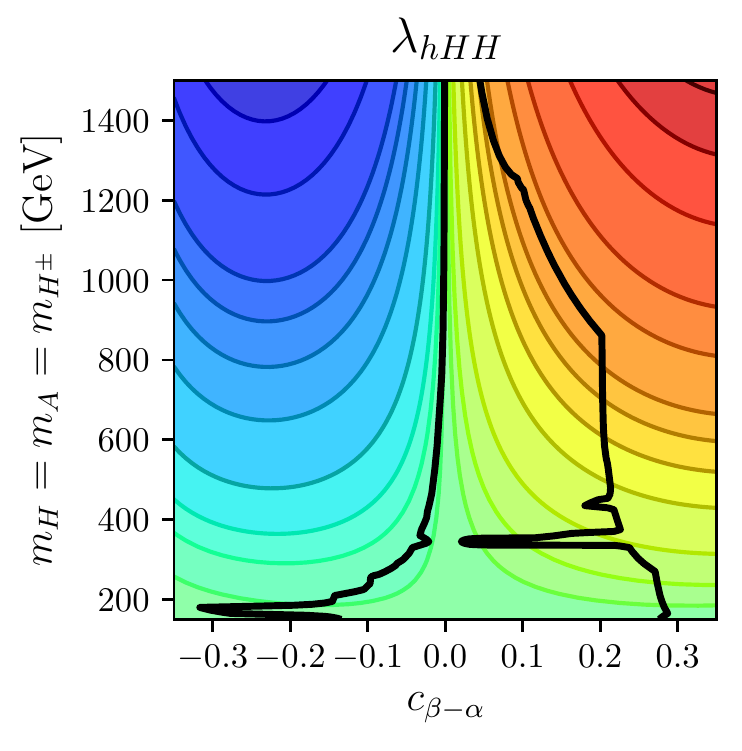}
		\includegraphics[height=0.25\textheight]{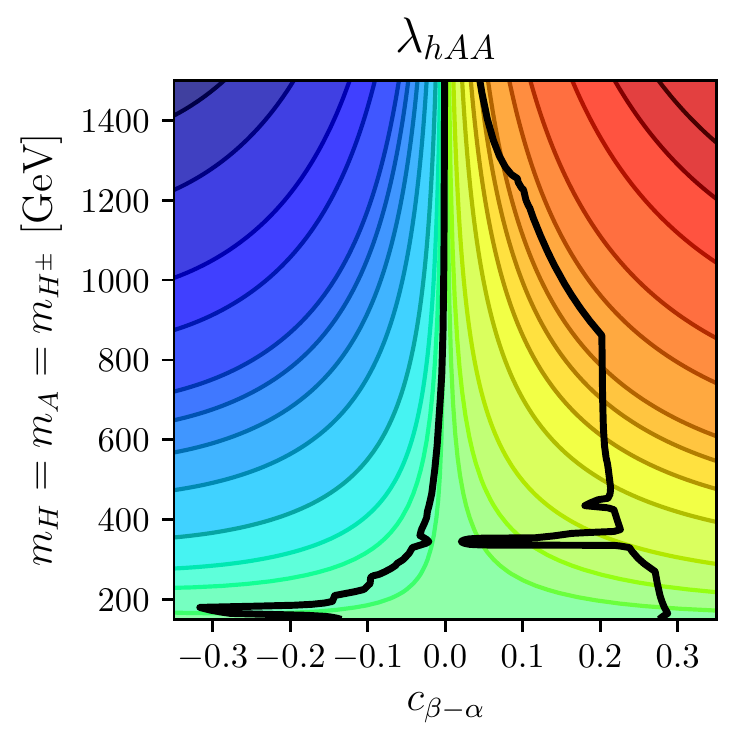}\includegraphics[height=0.25\textheight]{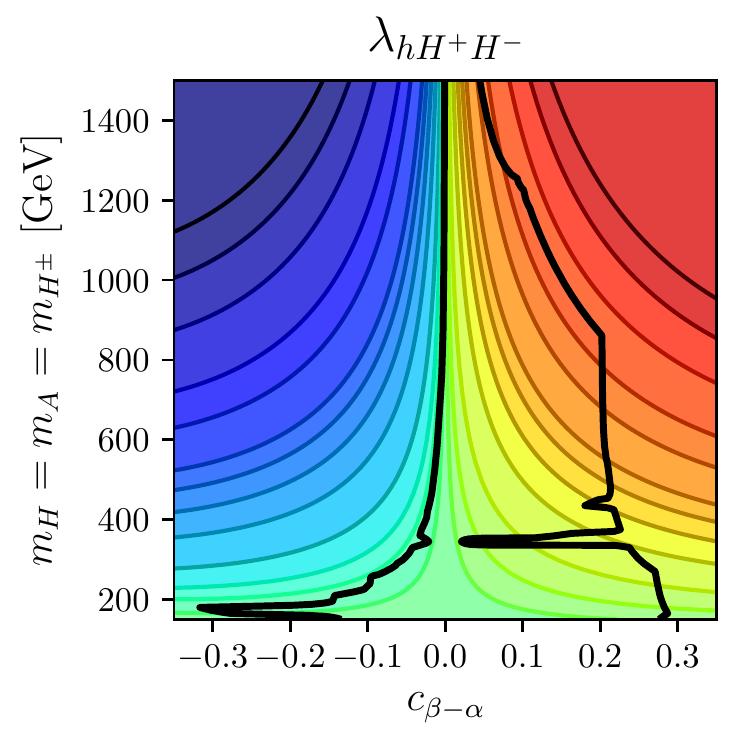}
	\end{subfigure}	
	\begin{subfigure}[b]{0.1\textwidth}
		\includegraphics[height=0.48\textheight]{heavycolorbar}
	\end{subfigure}
\caption{
{\bf (B)} \footnotesize{
Contour lines for triple Higgs couplings in the  2HDM type~I, scenario~C,
in the $(\CBA, m)$ plane for  $m = \MH = \MA = \MHp$,
$\msq = (\MH^2\cos^2\al)/(\tb)$ and $\tb = 10$.
\emph{Upper left:} $\lahhH$,
\emph{upper right:} $\lahHH$,
\emph{lower left:} $\lahAA$,
\emph{lower right:} $\lahHpHm$. 
The thick solid contour is as in \protect\reffi{fig:C1-cba-MHp}(A). 
}}
\label{fig:C1-cba-MHp-heavy}
\end{center}
\end{figure}

The results for the triple Higgs couplings involving heavy Higgs bosons are
shown in \reffi{fig:C1-cba-m122-heavy}(B), analogous to
\reffi{fig:C1-cba-tb-heavy}(B). As for $\lahhh$ the variation with $\msq$
(in the allowed interval) is relatively small. The intervals found in
this case are 
$\lahhH \sim \inter{-1}{0.3}$,
$\lahHH \sim \inter{-0.3}{7}$,
$\lahAA \sim \inter{-0.3}{7}$ and
$\lahHpHm \sim \inter{-0.5}{14}$.
It should be noted that due to the contribution from $m_{12}^2$ here
these couplings can also be slightly negative. 
As before, the maximum of $\lahhH$ is found for $\CBA\sim0.1$ whereas
for the other couplings, which can be of ${\cal O}(10)$,  the largest values are realized for the largest
departure of the alignment limit, and thus will be under scrutiny by the
next round of Higgs-boson rate measurements at the LHC.

\begin{figure}[p]
\begin{center}
	{\small 2HDM type II, scenario C}
	
	\includegraphics[height=0.25\textheight]{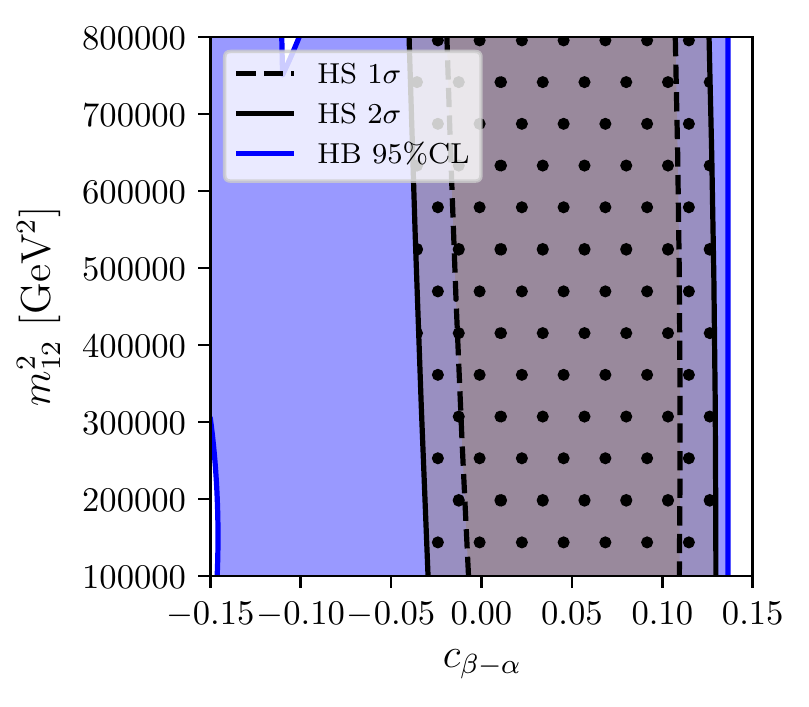}\includegraphics[height=0.25\textheight]{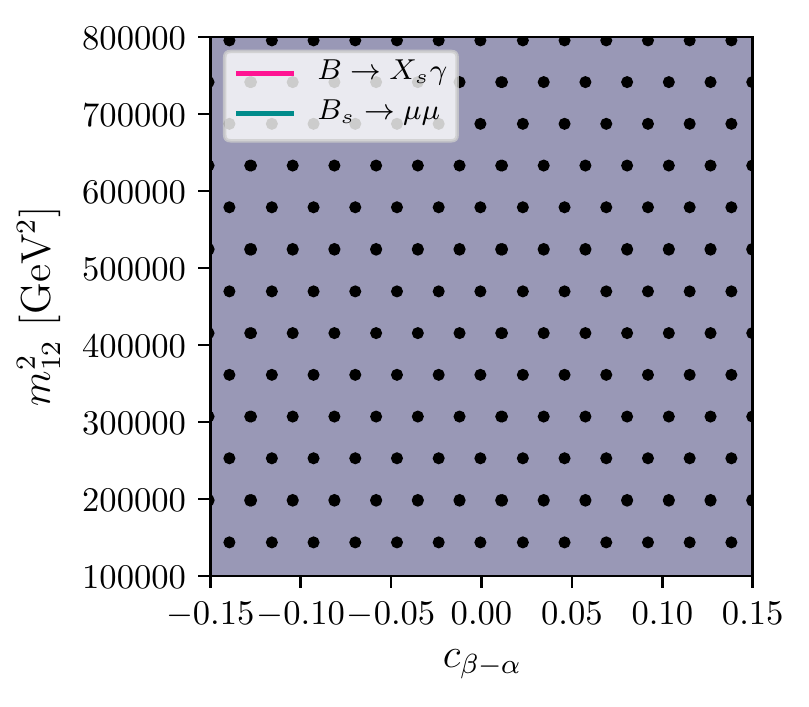}
	\includegraphics[height=0.25\textheight]{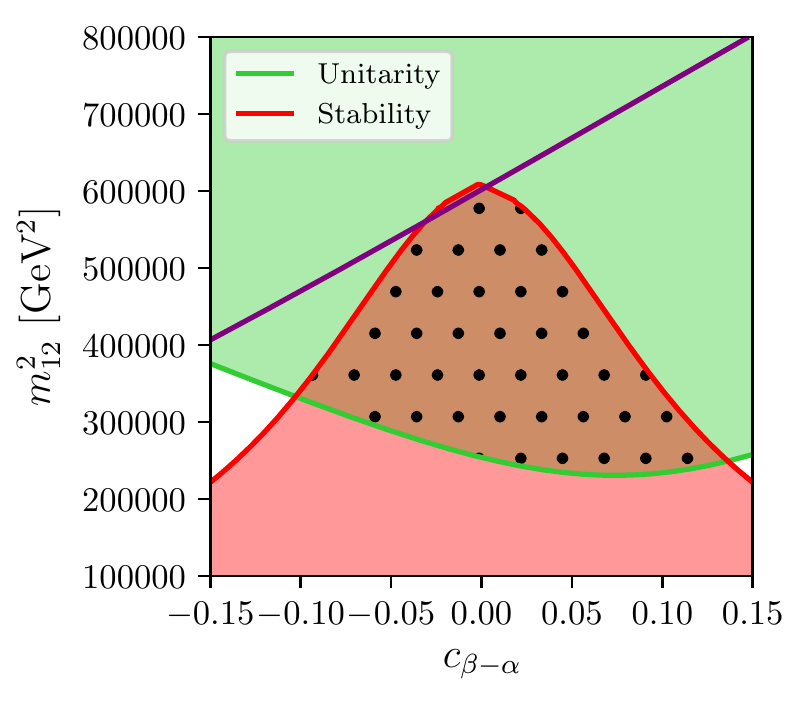}\includegraphics[height=0.25\textheight]{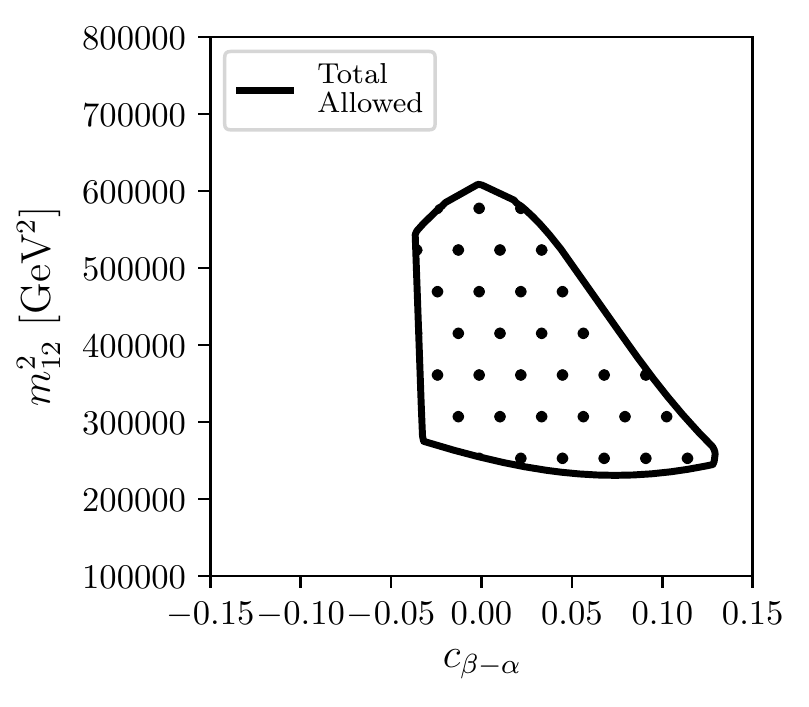}
	\includegraphics[height=0.4\textheight]{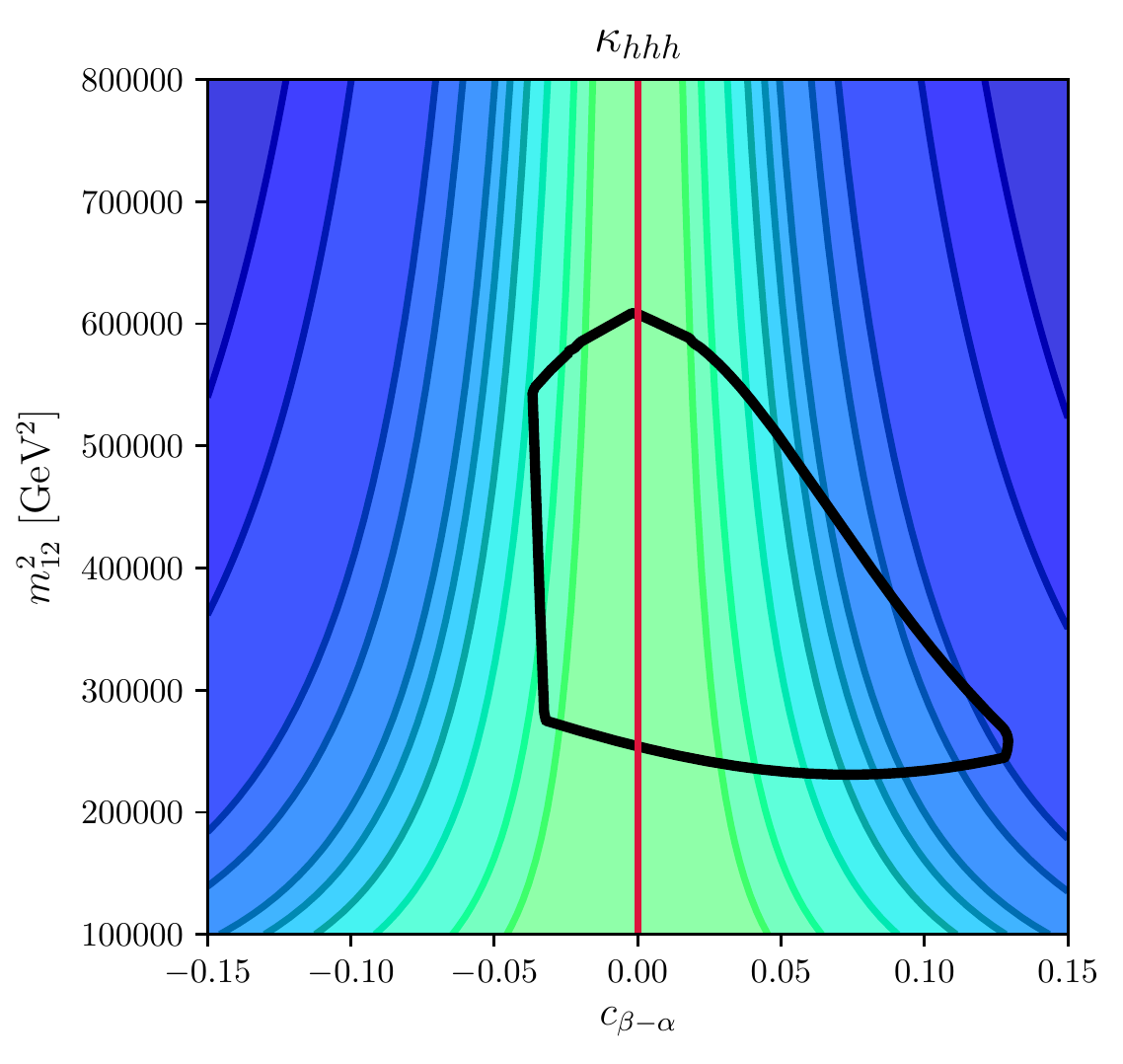}\includegraphics[height=0.4\textheight]{hhhcolorbar}
\caption{
{\bf (A)} \footnotesize{
Predictions for $\kala = \lahhh/\laSM$ in the  2HDM type~II, scenario~C,
in the $(\CBA, \msq)$ plane with $\MH = \MA = \MHp = 1100 \gev$
and $\tb = 0.9$.}
The description of the allowed regions is
as in \protect\reffi{fig:C1-cba-tb}(A).
Purple contour in the middle left plot satisfies the condition $\msq=(\MH^2\cos^2\al)/(\tb)$.
}
\label{fig:C2-cba-m122}
\end{center}
\end{figure}
\begin{figure}[t]\ContinuedFloat
\begin{center}
	{\small 2HDM type II, scenario C}
	
	\begin{subfigure}[b]{0.75\textwidth}
		\includegraphics[height=0.25\textheight]{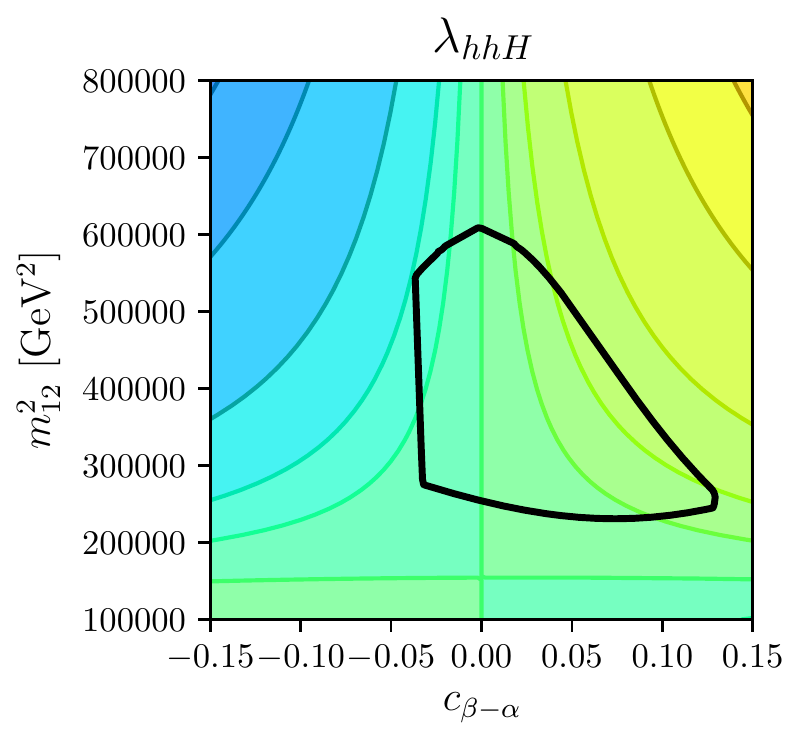}\includegraphics[height=0.25\textheight]{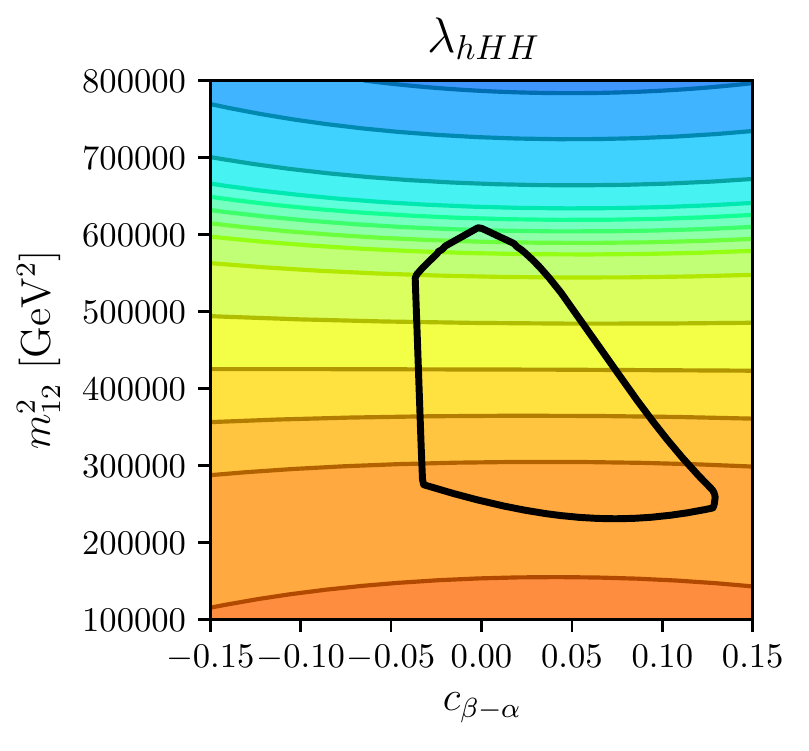}
		\includegraphics[height=0.25\textheight]{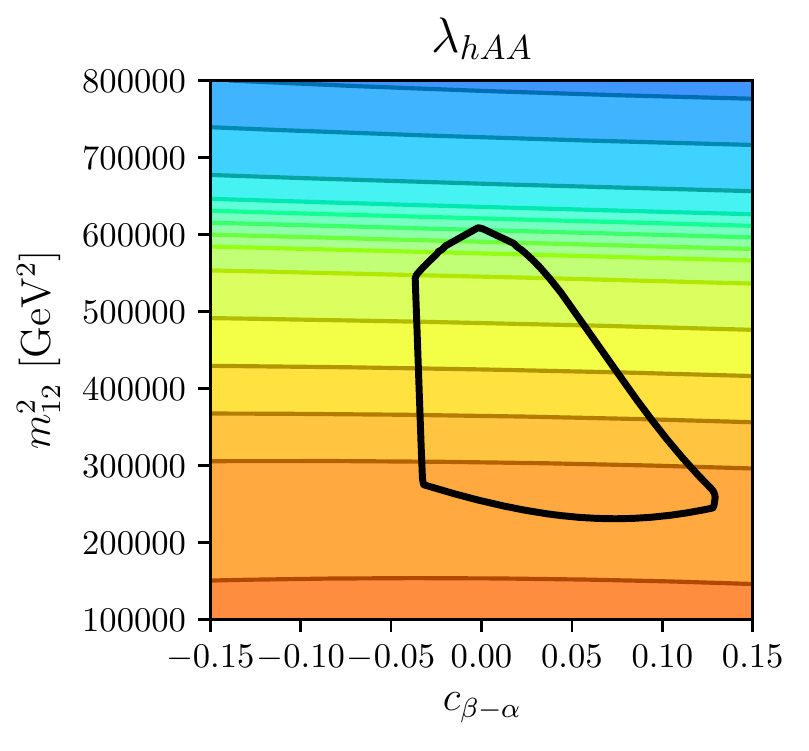}\includegraphics[height=0.25\textheight]{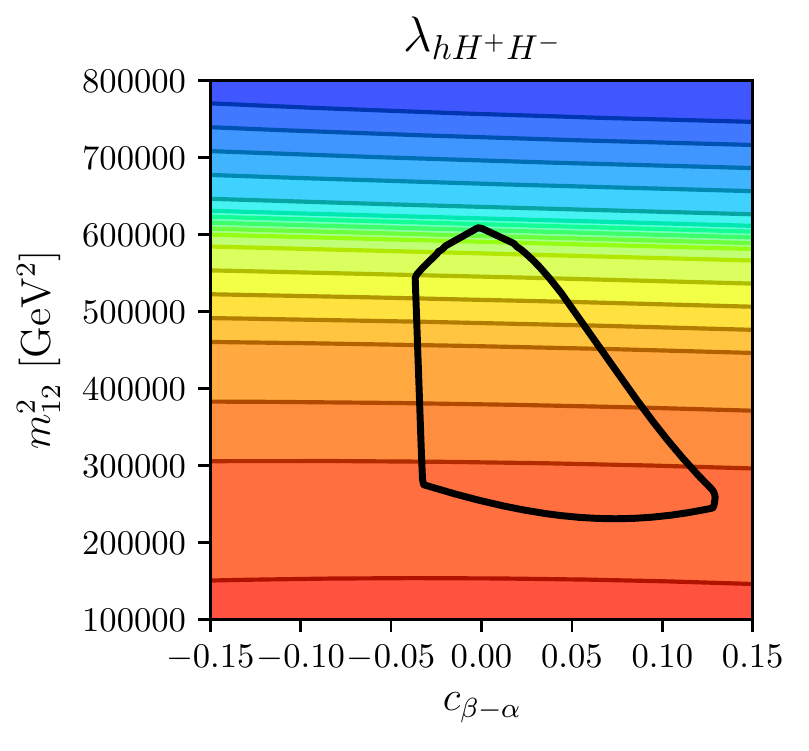}
	\end{subfigure}	
	\begin{subfigure}[b]{0.1\textwidth}
		\includegraphics[height=0.48\textheight]{heavycolorbar}
	\end{subfigure}
\caption{
{\bf (B)} \footnotesize{
Contour lines for triple Higgs couplings in the  2HDM type~II, scenario~C,
in the $(\CBA, \msq)$ plane for  $\MH = \MA = \MHp = 1100 \gev$,
and $\tb = 0.9$.
\emph{Upper left:} $\lahhH$,
\emph{upper right:} $\lahHH$,
\emph{lower left:} $\lahAA$,
\emph{lower right:} $\lahHpHm$. 
The thick solid contour is as in \protect\reffi{fig:C2-cba-m122}(A). 
}}
\label{fig:C2-cba-m122-heavy}
\end{center}
\end{figure}

\medskip
We finish our analysis of the scenario~C, type~I in \reffi{fig:C1-cba-MHp}, 
where we show the $(\CBA, \MH = \MA = \MHp)$ plane, and
where $\msq$ is fixed by \refeq{eq:m12special} to maximize
the regions allowed by unitarity and stability of the potential, and
with $\tb = 10$. The sequence and the color coding of the plots is the
same as in \reffi{fig:C1-cba-tb}. The upper left plot in
\reffi{fig:C1-cba-tb}(A) shows the areas allowed
by \HB\ and \HS, as discussed in \refses{sec:collider} and
\ref{sec:SMlike}. The \HB\ allowed area exhibits several spikes around 
$\MH = \MA = \MHp \sim 400 \gev$. Here the exclusion bounds are stemming
from the channel $gg \to A \to Zh \to l \bar l \, b \bar b$~\cite{CMS:2018xvc},
which exhibits several ``spikes'' which we identified as due to
statistical fluctuations in the experimental limits. 
The overall allowed area now exhibits positive {\em and} negative values
of $\CBA$ for low $\MHp<400 \gev$. For larger masses only positive values are
allowed, reaching slightly above $\CBA \sim 0.2$.

The values that can be reached by $\kala$, as shown in the lower plot
of \reffi{fig:C1-cba-MHp}(A), range from $\kala \sim 0.07$
for $\CBA \sim 0.1$ and large $\MHp$ close to $1200 \gev$ to about $\kala
\sim 1.2$ for the largest allowed $\CBA$ values and $\MHp \sim 300 \gev$.  
The ranges reached by the triple Higgs couplings involving at least one
heavy Higgs boson, as shown in \reffi{fig:C1-cba-MHp-heavy}(B), are found
to be
$\lahhH \sim \inter{-0.2}{1.6}$,
$\lahHH \sim \inter{-0.2}{12}$, 
$\lahAA \sim \inter{-0.2}{12}$ and
$\lahHpHm \sim \inter{-0.5}{24}$.
The largest values of $\lahhH$ are found for $\CBA\sim0.1$ and large $\MHp$ and for
the rest are found on the edge for larger $\CBA$ and 
$\MHp \gsim 800 \gev$.

\medskip
We finish our analysis of scenario~C with the $(\CBA, \msq)$ plane in
the 2HDM type~II for $\MH = \MA = \MHp = 1000 \gev$ and $\tb = 0.9$,
as presented in \reffi{fig:C2-cba-m122}. 
The sequence of the plots and the color coding are as in
\reffi{fig:C1-cba-m122}. The total allowed area is found, roughly between 
$\CBA \sim -0.05$ and $\CBA \lsim 0.1$, as well as
$\msq \gsim 2\cdot 10^5 \gev^2$ and $\msq \lsim 6 \cdot 10^5 \gev^2$.

The values that can be reached by  $\kala$, as shown in the lower plot
of \reffi{fig:C2-cba-m122}(A), range from $\kala \sim 0.0$
for $\CBA \sim 0.13$ and low $\msq$ to $\kala = 1$ for
the alignment limit.
The ranges reached by the triple Higgs couplings involving at least one
heavy Higgs boson, as shown in \reffi{fig:C2-cba-m122-heavy}(B), are found
to be
$\lahhH \sim \inter{-1}{1.4}$,
$\lahHH \sim \inter{-0.2}{12}$,
$\lahAA \sim \inter{-0.2}{12}$ and
$\lahHpHm \sim \inter{-0.4}{24}$.
Again negative values can be reached, due to the effects caused by $\msq$.
The largest values for $\lahHH$, $\lahHpHm$ and $\lahAA$ are found for
the lowest allowed $\msq$ values, and are nearly independent on $\CBA$.
In contrast, $\lahhH$ shows dependence on both variables where its
maximum is found around $\msq \sim 400000 \gev^2$ and $\CBA \sim 0.08$
and its minimum is found around $\msq \sim 550000 \gev^2$ and
$\CBA \sim -0.03$.
As for the 2HDM type~I, the phenomenological interpretation of these
intervals will be given in \refse{sec:impl}.



\subsection{Scenario A}
\label{sec:scenA}

We continue our numerical investigation by relaxing the conditions for the
heavy Higgs-boson masses and evaluate the triple Higgs-boson couplings in
scenario~A, as defined in \refse{sec:stu}, $\MA = \MHp \neq \MH$%
\footnote{Here and in the following we will denote this common mass as
$\MHp$.}, and $\Mh = 125 \gev$. 

\begin{figure}[p]
\begin{center}
	{\small 2HDM type I, scnario A, $\msq = (\MH^2\cos^2\al)/(\tb)$}
	\begin{subfigure}[b]{0.51\textheight}
		\includegraphics[width=0.48\textwidth]{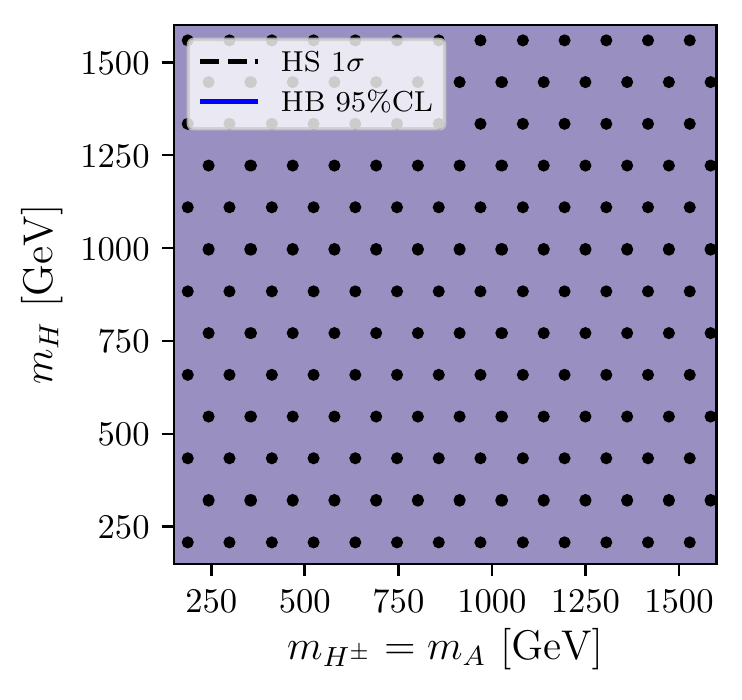}\includegraphics[width=0.48\textwidth]{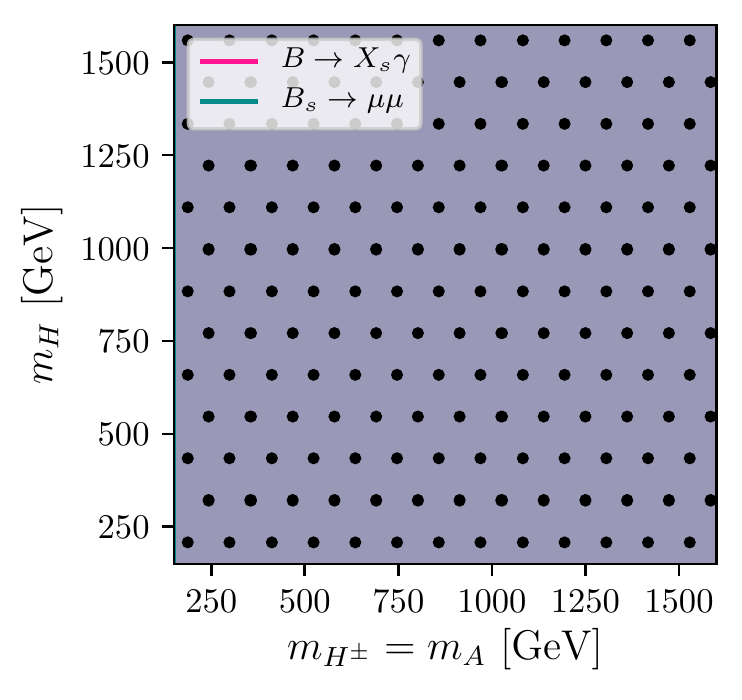}
		\includegraphics[width=0.48\textwidth]{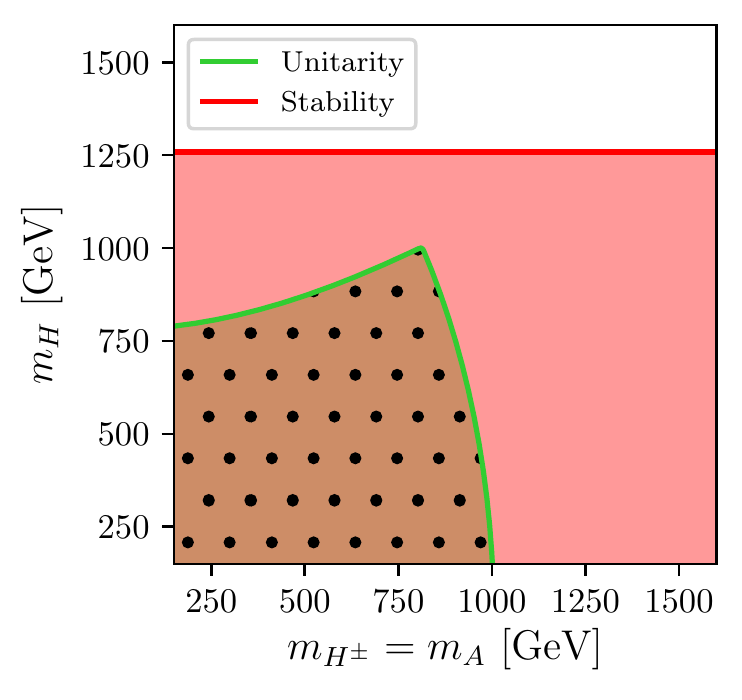}\includegraphics[width=0.48\textwidth]{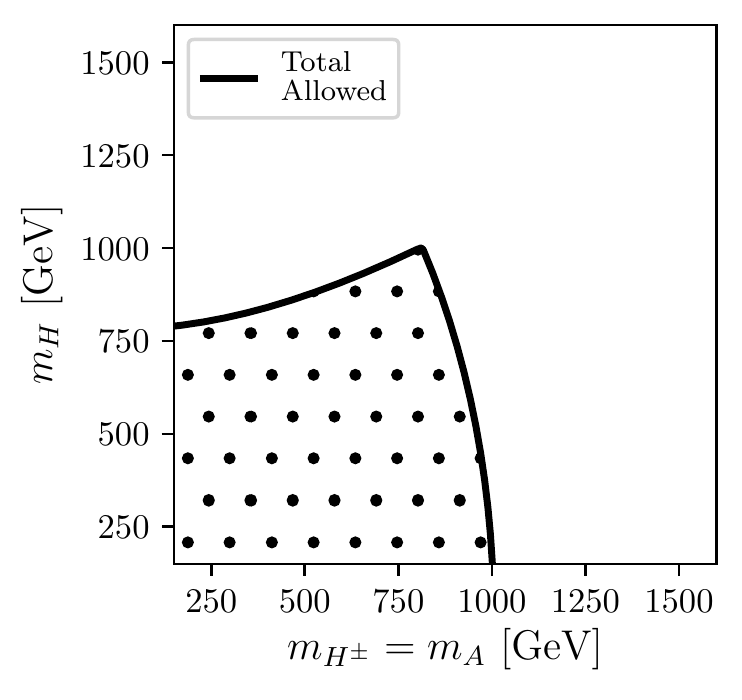}
		\includegraphics[width=0.48\textwidth]{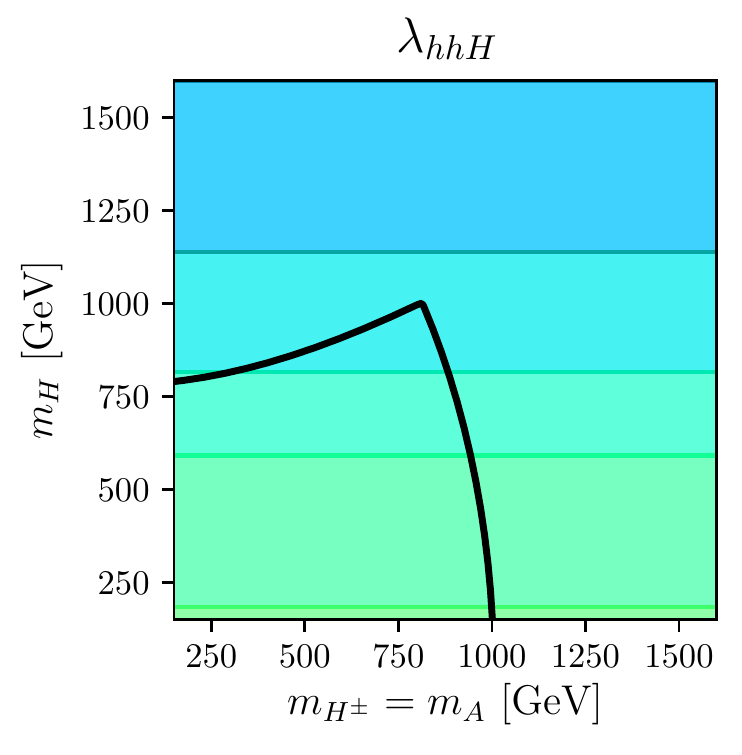}\includegraphics[width=0.48\textwidth]{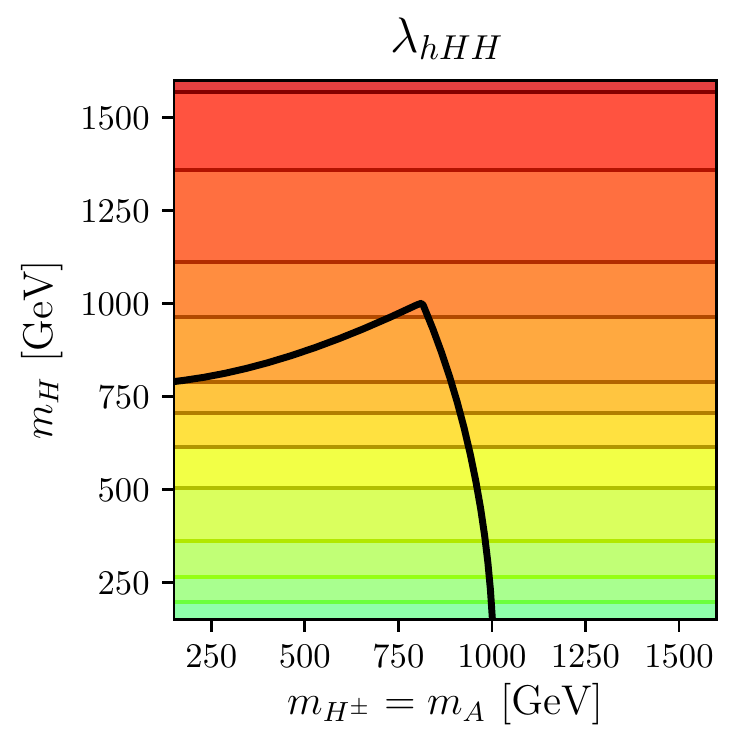}
		\includegraphics[width=0.48\textwidth]{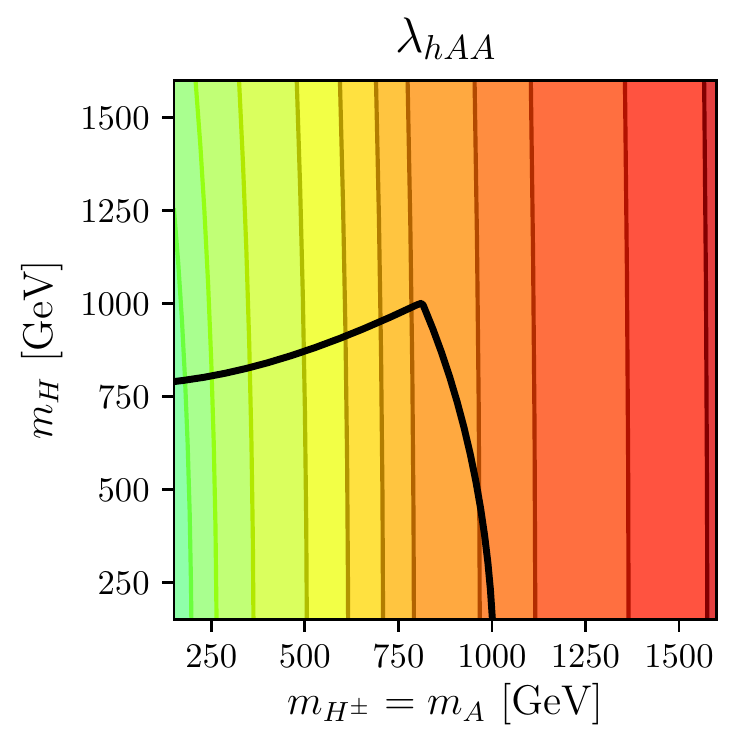}\includegraphics[width=0.48\textwidth]{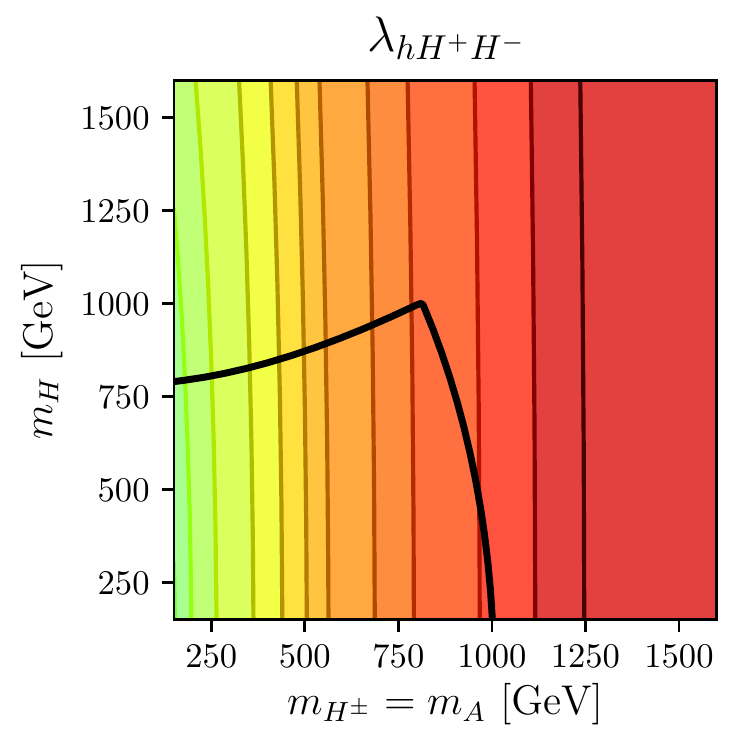}	
	\end{subfigure}	
	\begin{subfigure}[b]{0.06\textwidth}
		\includegraphics[height=0.48\textheight]{heavycolorbar}
	\end{subfigure}
\caption{\footnotesize{
Predictions for triple Higgs couplings in the  2HDM type~I, scenario~A,
in the $(\MHp = \MA, \MH)$ plane with $\CBA = 0.2$, $\tb = 10$ and
$\msq = (\MH^2\cos^2\al)/(\tb)$.
\emph{Upper four plots:} allowed regions 
as in \protect\reffi{fig:C1-cba-tb}(A). 
\emph{Third line left:} $\lahhH$,
\emph{third line right:} $\lahHH$,
\emph{lower left:} $\lahAA$,
\emph{lower right:} $\lahHpHm$. 
}}
\label{fig:A1-MHp-MH}
\end{center}
\end{figure}

In \reffi{fig:A1-MHp-MH} we present the $(\MHp = \MA, \MH)$ plane with 
$\msq = (\MH^2\cos^2\al)/(\tb)$, to maximize the parameter space allowed by
unitarity and stability of the Higgs potential, and $\CBA = 0.2$ and 
$\tb = 10$. The upper two rows show the various constraints, with the same
color coding as in \reffi{fig:C1-cba-tb}. One can see that the LHC
searches and measurements, as well as the flavor observables allow for the
whole plane. Unitarity and stability roughly select a square bounded from
above by $\MHp \sim \MH \sim 1000 \gev$. 
The results for $\lahhh$ are not explicitly shown, as they vary only very
weakly in the chosen scenario. The values reached are in the interval
$\kala \sim \inter{0.98}{1.02}$. 
The lower two rows in \reffi{fig:A1-MHp-MH} show the results for the triple 
Higgs couplings involving at least one heavy Higgs boson. The upper left plot
(of the two lower rows) shows $\lahhH$, which is independent of $\MHp$. Lowest
(highest) values are reached for high (low) values of $\MH$, following the
analytic result in \refeq{eq:hhH_phys}. They range from 0.02 to
-1.5.

\begin{figure}[p]
\begin{center}
	{\small 2HDM type II, scenario A, $\msq = (\MH^2\cos^2\al)/(\tb)$}
	\begin{subfigure}[b]{0.51\textheight}
		\includegraphics[width=0.48\textwidth]{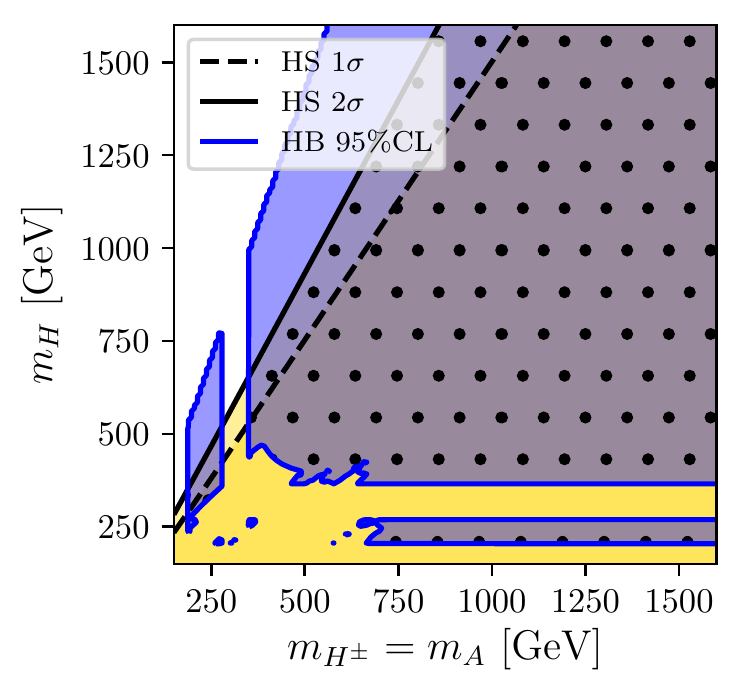}\includegraphics[width=0.48\textwidth]{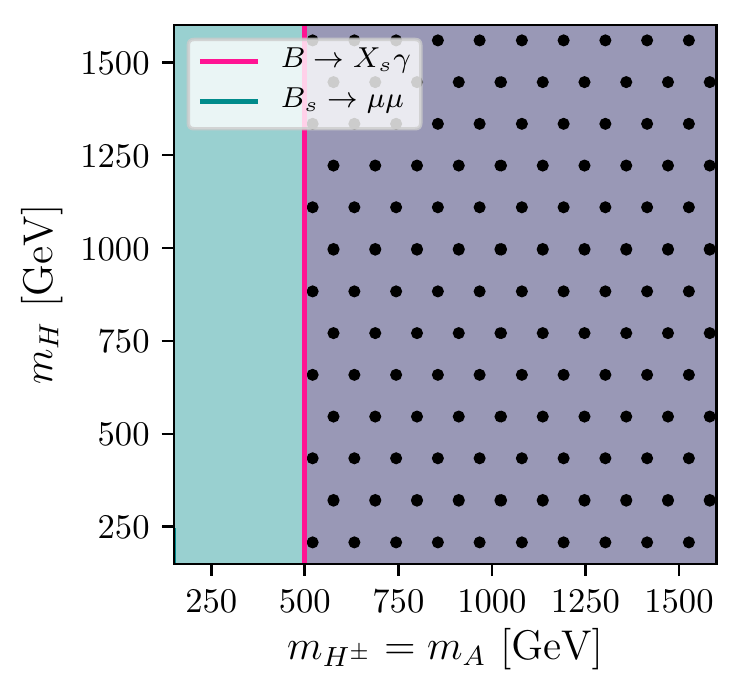}
		\includegraphics[width=0.48\textwidth]{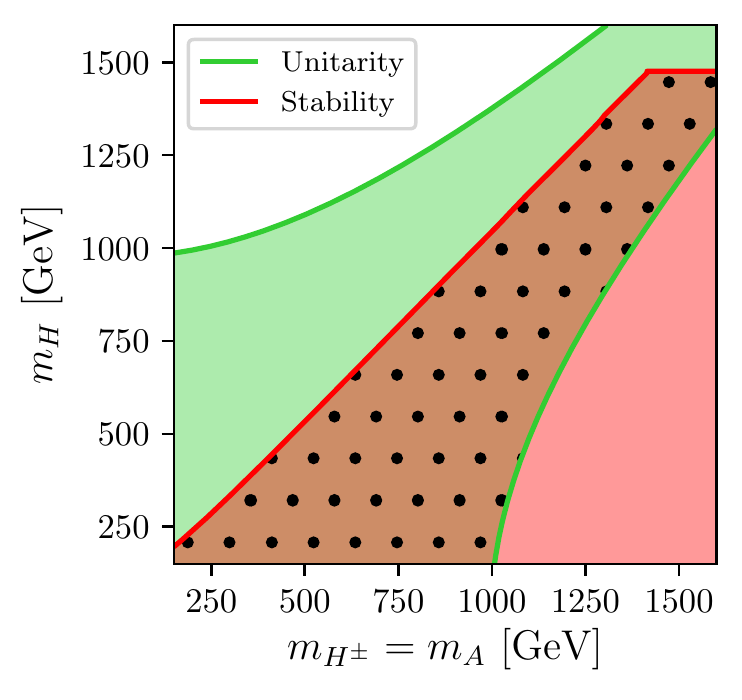}\includegraphics[width=0.48\textwidth]{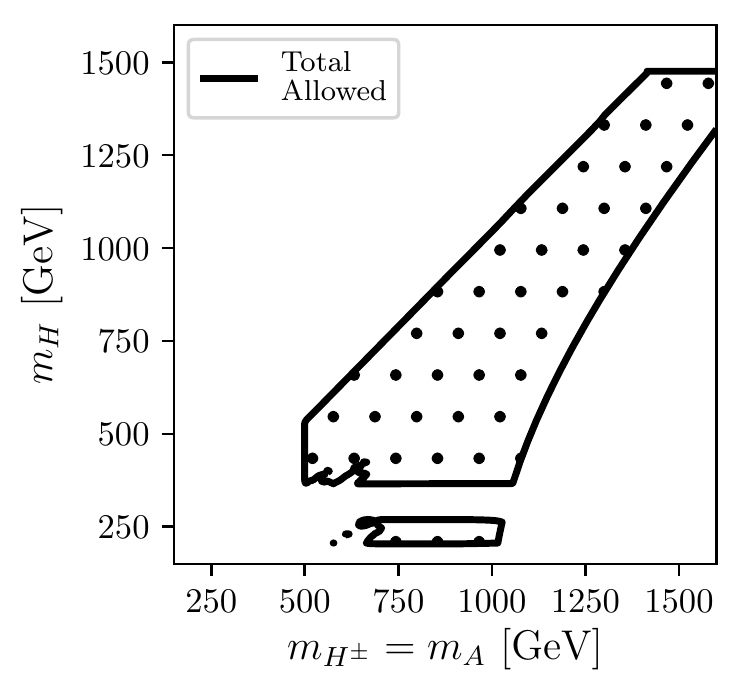}
		\includegraphics[width=0.48\textwidth]{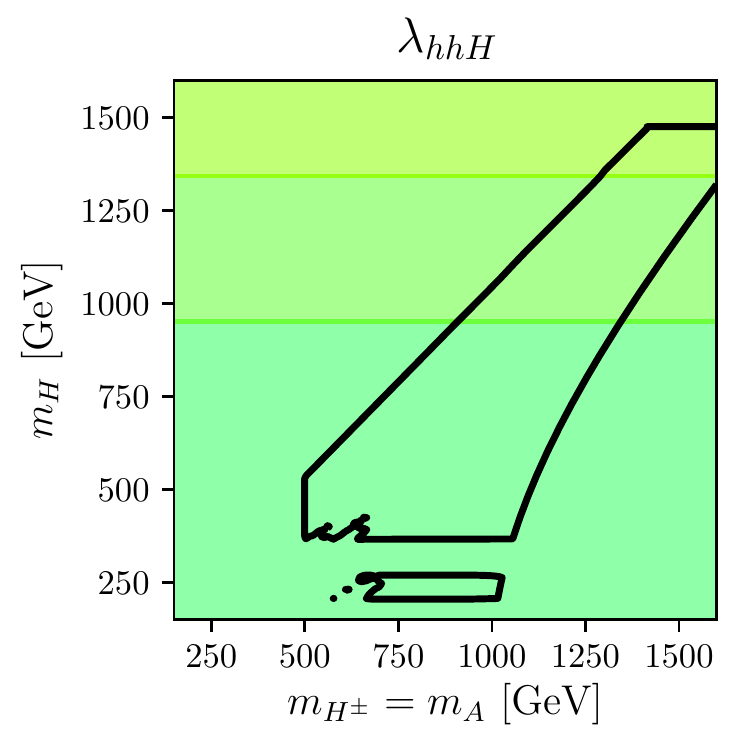}\includegraphics[width=0.48\textwidth]{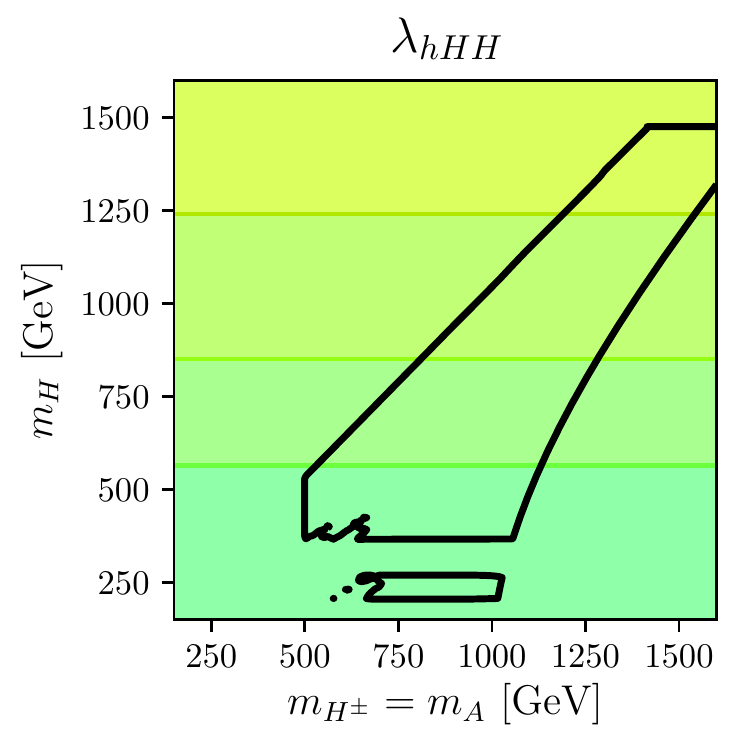}
		\includegraphics[width=0.48\textwidth]{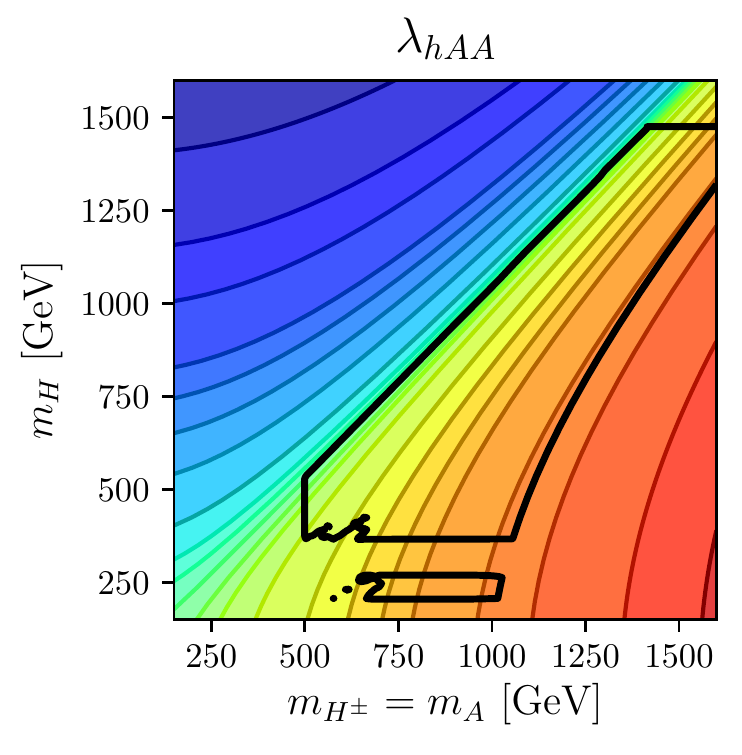}\includegraphics[width=0.48\textwidth]{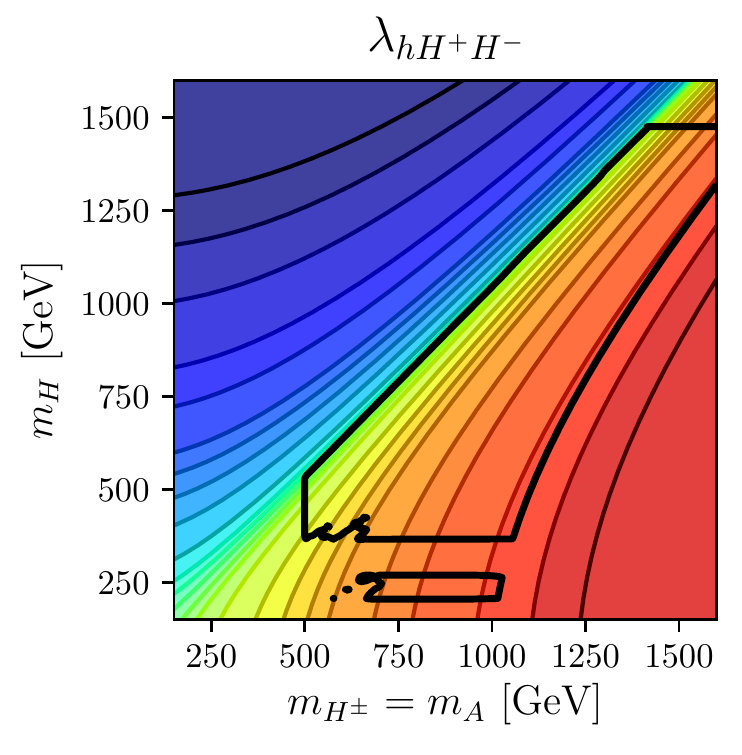}	
	\end{subfigure}	
	\begin{subfigure}[b]{0.06\textwidth}
		\includegraphics[height=0.48\textheight]{heavycolorbar}
	\end{subfigure}
\caption{\footnotesize{
Predictions for triple Higgs couplings in the  2HDM type~II, scenario~A,
in the $(\MHp = \MA, \MH)$ plane with $\CBA = 0.025$, $\tb = 6.5$ and
$\msq = (\MH^2\cos^2\al)/(\tb)$.
\emph{Upper four plots:} allowed regions
as in \protect\reffi{fig:C1-cba-tb}(A). 
\emph{Third line left:} $\lahhH$,
\emph{third line right:} $\lahHH$,
\emph{lower left:} $\lahAA$,
\emph{lower right:} $\lahHpHm$. 
}}
\label{fig:A2-MHp-MH}
\end{center}
\end{figure}

\begin{figure}[p]
\begin{center}
	{\small 2HDM type II, scenario A, $\msq = 100000 \gev^2$}
	\begin{subfigure}[b]{0.51\textheight}
		\includegraphics[width=0.48\textwidth]{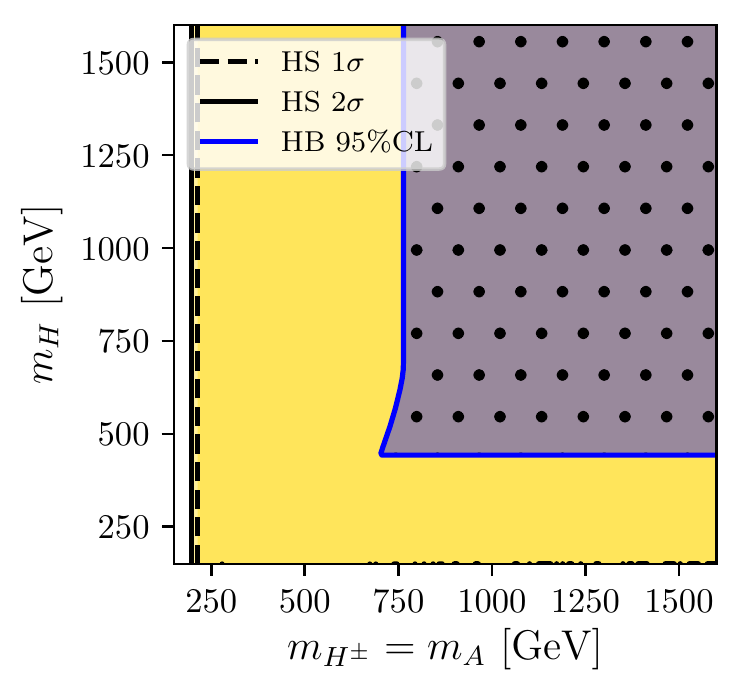}\includegraphics[width=0.48\textwidth]{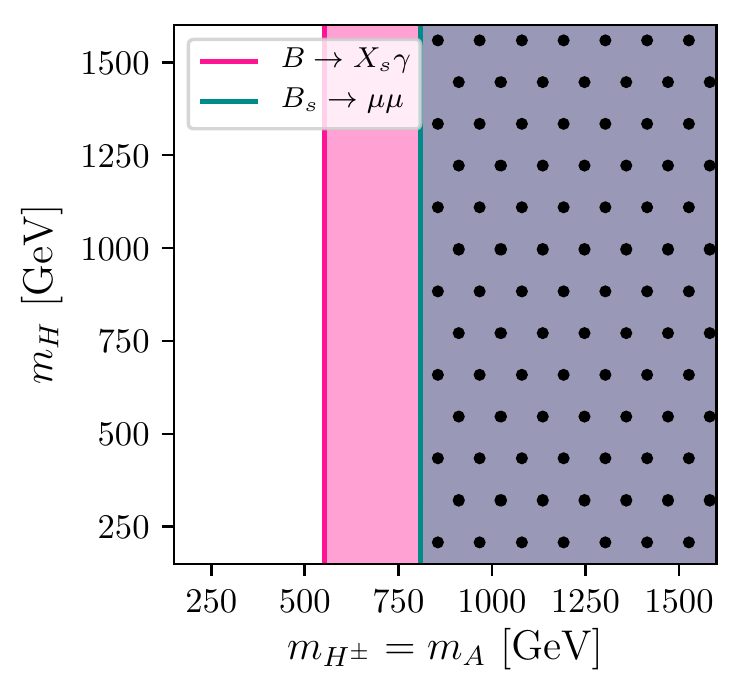}
		\includegraphics[width=0.48\textwidth]{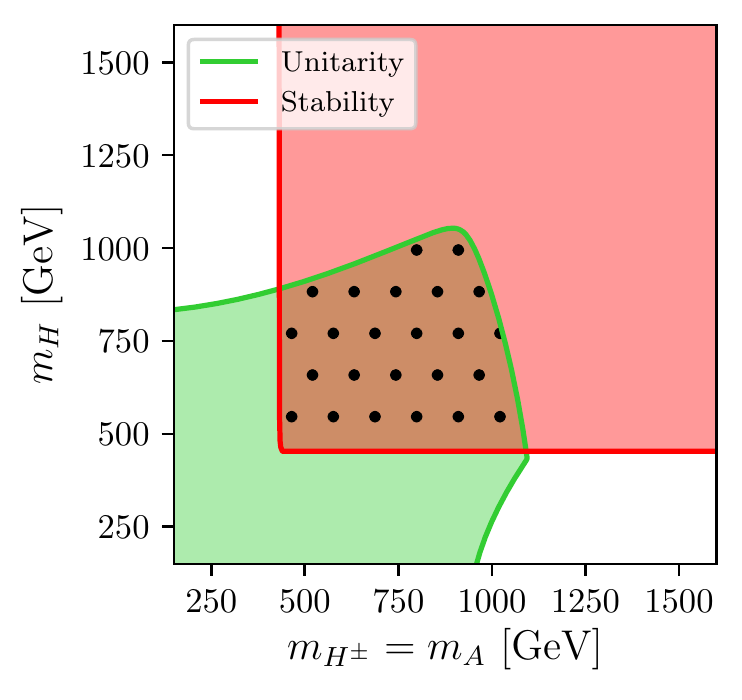}\includegraphics[width=0.48\textwidth]{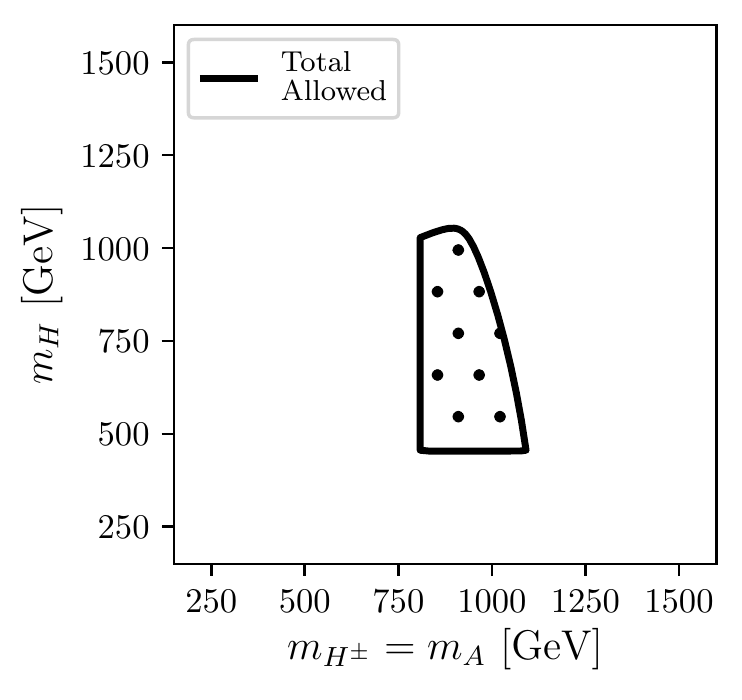}
		\includegraphics[width=0.48\textwidth]{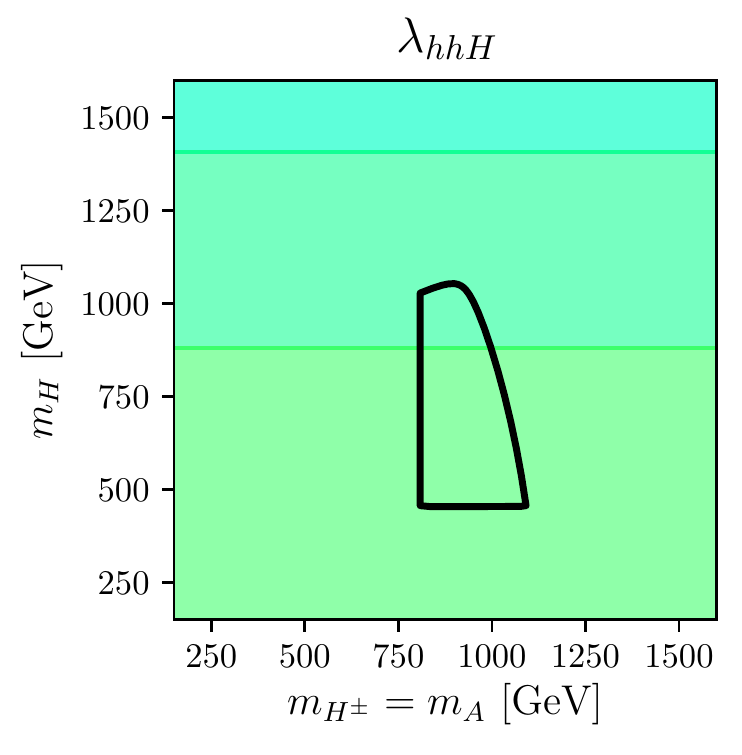}\includegraphics[width=0.48\textwidth]{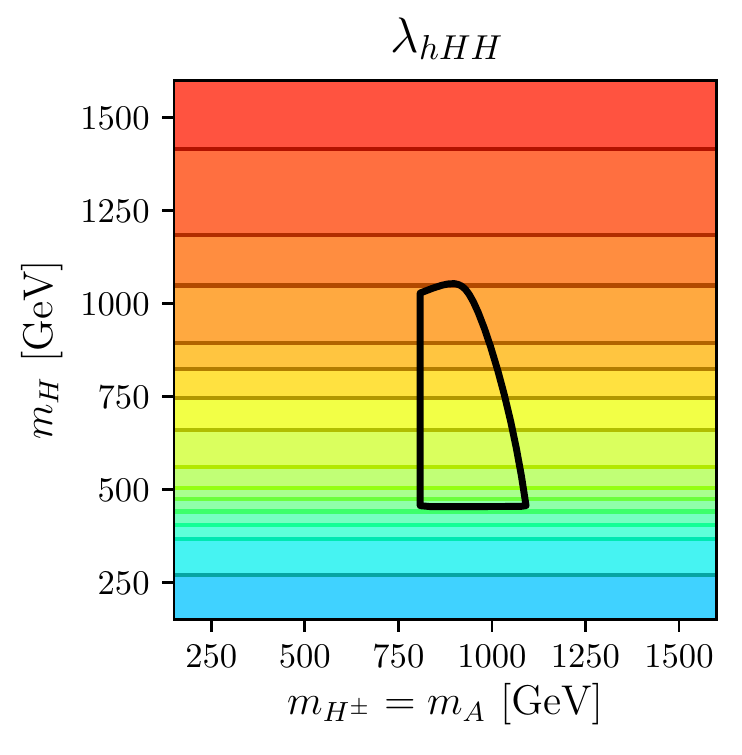}
		\includegraphics[width=0.48\textwidth]{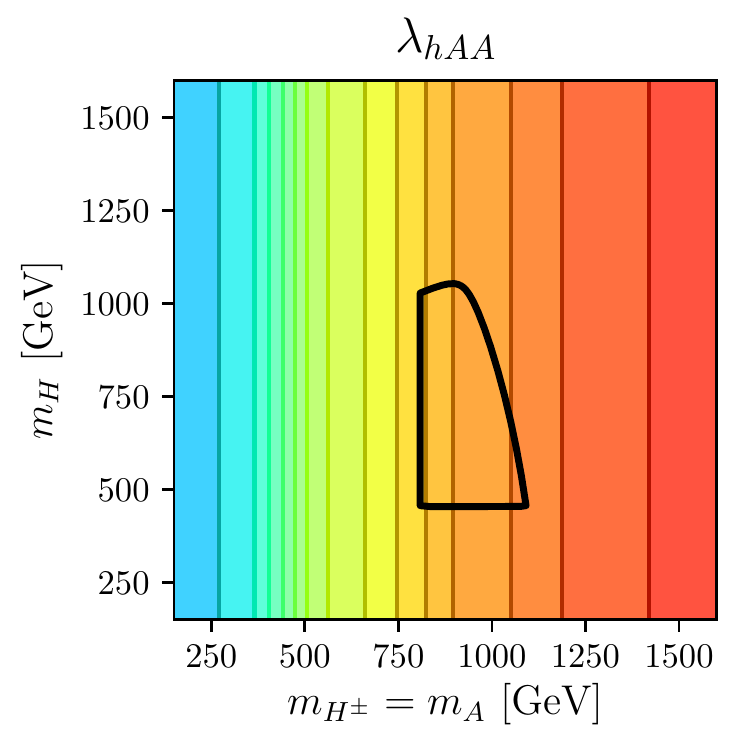}\includegraphics[width=0.48\textwidth]{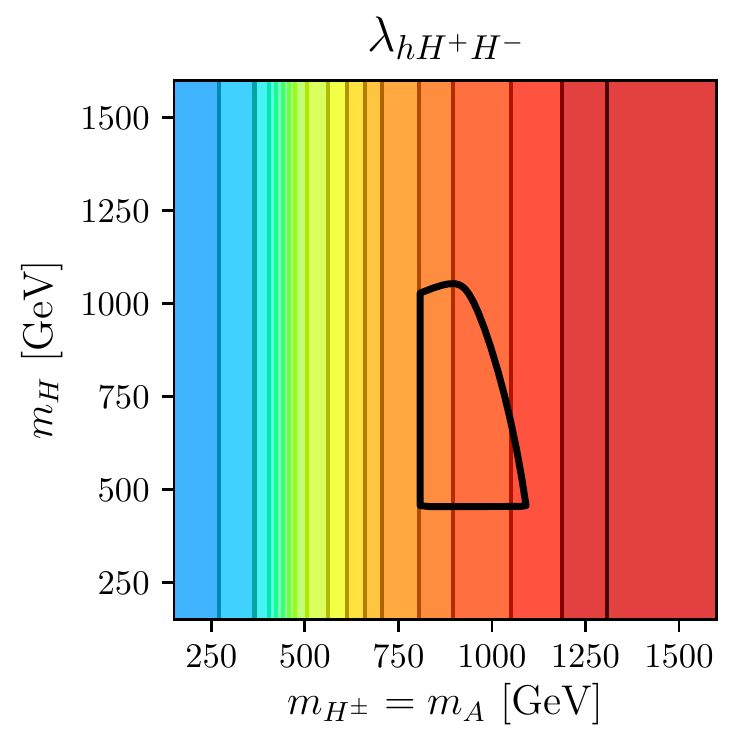}	
	\end{subfigure}	
	\begin{subfigure}[b]{0.06\textwidth}
		\includegraphics[height=0.48\textheight]{heavycolorbar}
	\end{subfigure}
\caption{\footnotesize{
Predictions for triple Higgs couplings in the  2HDM type~II, scenario~A,
in the $(\MHp = \MA, \MH)$ plane with $\CBA = 0.05$, $\tb = 0.9$ and
$\msq = 100000 \gev^2$.
\emph{Upper four plots:} allowed regions
as in \protect\reffi{fig:C1-cba-tb}(A). 
\emph{Third line left:} $\lahhH$,
\emph{third line right:} $\lahHH$,
\emph{lower left:} $\lahAA$,
\emph{lower right:} $\lahHpHm$. 
}}
\label{fig:A2-MHp-MH-lowtb}
\end{center}
\end{figure}

The upper right plot depicts the results for $\lahHH$, again independent of
$\MHp$. Here lowest (highest) values are reached for low (high) values of
$\MH$, following the analytic result in \refeq{eq:hHH_phys}. For $\lahHH$ they
range from 0.2 to 16. 
The lower row shows the results for $\lahAA$ (left) and $\lahHpHm$ (right),
which exhibit a similar behavior, see \refeq{eq:hAA_phys} and \refeq{eq:hHpHp_phys}. The values are nearly
independent of $\MH$, where lowest (highest) values are found for low (high) 
$\MA = \MHp$. They range from 0 to 16 for $\lahAA$, and from
0 to 32 for $\lahHpHm$. 
As in \refse{sec:scenC} we leave the phenomenological discussion to
\refse{sec:impl}.

Analogous results in the 2HDM type~II are presented in \reffi{fig:A2-MHp-MH},
with the color codings as in \reffi{fig:A1-MHp-MH}. As before $\msq$ is fixed by
$\msq = (\MH^2\cos^2\al)/(\tb)$. 
In order to maximize the results for the triple Higgs couplings we have chosen
$\CBA = 0.025$ and $\tb = 6.5$. 
The overall allowed region, as depicted in the upper two rows, can be found on the strip
roughly around the diagonal $\MHp = \MA \sim \MH$. 
The results for $\lahhh$ again vary only weakly in this region, and are found
in the interval $\kala \sim \inter{0.8}{1}$. 
The third row shows the results for $\lahhH$ (left) and $\lahHH$ (right),
which follow similar patterns and are independent of $\MHp$, see 
\refeq{eq:hhH_phys} and \refeq{eq:hHH_phys}. Lowest (highest) values are found at low (high) $\MH$,
ranging from 0 to 1.25 for $\lahhH$ and from 0.15 to
3 for $\lahHH$. 
The fourth row presents the results for $\lahAA$ (left) and $\lahHpHm$ (right),
which again follow similar patterns and are nearly independent of $\MH$, see 
\refeq{eq:hAA_phys} and \refeq{eq:hHpHp_phys}. The lowest values are
found at the diagonal $\MHp = \MA \sim \MH$, whereas the highest values
are found for the highest allowed $\MHp = \MA > \MH$ with a mass
splitting of about $250 - 300 \gev$. They range from 
0.4 to 16 for $\lahAA$ and from 0.8 to
32 for $\lahHpHm$. The phenomenological implications are discussed in
\refse{sec:impl}.

We finish our analysis in the scenario~A with the 2HDM type~II presented in
\reffi{fig:A2-MHp-MH-lowtb}, with the color codings as in \reffi{fig:A1-MHp-MH}. 
In comparison with the previous analysis 
we have chosen a relatively low value of $\tb = 0.9$, and fixed
$\msq = 100000 \gev^2$, while for $\CBA$ a relatively large value (for the
2HDM type~II) of $\CBA = 0.05$ was chosen. 
The overall allowed region, as depicted in the upper two rows can be found
roughly around $800 \gev < \MHp = \MA < 1100 \gev$  and $\MH$ between
$500 \gev$ and $1000 \gev$. 
As before, the results for $\lahhh$ vary only weakly in this region, and it 
takes values for $\kala \sim 0.9$ in the whole plane. 
The third row shows the results for $\lahhH$ (left) and $\lahHH$ (right), and
as before both are independent of $\MHp$, see 
\refeq{eq:hhH_phys} and \refeq{eq:hHH_phys}. $\lahhH$ exhibits a small variation between 
-0.14 to 0.23. $\lahHH$, on the other hand, can reach very large
values for large $\MH$, and it is found to be in the range of 0.3 and 15.
The fourth row presents the results for $\lahAA$ (left) and $\lahHpHm$ (right),
which again follow similar patterns and are independent of $\MH$, see 
\refeq{eq:hAA_phys} and \refeq{eq:hHpHp_phys}. The lowest (highest) values are found at the lowest (highest) allowed
values for  $\MHp = \MA \sim 800\,(1100) \gev$. They range from 
8 to 16 for $\lahAA$ and from 16 to
32 for $\lahHpHm$. The phenomenological implications are discussed in
\refse{sec:impl}.



\subsection{Scenario B}
\label{sec:scenB}

We finish our numerical investigation with the third scenario suggested
by the electroweak precision observables, scenario~B, as defined in
\refse{sec:stu}, $\MA \neq \MHp = \MH$%
\footnote{Here and in the following we will denote this common mass as
$\MHp$.}, and $\Mh = 125 \gev$. 

\begin{figure}[p]
\begin{center}
	{\small 2HDM type I, scenario B, $\msq = (\MH^2\cos^2\al)/(\tb)$}
	\begin{subfigure}[b]{0.51\textheight}
		\includegraphics[width=0.48\textwidth]{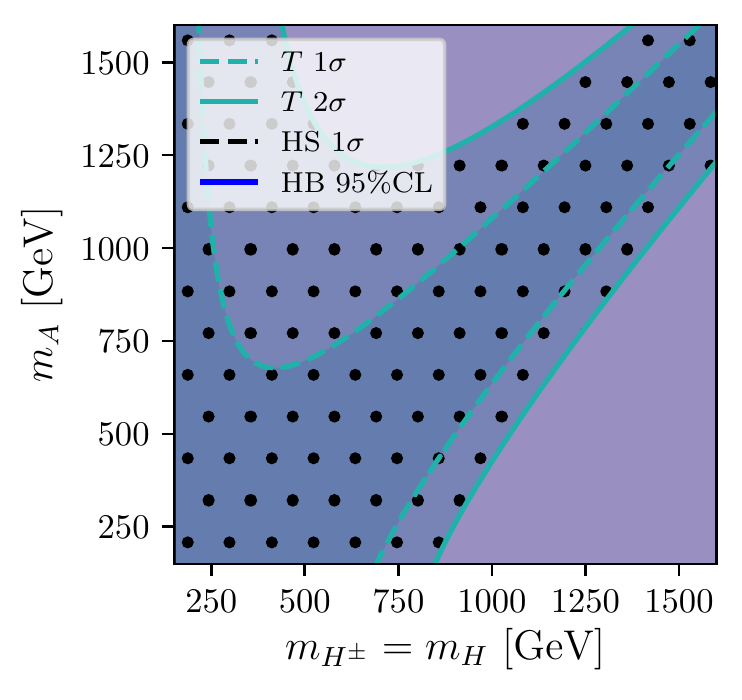}\includegraphics[width=0.48\textwidth]{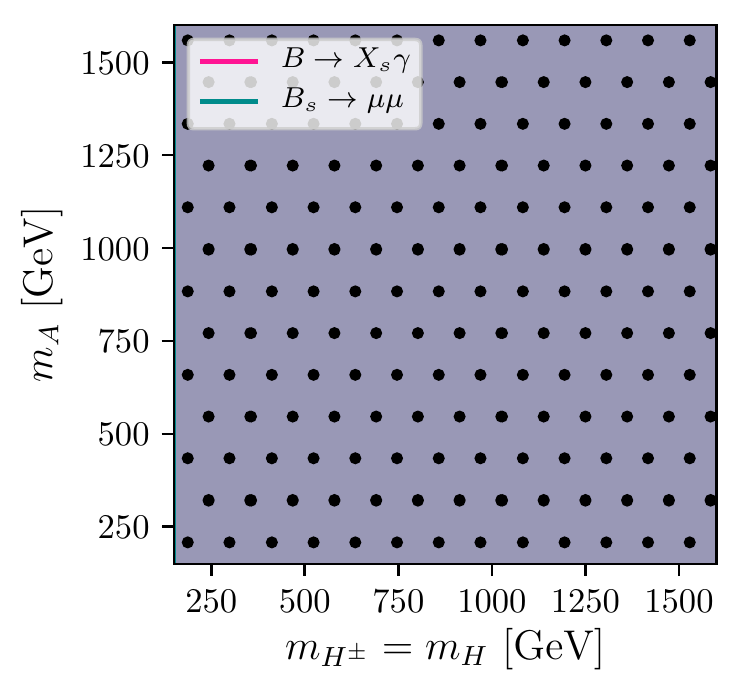}\vspace{-1.5em}
		\includegraphics[width=0.48\textwidth]{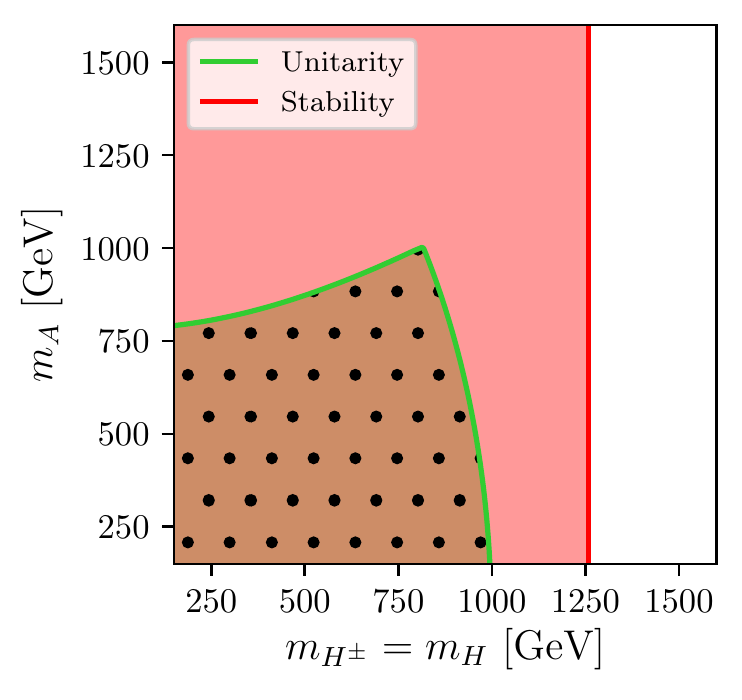}\includegraphics[width=0.48\textwidth]{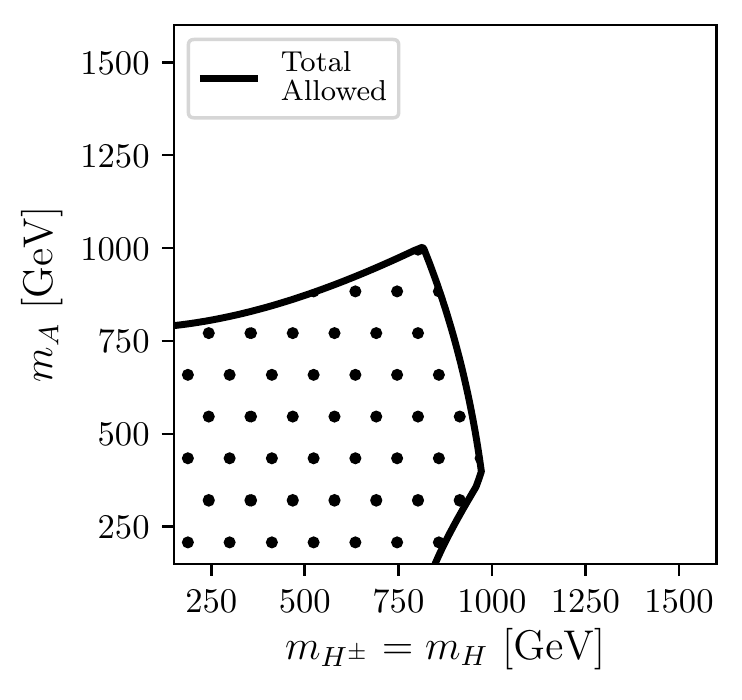}
		\includegraphics[width=0.48\textwidth]{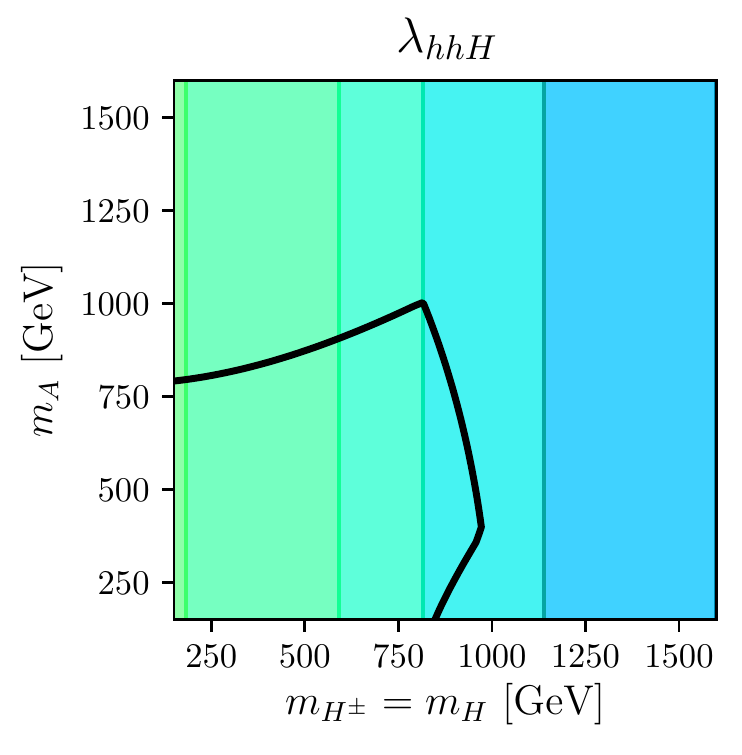}\includegraphics[width=0.48\textwidth]{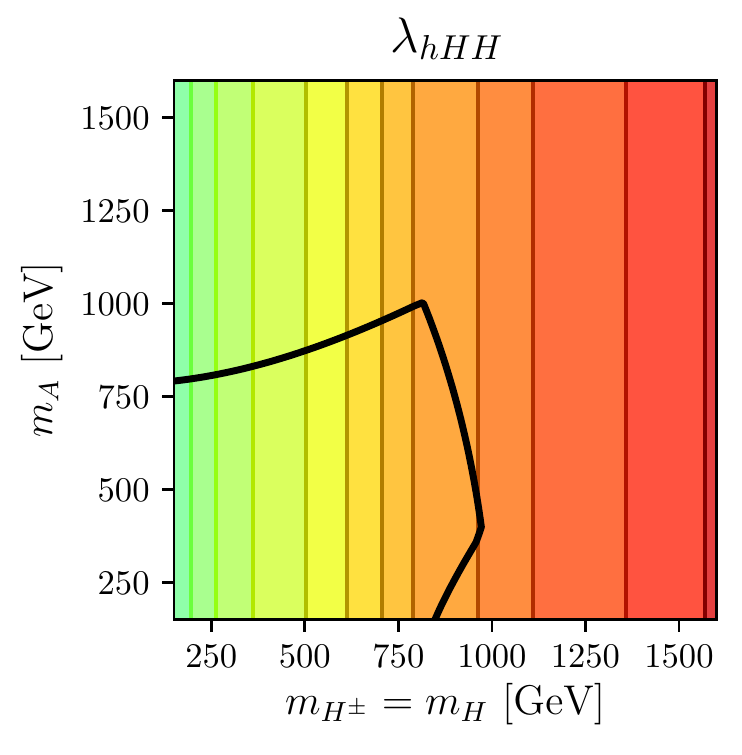}
		\includegraphics[width=0.48\textwidth]{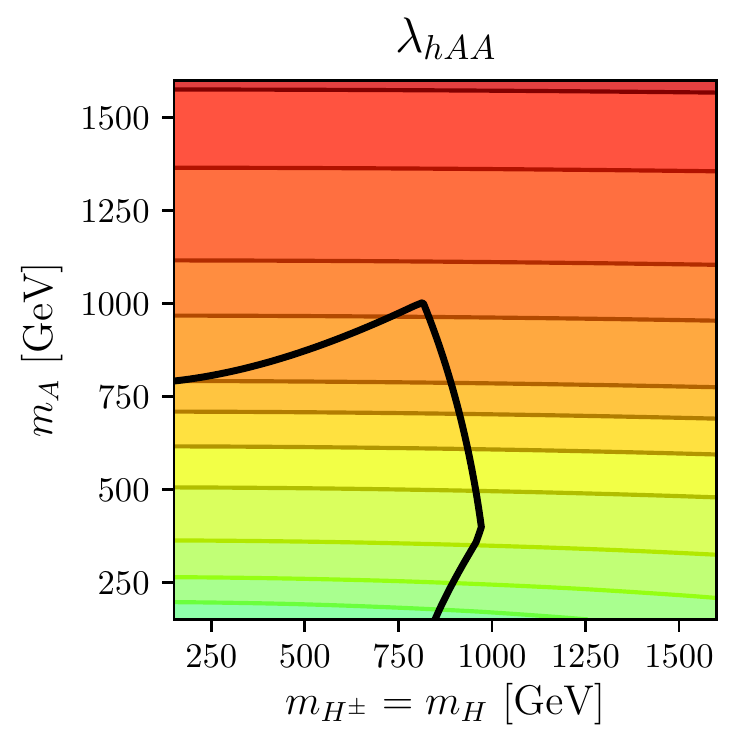}\includegraphics[width=0.48\textwidth]{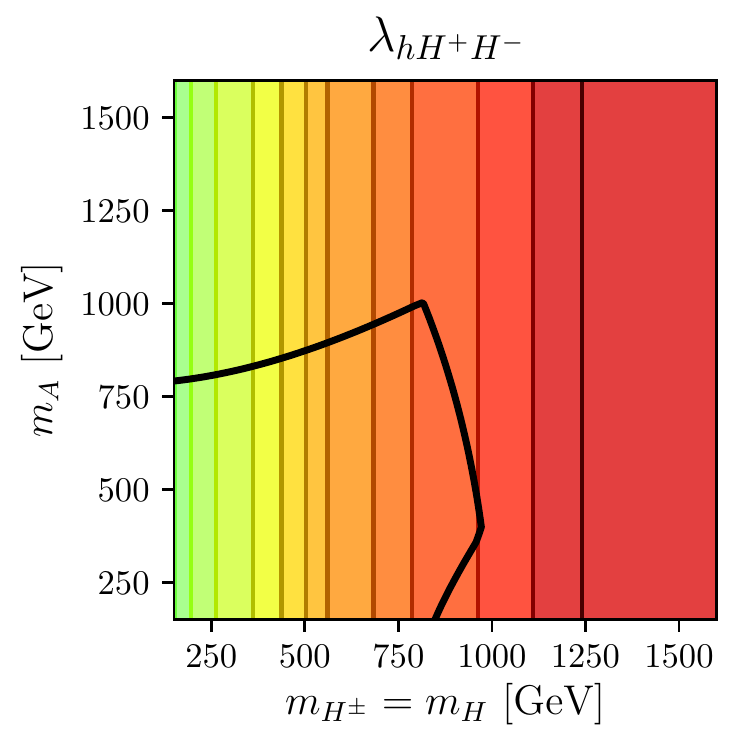}	
	\end{subfigure}	
	\begin{subfigure}[b]{0.06\textwidth}
		\includegraphics[height=0.48\textheight]{heavycolorbar}
	\end{subfigure}
\caption{\footnotesize{
Predictions for triple Higgs couplings in the  2HDM type~I, scenario~B,
in the $(\MHp = \MH, \MA)$ plane with $\CBA = 0.2$, $\tb = 10$ and
$\msq = (\MH^2\cos^2\al)/(\tb)$.
\emph{Upper four plots:} allowed regions
as in \protect\reffi{fig:C1-cba-tb}(A).
The light blue lines in
the upper left plot correspond to the 1$\sigma$ (dashed) and 2$sigma$
(solid) allowed regions by the $T$ parameter.
\emph{Third line left:} $\lahhH$,
\emph{third line right:} $\lahHH$,
\emph{lower left:} $\lahAA$,
\emph{lower right:} $\lahHpHm$. 
}}
\label{fig:B1-MHp-MA}
\end{center}
\end{figure}

\begin{figure}[p]
\begin{center}
	{\small 2HDM type II, scenario B, $\msq = 100000 \gev^2$}
	
	\begin{subfigure}[b]{0.51\textheight}
		\includegraphics[width=0.48\textwidth]{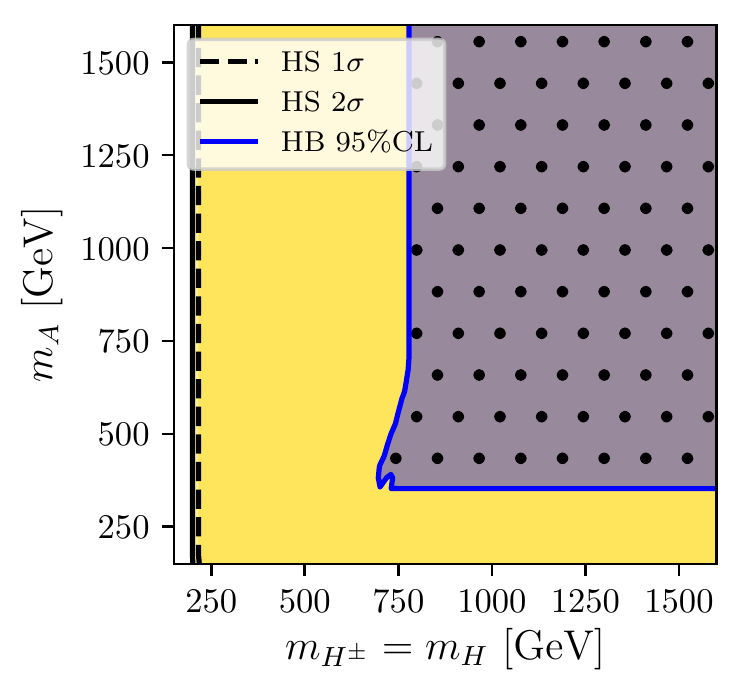}\includegraphics[width=0.48\textwidth]{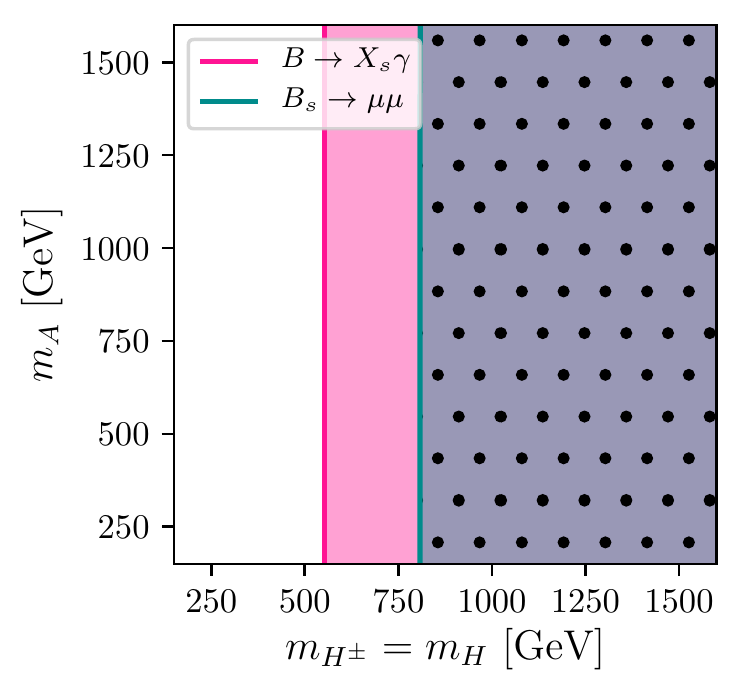}
		\includegraphics[width=0.48\textwidth]{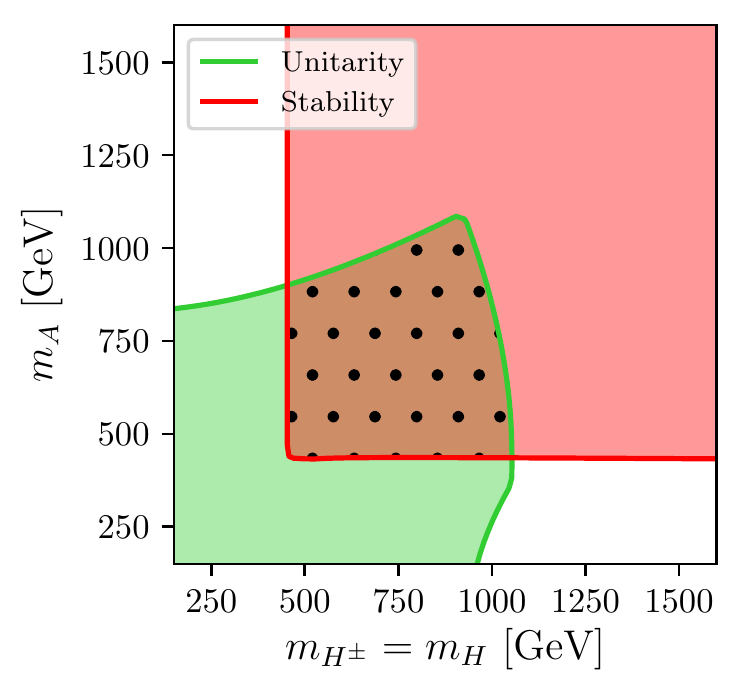}\includegraphics[width=0.48\textwidth]{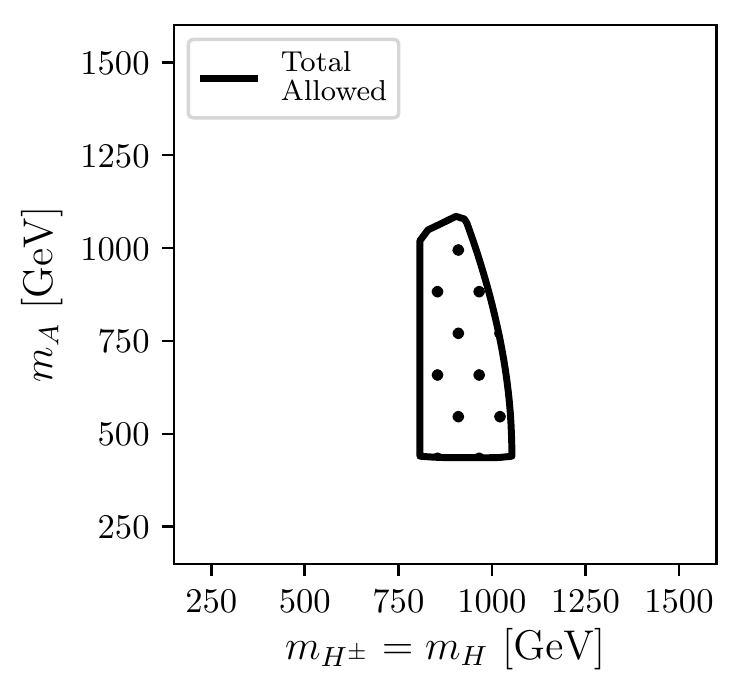}
		\includegraphics[width=0.48\textwidth]{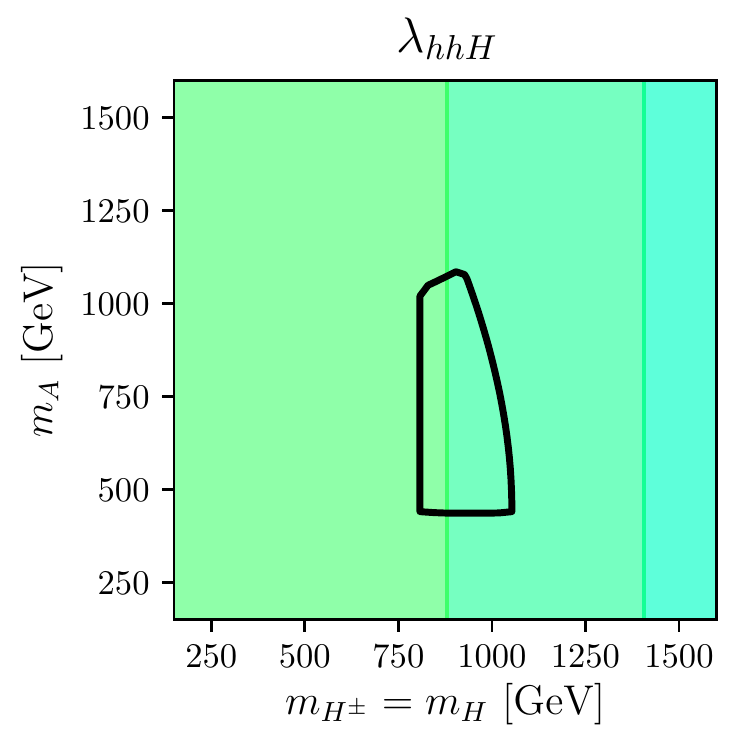}\includegraphics[width=0.48\textwidth]{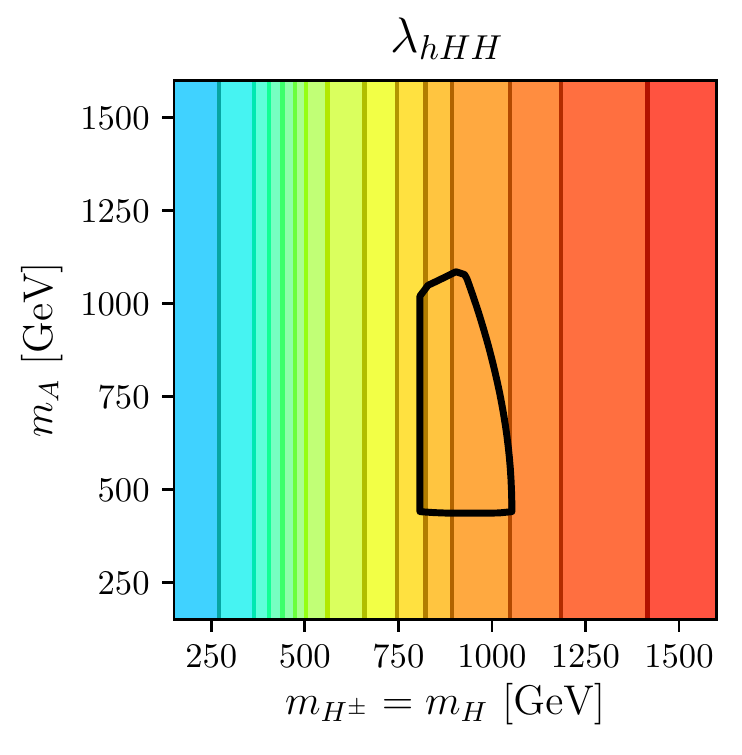}
		\includegraphics[width=0.48\textwidth]{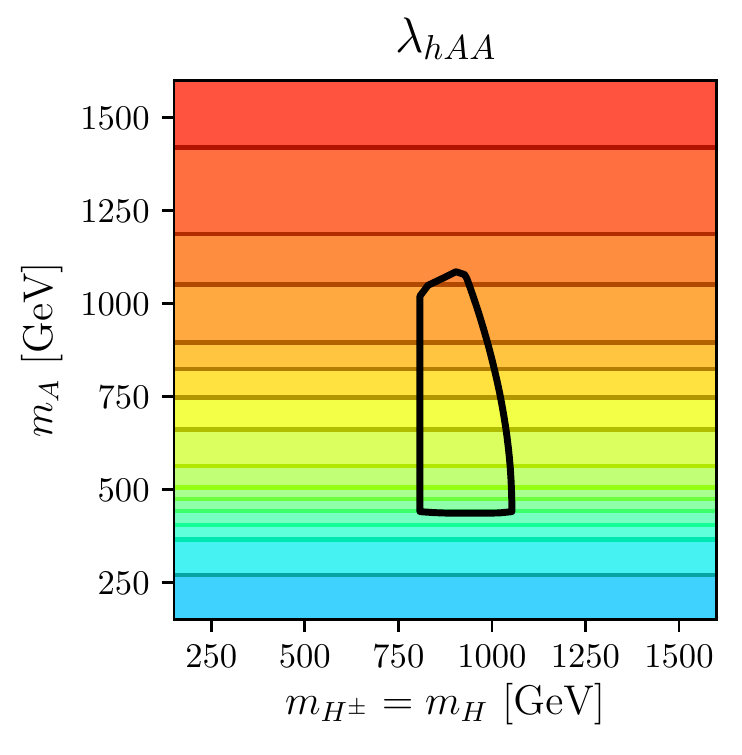}\includegraphics[width=0.48\textwidth]{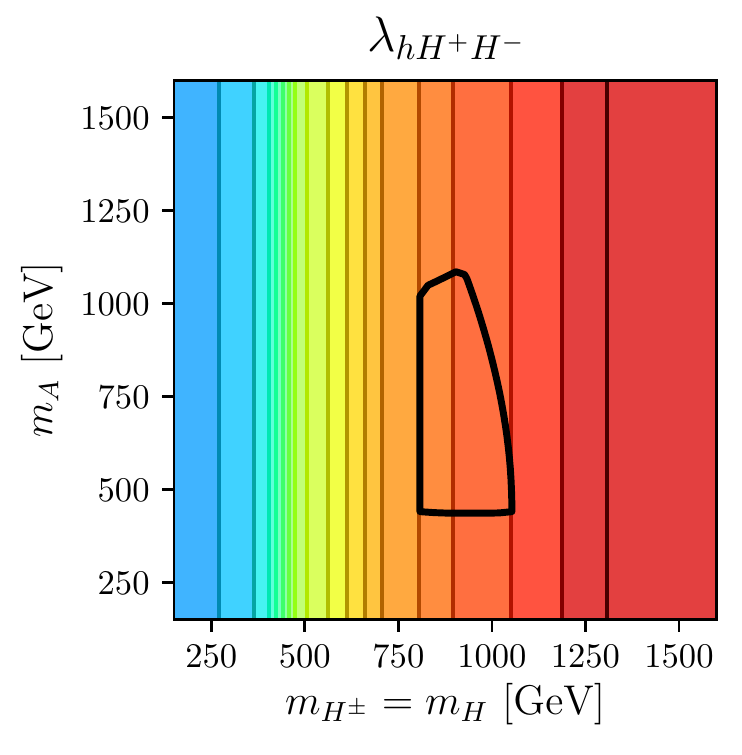}	
	\end{subfigure}	
	\begin{subfigure}[b]{0.06\textwidth}
		\includegraphics[height=0.48\textheight]{heavycolorbar}
	\end{subfigure}
\caption{\footnotesize{
Predictions for triple Higgs couplings in the  2HDM type~II, scenario~B,
in the $(\MHp = \MH, \MA)$ plane with $\CBA = 0.05$, $\tb = 0.9$ and
$\msq = 100000 \gev^2$.
\emph{Upper four plots:} allowed regions
as in \protect\reffi{fig:C1-cba-tb}(A). 
\emph{Third line left:} $\lahhH$,
\emph{third line right:} $\lahHH$,
\emph{lower left:} $\lahAA$,
\emph{lower right:} $\lahHpHm$. 
}}
\label{fig:B2-MHp-MA}
\end{center}
\end{figure}

In \reffi{fig:B1-MHp-MA} we present the $(\MHp = \MH, \MA)$ plane with 
the other parameters chosen as in the corresponding scenario~A, with
$\msq = (\MH^2\cos^2\al)/(\tb)$ to maximize the parameter space allowed by
unitarity and stability of the Higgs potential, and for $\CBA = 0.2$ and 
$\tb = 10$. The upper two rows show the various constraints, with the same
color coding as in \reffi{fig:C1-cba-tb}. Besides, in the upper left
plot, where we indicate the regions allowed by the LHC measurements, we
also indicate the bound arising from the EWPO, see
\reffi{fig:EWPO}. While the LHC measurements of the SM-like Higgs boson
as well as the direct searches for BSM Higgs bosons do not yield
restrictions in the parameter space, the EWPO favor a broad region
roughly around the diagonal $\MHp = \MH \sim \MA$. As in the
corresponding scenario~A, 
unitarity and stability roughly select a square bounded from
above by $\MA \sim \MH \sim 1000 \gev$. 
The results for $\lahhh$ are again not explicitly shown, as they vary only very
weakly in the chosen scenario. The values reached are in the interval
$\kala \sim \inter{0.98}{1.03}$. 
The lower two rows in \reffi{fig:B1-MHp-MA} show the results for the triple 
Higgs couplings involving at least one heavy Higgs boson. The upper left plot
(of the two lower rows) shows $\lahhH$, which is independent of $\MA$. Lower
(higher) values are reached for high (low) values of $\MHp$, following the
analytic result in \refeq{eq:hhH_phys}. They range from -1.5 to
1. 
The upper right plot depicts the results for $\lahHH$, again independent of
$\MA$. Here, the lowest (highest) values are reached for the lowest (highest) allowed values of
$\MHp$, following the analytic result in \refeq{eq:hHH_phys}. For $\lahHH$ they
range from 0.2 to 15. 
The lower row shows the results for $\lahAA$ (left) and $\lahHpHm$
(right), where the latter exhibits a similar behavior as $\lahHH$, see
\refeq{eq:hAA_phys} and \refeq{eq:hHpHp_phys}. The values of $\lahAA$ ($\lahHpHm$) are nearly 
independent of $\MHp$ ($\MA$), where the lowest (highest) values are found
for the lowest (highest) allowed values of $\MA$ ($\MHp = \MH)$. 
They range from 0.2 to 16 for $\lahAA$, and from
0.5 to 30 for $\lahHpHm$. 
As in \refse{sec:scenC}, 
we leave the phenomenological
discussion to \refse{sec:impl}.

The final scenario analyzed is scenario~B analogous to the last example
in scenario~A as presented in
\reffi{fig:B1-MHp-MA}, with the color codings as in \reffi{fig:A1-MHp-MH}. 
As in scenario~A we have chosen a relatively low value of $\tb = 0.9$,
and fixed $\msq = 100000 \gev^2$, while for $\CBA$ a relatively large
value (for the 2HDM type~II) of $\CBA = 0.05$ was chosen. 
The overall allowed region, as depicted in the upper two rows can be found
roughly around $\MHp = \MH \sim 900 \gev$ and $\MA$ between $500 \gev$ 
and $1000 \gev$, analogous to the corresponding scenario~A. 
As before, the results for $\lahhh$ vary only weakly in this region, and it 
leads to values of $\kala \sim 0.9$ in the whole plane. 
The lower two rows in \reffi{fig:B2-MHp-MA} show the results for the triple 
Higgs couplings involving at least one heavy Higgs boson. The upper left plot
(of the two lower rows) shows $\lahhH$, which is independent of $\MA$
and varies only weakly in the parameter plane. Lower
(higher) values are reached for high (low) values of $\MHp$, following the
analytic result in \refeq{eq:hhH_phys}. They range from -0.14 to
0. 
The upper right plot depicts the results for $\lahHH$, again independent of
$\MA$. Here, the lowest (highest) values are reached for the lowest (highest) allowed values of
$\MHp$, following the analytic result in \refeq{eq:hHH_phys}. For $\lahHH$ they
range from 8 to 15. 
The lower row shows the results for $\lahAA$ (left) and $\lahHpHm$
(right), where the latter exhibits a similar behavior as $\lahHH$, see
\refeq{eq:hAA_phys} and \refeq{eq:hHpHp_phys}. The values of $\lahAA$ ($\lahHpHm$) are 
independent of $\MHp$ ($\MA$), where the lowest (highest) values are found
for the lowest (highest) allowed values of $\MA$ ($\MHp = \MH)$. 
They range from 0.14 to 16 for $\lahAA$, and from
15 to 30 for $\lahHpHm$. 

\medskip

Finally, to close the numerical results section, we present in
\refta{tab:benchmarks} some examples of interesting configurations
that maximize the size of the triple Higgs couplings. 
In the Yukawa type~I, all examples have $\tb>1$ while for type~II all
the points are around $\tb\sim1$. This is mainly due to the
constraints from the LHC data, because it is easier to accommodate
a SM-like Higgs in those regions (see \refse{sec:collider}). 
In addition, particularly in type~I, flavor observables disallow
low values of $\tb$.
In type I we recover a larger allowed parameter region by
choosing $\msq$ according to \refeq{eq:m12special}, 
especially for the larger values of $\tb$, which are
easier in conflict with the theoretical constraints, as we discussed in
\refse{sec:theo}. Furthermore, in type~II the tight constraint from
$B\to X_s\gamma$ that sets $\MHp \gsim 500 \gev$ should be also satisfied.  
However, as we have discussed in the previous subsections, the main
constraint that prevent from obtaining large triple Higgs
couplings are the theoretical constraints. 

For both types, I and II, points with larger triple Higgs
couplings are also the ones with the heavier Higgs masses around
1~TeV (where we have not explored values above $\sim 1.6 \tev$).
The only exception to this are $\lahhh$ and $\lahhH$.
In the case of $\lahhh$ in the alignment limit the SM value is
reproduced. On the other hand, $\lahhH$ is proportional to
$\CBA$, see Appendix~\ref{appendix:FR}.
Consequently, their extrema are both found outside the
alignment limit.
As a consequence those points are stronger tested and
possibly ``easier'' excluded
in the future by more precise measurements.
Overall, and particularly for type~II, $\kala$ is close to
unity. While this does not correspond to large enhancements of
di-Higgs production, the deviations are still large enough to be
tested at future colliders, see \refse{sec:hhh-exp}.
On the other hand, it is
possible to find large allowed values of couplings involving more
than one heavy Higgs boson near the alignment limit. Those triple
Higgs couplings always have positive and large values for
all scenarios~A, B and~C.  In fact, they can be larger in scenarios A
and B due to the allowed splitting of two of the masses, as we have seen
in \refses{sec:scenA} and \ref{sec:scenB}. In those cases, the larger
mass is the one of the Higgs boson that appears in the vertex.
Overall, we find values
of ${\cal O}(10)$, with the maximum 
value corresponding to $\lahHpHm\sim 30$ in both type~I and~II.

\begin{table}[t!]
\begin{center}
{\footnotesize 
Yukawa type I\\
\begin{tabular}{c|c|c|c|c|c|c|c|c|c|c}
{$m_{H}$} & {$m_{A}$} & {$m_{H^{\pm}}$} & {$\tan\be$} & {$c_{\be-\al}$} & {$m_{12}^{2}$} & {\kala} & {\lahhH} & {\lahHH} & {\lahAA} & {\lahHpHm} \tabularnewline
\hline 
750 & 750 & 750 & 5.5 & 0.25 & \refeq{eq:m12special} & \textbf{-0.4} & 0.4 & 7 & 6 & 12 \tabularnewline
1000 & 1000 & 1000 & 7.5 & 0.2 & \refeq{eq:m12special} & \textbf{-0.3} & 0.1 & 13 & 12 & 24 \tabularnewline
650 & 650 & 650 & 6.0 & 0.2 & \refeq{eq:m12special} & 0.1 & 0.5 & 4 & 4 & 8 \tabularnewline
300 & 300 & 300 & 15.0 & 0.25 & \refeq{eq:m12special} & \textbf{1.5} & -0.6 & 2 & 2 & 5 \tabularnewline
400 & 400 & 400 & 12.5 & 0.2 & 12500 & 1.2 & -0.4 & 3 & 3 & 6 \tabularnewline
600 & 600 & 600 & 10.0 & 0.2 & \refeq{eq:m12special} & 1.0 & -0.5 & 6 & 6 & 12 \tabularnewline
$\ast$ 1500 & 1500 & 1500 & 2.0 & -0.025 & 820000 & 0.8 & \textbf{-1.2} & 3 & 3 & 6 \tabularnewline
\hline 
650 & 400 & 400 & 12.0 & 0.15 & \refeq{eq:m12special} & 0.9 & -0.3 & 6 & 2 & 4 \tabularnewline
300 & 600 & 600 & 2.5 & 0.1 & 5000 & 1.0 & 0.0 & 1 & 6 & 12 \tabularnewline
300 & 600 & 600 & 12.5 & 0.2 & \refeq{eq:m12special} & 1.1 & -0.2 & 2 & 6 & 12 \tabularnewline
$\ast$ 700 & 1200 & 1200 & 2.0 & 0.0 & \refeq{eq:m12special} & 1.0 & 0.0 & 0.0 & \textbf{16} & \textbf{32} \tabularnewline
\hline
700 & 1000 & 700 & 7.0 & 0.2 & \refeq{eq:m12special} & 0.3 & 0.2 & 6 & 14 & 11 \tabularnewline
350 & 600 & 350 & 10.0 & 0.2 & \refeq{eq:m12special} & 1.0 & -0.1 & 2 & 6 & 4 \tabularnewline
600 & 350 & 600 & 10.0 & 0.2 & \refeq{eq:m12special} & 1.0 & -0.5 & 6 & 2 & 11 \tabularnewline
\end{tabular}
}

\medskip
\footnotesize{
Yukawa type II\\
\begin{tabular}{c|c|c|c|c|c|c|c|c|c|c}
{$m_{H}$} & {$m_{A}$} & {$m_{H^{\pm}}$} & {$\tan\be$} & {$c_{\be-\al}$} & {$m_{12}^{2}$} & {\kala} & {\lahhH} & {\lahHH} & {\lahAA} & {\lahHpHm} \tabularnewline
\hline
1100 & 1100 & 1100 & 0.9 & 0.13 & 260000 & \textbf{-0.1} & 0.9 & 11 & 11 & 23 \tabularnewline
1500 & 1500 & 1500 & 0.8 & 0.05 & 775000 & 0.5 & \textbf{1.7} & 11 & 11 & 21 \tabularnewline
600 & 600 & 600 & 1.5 & 0.02 & 25000 & \textbf{1.0} & 0.0 & 5 & 5 & 10 \tabularnewline
\hline 
1150 & 1000 & 1000 & 0.95 & 0.025 & 210000 & \textbf{1.0} & 0.1 & \textbf{15} & 10 & 19 \tabularnewline
400 & 600 & 600 & 1.5 & 0.04 & 10000 & \textbf{1.0} & 0.0 & 2 & 6 & 11 \tabularnewline
\hline
1350 & 1000 & 1350 & 0.9 & 0.05 & 460000 & 0.7 & 0.8 & \textbf{15} & 1 & 30 \tabularnewline
600 & 400 & 600 & 1.5 & 0.05 & 8000 & \textbf{1.0} & -0.1 & 6 & 2 & 12 \tabularnewline
\end{tabular}
}

\caption{Examples in the 2HDM for parameter inputs that present large size of some 
triple Higgs couplings allowed by current constraints in the Yukawa type I (top table) and Yukawa type II (bottom table). Values in bold are the ones that are
close to the maximum deviation from the SM for $\kala$ and the
absolute extremum for the other couplings found in our study. In each table, horizontal lines
distinguish between scenario C (top), scenario A (center) and scenario B (bottom) defined 
in \refse{sec:stu}. The masses $\MH$, $\MA$ and $\MHp$ are expressed in GeV 
and $\msq$ is expressed in GeV$^2$. Points marked with an asterisk ($\ast$) are also allowed in type II.}
\label{tab:benchmarks}
\end{center}
\end{table}

In \refta{tab:benchmarks} we also include points with smaller Higgs
masses that also yield interesting sizes of the triple Higgs
couplings. Due to the relatively smaller masses, these points are
better kinematically accessible. These last kind of points are
{presumably the easiest to probe at future colliders.
For these more moderate, but potentially more accessible masses, we
find triple Higgs couplings with half the size w.r.t. the maximum values.

Finally, before ending this section we would like to make some comments
about the viability of our 2HDM scenarios to produce a strong first order
EW phase transition (FOPT), which is needed for EW Baryogenesis. 
In \cite{Dorsch:2013wja} it has been pointed out that in the 2HDM a
FOPT is correlated with large values of the Higgs couplings $\lambda_4$ and/or $\lambda_5$.
This issue of the preferred  2HDM Higgs self couplings favoring a FOPT
and the correlated heavy Higgs mass region for $m_A$ and $m_H$ 
together with the preferences in the other relevant 2HDM parameters like 
$c_{\beta-\alpha}$, $\tan\beta$  and $m_{12}^2$ have also been explored in  
\cite{Dorsch:2014qja} and \cite{Dorsch:2017nza}. In order to study the 
strength of the EW phase transition for our 2HDM scenarios, which 
exhibit large Higgs self-couplings, we follow \cite{Dorsch:2017nza}. 
Instead of the more standard method using the thermal 1-loop effective 
potential (as done, for instance, in \cite{Dorsch:2014qja}) the method of 
\cite{Dorsch:2017nza} correlates this strength  with the zero 
temperature vacuum energy difference of the 2HDM with respect to 
the SM. Finding this difference, $\Delta \mathcal{F}_0=\mathcal{F}_0-\mathcal{F}_0^{\mathrm{SM}}$, 
usually normalised to the SM value  $\Delta \mathcal{F}_0/\mathcal{F}_0^{\mathrm{SM}}$, 
to be in the range $\Delta \mathcal{F}_0/\mathcal{F}_0^{\mathrm{SM}} \leq -0.34$
provides, accordingly to \cite{Dorsch:2017nza},  a good indicator of a FOPT. 
Since this topic is clearly beyond the scope of this work, we
have just evaluated here this estimate of 
$\Delta \mathcal{F}_0/\mathcal{F}_0^{\mathrm{SM}}$
for the points listed in \refta{tab:benchmarks}\footnote{We warmly thank 
Jose Miguel No for his invaluable help in our investigation of the FOPT, and for 
providing us with his private code for the estimate of 
$\Delta \mathcal{F}_0/\mathcal{F}_0^{\mathrm{SM}}$.}. We find that all 
the points in this table give large negative values (except the ones in 
the third and seventh rows leading to differences of around -0.23) fulfilling 
$\Delta \mathcal{F}_0/\mathcal{F}_0^{\mathrm{SM}} \leq -0.34$. 
Therefore, we conclude that our 2HDM scenarios leading to large 
Higgs self-couplings appear to favor a FOPT.


\subsection{Possible implications for future collider measurements}
\label{sec:impl}

We now turn to the phenomenological implications of the allowed ranges
found for the various triple Higgs couplings, as discussed in the
previous subsections. As an overall result we find that the allowed
intervals for the various triple Higgs couplings depend only weakly on
the chosen EWPO scenario~A, B or~C. However, the 2HDM type~I exhibits a
substantially stronger variation in $\lahhh$ than type~II. This is
mostly owed to the larger allowed deviation from the alignment limit,
see \refses{sec:collider}, \ref{sec:SMlike}. In this section we will
concentrate on the anticipated impact of the triple Higgs couplings
on the various di-Higgs production cross sections, where we leave a full
phenomenological analysis for future work~\cite{2hdm-hhh-col}.

For $\lahhh$ we roughly find allowed intervals of $\inter{-0.5}{1.5}$ in
the 2HDM type~I and $\inter{0}{1}$ in type~II. While the production of
two SM-like Higgs bosons, both at $pp$ and at $e^+e^-$ colliders depends
already at the tree-level on $\lahhh$ and $\lahhH$, the dependence on
$\lahhh$ is expected to be substantially stronger due to the propagator
suppression with the inverse of $\MH^2$ of $\lahhH$. Consequently, over
the possible parameter range of $\lahhh$ the HL-LHC is not expected to
yield a precision on $\kala$ better than $35\%$, and a deviation from
$\lahhh = 0$ can not be established better than $\sim
2\,\sig$. Comparing the HL-LHC to the ILC500, the HL-LHC performs better
(worse) than the ILC500 for $\kala \lsim (\gsim)\, 0.5$, where both
intervals are still allowed in both types of the 2HDM. In other words,
the HL-LHC results in comparison with the ILC500 may look a bit better
than anticipated for $\kala = 1$. However, in this
comparison it must be kept in mind that the HL-LHC analysis is based
on the variation of the Higgs triple coupling only, whereas for the
ILC500 (at $\kala = 1$) it has been shown that the analysis holds also
for a variation of all Higgs-boson couplings within their anticipated
experimental accuracies. Furthermore, deviations below $\kala \sim 0.5$ 
are realized for larger deviations from the alignment limit 
and may thus be tested in the next round of Higgs rate measurements at
the LHC. Combining the ILC500 measurements with the final stage of the
ILC1000, the Linear Collider shows a substantially better result than
the HL-LHC for all the allowed $\lahhh$ parameter space. Only around a
vanishing trilinear Higgs coupling similar precisions are anticipated
(but the above mentioned caveat of the differences in the HL-LHC and ILC
analyses still holds). 

The phenomenological implications of the allowed ranges for $\lahhH$ 
are twofold. This coupling can enhance or suppress the contribution of the
off-shell heavy Higgs in the $hh$ production, which, however, are generally
suppressed as mentioned above. On the other hand, a very large
enhancement of this coupling would yield a relatively large cross
section for $hH$ production. However, we find that large values of
$\lahhH$ are not allowed taking all existing experimental and
theoretical constraints into account. 

The triple Higgs couplings involving two heavy Higgs bosons, $\lahHH$,
$\lahAA$ and $\lahHpHm$ can have a very strong impact on the heavy
di-Higgs production and possibly facilitate the discovery of such heavier
Higgs bosons (see, e.g.,~\cite{Chen:2019pkq,Kling:2020hmi}).
Roughly independent of the EWPO scenario and the 2HDM
type, we find values of up to~15, 16 and~32, respectively. 
Here it must be kept in mind that the larger values of a triple Higgs coupling of $h$ to two 
heavy Higgs bosons are realized for larger values of the respective heavy
Higgs-boson mass. 
Consequently, the effects of the large coupling and the heavy mass
always go in opposite directions. 
A detailed study will be left to future work~\cite{2hdm-hhh-col}. 

Finally, we would like to comment briefly about the projections after HL-LHC.
As we have discussed above, the requirement of having the properties of the
light $\cp$-even Higgs-boson, $h$, in agreement with the LHC rate measurements,
see \refse{sec:SMlike}, restricts in particular the possible deviation
of the Higgs-boson sector from the alignment limit. The Higgs-boson rate
measurements will improve significantly at the HL-LHC \cite{Cepeda:2019klc}. If the HL-LHC
does not find any significant deviation from the SM predictions, this would
restrict further the possible deviations from the alignment limit and thus
in particular the deviations of $\kala$ from unity (such an analysis, however,
goes beyond the scope of our paper). If, on the other hand, the HL-LHC would
observe a deviation from the SM predictions, a new fit around the then preferred
values would have to be performed. The then possible sizes of the
triple Higgs couplings will strongly depend on the hypothetical future preferred
(non-SM) Higgs-boson sector parameters (again, such an analysis
goes far beyond the scope of our paper).


\section{Conclusions}
\label{sec:conclusions}

An important task at future colliders is the measurement of the triple Higgs
coupling $\lambda_{hhh}$. Depending on its size relative to the SM value,
certain collider options result in a higher experimental accuracy. 
Similarly, large values of triple Higgs couplings involving heavy Higgs
bosons can lead to enhanced production cross sections of BSM Higgs
bosons.

Within the framework of Two Higgs Doublet Models (2HDM) type~I and~II we
investigate the allowed ranges for all triple Higgs couplings involving at
least one light, SM-like Higgs boson. We take into account all relevant
theoretical and experimental constraints. From the theory side these
comprise unitarity, and stability conditions. From the
experimental side we require agreement with the direct BSM Higgs-boson
searches, as well as with measurements of the SM-like Higgs-boson 
rate as measured at the LHC.
We furthermore require agreement with  flavor observables and
electroweak precision data (where the $T$~parameter plays the most
important role). In this context we investigate more extensively
the dependence of several of these constraints on the 
soft $Z_2$-breaking parameter, $\msq$. Here we find that large values of
this parameter can affect notably the allowed parameter space,
especially in the region of large $\tb$. 

For theoretical constraints $\msq$ plays a key role:
lower (higher) values are favored by the tree-level stability (unitarity)
  constraint, and the size of the intersection region is thus controlled by $\msq$.
Thus, to enlarge the allowed region by both unitarity and stability we have used
\refeq{eq:m12special} on several occasions.

Regarding the experimental constraints,
BSM Higgs boson searches and measurements of the 125 GeV Higgs boson
at the LHC
can also be sensitive to the effects of $\msq$ in the scalar sector such like the 
$h \to \gamma \gamma$ decay (via the $hH^+H^-$ vertex)
or the production of a heavy BSM boson
that decays to two 125~GeV Higgs bosons, specially in the range of low
masses.
On the other hand, the triple Higgs couplings $\lambda_{hH^+H^-}$ and $\lambda_{HH^+H^-}$
also enter in  the 2HDM prediction for $B_s\to\mu^+\mu^-$ via the $h$ and $H$
Higgs penguins contributions with charged Higgs bosons in the loops, 
and they can be relevant (see also~\cite{Cheng:2015yfu}). 
The largest effect from $m_{12}^2$ in $B_s\to\mu^+\mu^-$ is due to 
$\lambda_{HH^+H^-}$ in the region of large $\tan \beta$ 
and low $\MHp$ and, therefore, this region is correspondingly 
constrained by the $B_s\to\mu^+\mu^-$ data.

Based on a parameter scan we investigated several mass and parameter
planes. We demanded agreement with the above given constraints and
evaluated the maximum and minimum values of the various triple Higgs
couplings.
For the SM-type triple Higgs coupling w.r.t.\ its SM value,
$\kala = \lahhh/\laSM$, we roughly find allowed intervals of
$\inter{-0.5}{1.5}$ in the 2HDM type~I and $\inter{0}{1}$ in type~II.
The production of
two SM-like Higgs bosons, both at $pp$ and at $e^+e^-$ colliders depends
already at the tree-level strongly on $\lahhh$. Consequently, over
the possible parameter range of $\lahhh$ the HL-LHC is not expected to
yield a precision on $\kala$ better than $35\%$, and a deviation from
$\lahhh = 0$ can not be established better than $\sim
2\,\sig$. Comparing the HL-LHC to the ILC500, the HL-LHC performs better
(worse) than the ILC500 for $\kala \lsim (\gsim)\, 0.5$.
Combining the ILC500 measurements with the final stage of the
ILC1000, the Linear Collider shows a substantially better result than
the HL-LHC for all the values of $\lahhh$ in the allowed intervals that we have found.

The production of two light Higgs bosons can also depend on $\lahhH$ in
the 2HDM. In this case, 
the prediction in the alignment limit is $\lahhH=0$, but here we
reach the maximum (minimum) value around $\CBA=\pm 0.05$. We find
that the total allowed interval of this coupling is $\inter{-1.4}{1.5}$
for type I and $\inter{-1.6}{1.8}$ for type II.

Concerning the triple Higgs couplings involving two heavy 2HDM Higgs
bosons, we find large allowed values for both 2HDM type~I and~II.
For $\lahHH$, $\lahAA$ and $\lahHpHm$ 
we find maximum values of up to~15, 16 and~32, respectively.
These triple Higgs couplings can have a very strong impact on the heavy
di-Higgs production at $pp$ and $e^+e^-$ colliders. Large coupling values can
possibly facilitate the discovery of such heavier Higgs bosons.  
However, it must be kept in mind that the larger values of triple Higgs couplings of $h$ with two
heavy Higgs bosons are realized for larger values of the respective heavy
Higgs-boson mass. 
Consequently, the effects of the large coupling and the heavy mass
always go in opposite directions. 
A detailed analysis of the various production cross sections will be
analyzed elsewhere~\cite{2hdm-hhh-col}.


\subsection*{Acknowledgements}

\begingroup 
We thank T.~Stefaniak for assistance with \HB\ and \HS.
We also thank A.~Arnan and A.~Pich for helpful discussions
regarding the correct implementation of the Higgs penguin contributions
in $B_s \to \mu^+\mu^-$. The present work has received financial support from the `Spanish Agencia Estatal de Investigaci\'on'' (AEI) and the EU
``Fondo Europeo de Desarrollo Regional'' (FEDER) through the project
FPA2016-78022-P and from the grant IFT
Centro de Excelencia Severo Ochoa SEV-2016-0597.
The work of S.H.\ was also supported in part by the
MEINCOP Spain under contract FPA2016-78022-P and in part by the “Spanish
Red Consolider MultiDark” FPA2017-90566-REDC. 
This work has also received partial funding/support from the European Union’s Horizon 2020 research and innovation programme under the Marie Skłodowska-Curie grant agreement No 674896.
The work of F.A. was also supported by the Spanish Ministry of Science and Innovation via a FPU grant with code FPU18/06634.
\endgroup



\begin{appendices}
\section{Feynman Rules}
\label{appendix:FR}
In this appendix we present the Feynman rules, obtained with the
Mathematica package \cite{Alloul:2013bka}, of the considered triple
  Higgs couplings in the $\la_i$ basis, defined in
\refeq{eq:original_inputs}, and in the physical basis, defined in
\refeq{eq:inputs}. The relation of these Feynman rules with the
dimensionless couplings $\la_{hh_ih_j}$ that have been studied in
this work is given in \refeq{eq:lambda}.
\\

\includegraphics{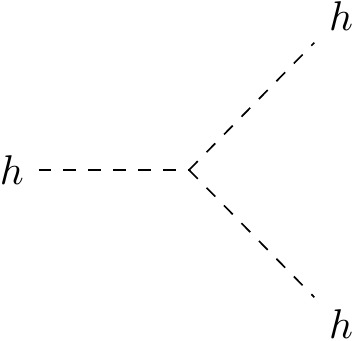}

\noindent$\la_i$ basis:
\begin{flalign} \label{eq:hhh_lambda}
	&=  3iv \Big\{ \la_1c_{\be } s_{\al }^3-\la _2 c_{\al }^3 s_{\be }
	+\left(\la_3+\la _4+\la_5\right) \left(c_{\al }^2 c_{\be } s_{\al }-c_{\al } s_{\al }^2 s_{\be}  \right) \Big\}. &&
\end{flalign}
Physical basis:
\begin{flalign}\label{eq:hhh_phys}
	&= -\frac{3 i}{v} \bigg\{m_h^2 s_{\be -\al }^3 
	+\left(3 m_h^2-2 \bar{m}^2\right) c_{\be -\al }^2 s_{\be -\al }
	+2 \cot 2 \be \left(m_h^2-\bar{m}^2\right) c_{\be -\al }^3\bigg\}. &&
\end{flalign}

\includegraphics{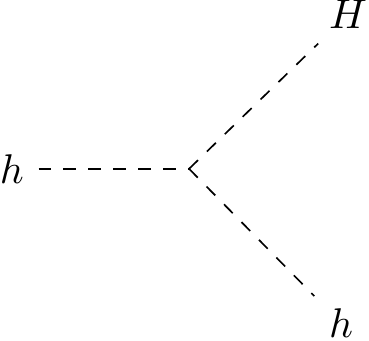}

\noindent$\la_i$ basis:
\begin{flalign}\label{eq:hhH_lambda}
	&= - iv \Big\{  3 \la _1  c_{\al } c_{\be } s_{\al }^2 + 3 \la _2 c_{\al }^2 s_{\al } s_{\be }
	+\left(\la_3+\la _4+\la_5\right) \left(c_{\al }^3 c_{\be }-2 c_{\al }^2 s_{\al }
   s_{\be }-2 c_{\al } c_{\be } s_{\al }^2+s_{\al }^3 s_{\be
   }\right)\Big\}. &&
\end{flalign}
Physical basis:
\begin{flalign}\label{eq:hhH_phys} \notag
	&= \frac{i \CBA}{v} \bigg\{  \left(2\Mh^2+\MH^2-4\mbar\right)\SBA^2
	+2\cot{2\be}\left(2\Mh^2+\MH^2-3\mbar\right) \SBA\CBA \\
	&-\left(2\Mh^2+\MH^2-2\mbar \right) \CBA^2
	 \bigg\}. &&
\end{flalign}

\includegraphics{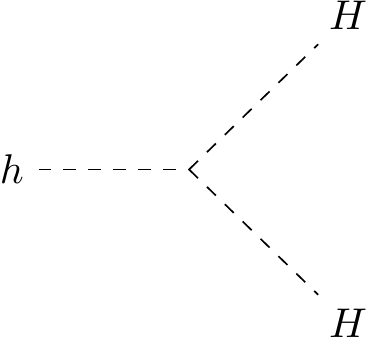}

\noindent$\la_i$ basis:
\begin{flalign}\label{eq:hHH_lambda}
	&=  iv \Big\{  
	3 \la _1 c_{\al }^2 c_{\be } s_{\al }-3 \la _2 c_{\al }
   s_{\al }^2 s_{\be }+\left(\la _3+\la _4+\la _5\right)
   \left(-c_{\al }^3 s_{\be }-2 c_{\al }^2 c_{\be }
   s_{\al }+2 c_{\al } s_{\al }^2 s_{\be }+c_{\be } s_{\al
   }^3\right)\Big\}. &&
\end{flalign}
Physical basis:
\begin{flalign} \label{eq:hHH_phys} \notag
&= -\frac{i\SBA}{v} \bigg\{ \left( \Mh^2+2\MH^2-2\mbar \right)\SBA^2
+2\cot2\be\left( \Mh^2+2\MH^2-3\mbar \right)\CBA\SBA \\
& -\left( \Mh^2+2\MH^2-4\mbar \right)\CBA^2
\bigg\}. &&
\end{flalign}

\includegraphics{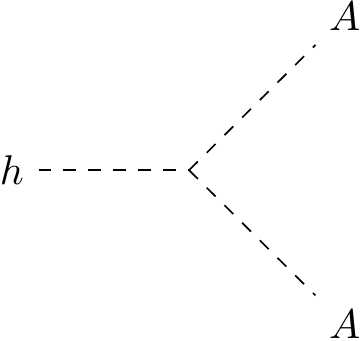}

\noindent$\la_i$ basis:
\begin{flalign} \label{eq:hAA_lambda}
	&=  i v \Big\{ 
	\la _1 c_{\be } s_{\al } s_{\be }^2-\la _2 c_{\al } c_{\be }^2 s_{\be }+\left(\la _3+\la _4\right)
   \left(c_{\be }^3 s_{\al }-c_{\al } s_{\be }^3\right)+\la
   _5 \left(-c_{\be }^3s_{\al }+2 c_{\al } c_{\be
   }^2 s_{\be }-2 c_{\be } s_{\al } s_{\be }^2+c_{\al }
   s_{\be }^3\right) \Big\}. &&
\end{flalign}
Physical basis:
\begin{flalign}\label{eq:hAA_phys}
	&=-\frac{i}{v} \bigg\{ \left(m_h^2 + 2 m_A^2-2 \bar{m}^2\right)s_{\be -\al }+2 \cot 2 \be \left(m_h^2-\bar{m}^2\right) c_{\be -\al } \bigg\}. &&
\end{flalign}

\includegraphics{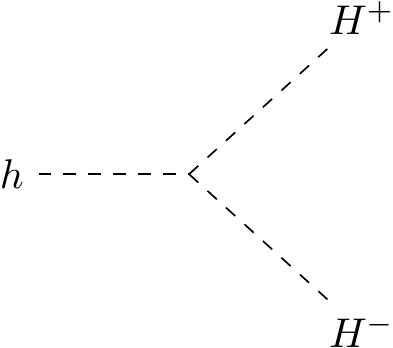}

\noindent$\la_i$ basis:
\begin{flalign}\label{eq:hHpHp_lambda}
	& =  i v \Big\{ 
	\la _1 c_{\be } s_{\al } s_{\be }^2 - \la _2 c_{\al } c_{\be }^2 s_{\be }
	+\la _3 \left(c_{\be }^3 s_{\al }-c_{\al } s_{\be }^3\right)
	+ \left(\la_4+\la_5\right) \left(c_{\al } c_{\be }^2 s_{\be }-c_{\be } s_{\al }   s_{\be }^2\right) \Big\}. &&
\end{flalign}
Physical basis:
\begin{flalign}\label{eq:hHpHp_phys}
	& = -\frac{i}{v} \bigg\{\left(m_h^2+2 m_{H^\pm }^2-2 \bar{m}^2\right)s_{\be -\al } + 2 \cot 2 \be \left(m_h^2-\bar{m}^2\right) c_{\be -\al }\bigg\}. &&
\end{flalign}

\newpage
\end{appendices}


\end{document}